\pdfoutput=1 
\newcommand*{\ATLASLATEXPATH}{latex/}

\PassOptionsToPackage{english}{babel} 

\documentclass[UKenglish, texlive=2016, cernpreprint]{\ATLASLATEXPATH atlasdoc}

\usepackage{\ATLASLATEXPATH atlaspackage}
\usepackage{\ATLASLATEXPATH atlasbiblatex}

\usepackage{\ATLASLATEXPATH atlascontribute}

\usepackage{\ATLASLATEXPATH atlasphysics}

\usepackage{amssymb}

\usepackage{multirow}

\graphicspath{{logos/}{figures/}}

\usepackage{PhotonIDPaper-defs}

\usepackage{datetime}
\usepackage{tabularx,ragged2e}

\AtlasTitle{Measurement of the photon identification efficiencies with the ATLAS detector using LHC Run 2 data collected in 2015 and 2016}

\AtlasAbstract{The efficiency of the photon identification criteria in the ATLAS detector is measured using 36.1~fb$^{-1}$ to 36.7~fb$^{-1}$ of $pp$ collision data at $\sqrt{s}$~=~13~\TeV\ collected in 2015 and 2016. The efficiencies are measured separately for converted and unconverted isolated photons, in four different pseudorapidity regions, for transverse momenta between 10~\GeV\ and 1.5~\TeV. The results from the combination of three data-driven techniques are compared with the predictions from simulation after correcting the variables describing the shape of electromagnetic showers in simulation for the average differences observed relative to data.  Data-to-simulation efficiency ratios are determined to account for the small residual efficiency differences. These factors are measured with uncertainties between 0.5\% and 5\% depending on the photon transverse momentum and pseudorapidity. The impact of the isolation criteria on the photon identification efficiency, and that of additional soft $pp$ interactions, are also discussed.  The probability of reconstructing an electron as a photon candidate is measured in data, and compared with the predictions from simulation.  The efficiency of the reconstruction of photon conversions is measured using a sample of photon candidates from $Z\to\mu\mu\gamma$ events, exploiting the properties of the ratio of the energies deposited in the first and second longitudinal layers of the ATLAS electromagnetic calorimeter.}

\author{The ATLAS Collaboration}

\AtlasRefCode{PERF-2017-02}

\PreprintIdNumber{CERN-EP-2018-216}

\AtlasJournal{Eur. Phys. J. C}

\AtlasDate{12$^{\mathrm{th}}$ March, 2019}

\AtlasJournalRef{Eur. Phys. J. C 79 (2019) 205} 

\AtlasDOI{DOI:10.1140/epjc/s10052-019-6650-6} 

\arXivId{1810.05087}

\AtlasJournal{EPJC}

\hypersetup{pdftitle={ATLAS document},pdfauthor={The ATLAS Collaboration}}

\begin{document}

\maketitle

\section{Introduction}
\label{sec:intro}

Processes with prompt photons in the final state, occurring in proton--proton collisions at the Large Hadron Collider (LHC), play a central role in the ATLAS physics programme. They encompass all phenomena where photons do not originate from hadron decays. These range from non-resonant QCD production, where prompt photons are produced in association with jets or in pairs with cross sections of the order of tens of nanobarns or picobarns respectively, to rarer processes where prompt photons arise from the decay of a heavy particle. The study of QCD prompt photon production at the LHC and the measurement of the corresponding production cross sections allow a test of perturbative and non-perturbative regimes of QCD, and can provide useful information about the proton's parton distribution functions (PDFs) (see for instance Ref.~\cite{STDM-2016-08} for a first measurement at $\sqrt{s}$ = 13 \TeV).  The excellent capability of the ATLAS detector to reconstruct, identify and calibrate prompt photons has proved fundamental to the discovery of the Higgs boson and to the precision measurement of its properties with the $H\to\gamma\gamma$ decay \cite{Aaboud:2018xdt,Aaboud:2018wps}. Similarly, prompt photons are paramount to several searches for phenomena beyond the Standard Model (SM), where they would come from the decay of various new heavy states \cite{Aaboud:2017yyg,SUSY-2016-04,EXOT-2015-12}.

The identification of prompt photons in hadronic collisions is particularly challenging, since the overwhelming majority of reconstructed photon candidates arise from background non-prompt photons from hadron decays in jets, while a smaller fraction of fake candidates are associated with hadrons that deposit significant energy in the electromagnetic calorimeter, mimicking that of real photons.

Prompt photons are identified in the ATLAS experiment by means of selections on quantities describing the shape and properties of the associated electromagnetic showers, and by requiring them to be isolated from other particles in the event. These selections are separately optimised for those photon candidates that convert into an electron--positron pair before reaching the electromagnetic calorimeter (\emph{converted} photons), and those photon candidates that are not associated with a conversion (\emph{unconverted} photons). As already observed using LHC data in Run~1 \cite{PERF-2013-04}, the efficiency of the selection criteria is modelled by Monte Carlo (MC) simulation with only $\mathcal{O}(2$--$5\%)$ accuracy, being mostly limited by an imperfect description of the electromagnetic shower development in the detector. The photon identification efficiency can on the other hand be measured in data with a precision of $\mathcal{O}(1\%)$. Corrections are thus applied to the MC-simulated samples in order to guarantee the highest possible accuracy for photon measurements.

In this paper, the reconstruction and identification of photons by the ATLAS detector are described, and the measurements of the identification efficiency using 36.1~\ifb\ to 36.7~\ifb\ of $pp$ collisions collected at $\sqrt{s}$~=~13~\TeV\ in 2015 and 2016 are reported. These measurements are based on the techniques developed for the photon identification efficiency measurement performed with $\sqrt{s}$~=~7~\TeV\ and 8~\TeV\ data \cite{PERF-2013-04}, while addressing the different beam conditions at $\sqrt{s}$~=~13~\TeV, and exploiting the larger integrated luminosity. The measurements reported in this paper focus on the identification criteria optimised for the data-taking period at $\sqrt{s}$~=~13~\TeV\, that have been revisited relative to those used for the $\sqrt{s}$~=~8~\TeV\ run, and reported in Ref.~\cite{PERF-2013-04}, in order to better cope with the larger average number of interactions per beam bunch crossing~$\mu$.

To overcome the difficulties arising from the absence of a single, pure control sample of prompt-photons over a large range of transverse momentum, three different data-driven techniques are used. The first method selects photons from radiative decays of the $Z$ boson, i.e. $Z\to\ell\ell\gamma$. The second one extrapolates photon properties from electrons and positrons\footnote{In the rest of this paper, \textit{electrons} will be used to refer to both electrons and positrons.} from $Z$ boson decays, by exploiting the similarity of the photon and electron interactions in the ATLAS electromagnetic calorimeter. The third approach exploits a track-based measure of photon isolation to determine the fraction of background present in a sample of isolated photon candidates. Each of these techniques can measure the photon identification efficiency in overlapping parts of complementary \ET\ regions\footnote{The photon transverse momentum \ET\ is defined in footnote\textsuperscript{\ref{NoteCoordinates}}.}. The combination of different measurements in the overlapping regions further improves the photon efficiency precision, which is measured for candidates with transverse momentum ranging from about 10~\GeV\ to about 1.5~\TeV.

The measurement of the rate of misidentification of electrons as photon candidates, as well as the results of a novel technique for measuring the efficiency of reconstructing a photon conversion, only deployed for the $\sqrt{s}$~=~13~\TeV\ data taking, are reported.

This paper is organised as follows. An overview of the ATLAS detector is provided in Section~\ref{sec:ATLAS}. The photon reconstruction and identification algorithms used in ATLAS for the data taking at $\sqrt{s}$~=~13~\TeV\ are detailed in Section~\ref{sec:PhotonRecoID}, highlighting the differences relative to the reconstruction procedure reported in Ref.~\cite{PERF-2013-04} and the properties of the identification criteria optimised for the $\sqrt{s}$~=~13~\TeV\ data taking. Section~\ref{sec:DataMC} summarizes the collision and simulated data samples used for the various measurements, and describes the corrections applied to the simulated photon shower shapes in order to improve agreement with the data. In Section~\ref{sec:measurements} the three data-driven approaches to the measurement of the photon identification efficiency are described, listing their respective sources of uncertainty and the precision reached in the relevant \ET\ ranges. Their combination in the overlapping \ET\ regions is presented, as well as a comparison of the combined data-driven photon identification efficiency with the MC predictions. The impact of the isolation criteria on the photon identification efficiency, and that of additional soft $pp$ interactions, is also discussed.  The measurement of the rate of misidentification of electrons as photon candidates is reported in Section~\ref{sec:EGammaFakeRate}.  The efficiency of reconstructing a photon conversion is summarised in Section~\ref{sec:ConvRecoEff}.

\section{The ATLAS detector}
\label{sec:ATLAS}

The ATLAS experiment  \cite{PERF-2007-01} uses a multipurpose particle detector with approximately forward--backward symmetric cylindrical geometry and nearly $4\pi$ coverage in solid angle.\footnote{\label{NoteCoordinates}ATLAS uses a right-handed coordinate system with its origin at the nominal interaction point (IP) in the centre of the detector and the $z$-axis along the beam pipe. The $x$-axis points from the IP to the centre of the LHC ring, and the $y$-axis points upward. Cylindrical coordinates ($r$, $\phi$) are used in the transverse plane, $\phi$ being the azimuthal angle around the beam pipe. The pseudorapidity is defined in terms of the polar angle $\theta$ as $\eta = -\ln \tan(\theta/2)$. Angular distance is defined as $\Delta R = \sqrt{(\Delta\eta)^2 + (\Delta\phi)^2}$. The transverse momentum of the photon candidates is defined as $\ET$ = $E/\cosh(\eta)$, where $E$ is the candidate energy.} It consists of an inner tracking system surrounded by a thin superconducting solenoid producing a 2 T axial magnetic field, electromagnetic and hadronic calorimeters, and a muon spectrometer incorporating three large toroid magnet assemblies.

The inner detector (ID) tracking system provides position measurements for charged particles in the range $|\eta|<2.5$ by combining information from three subdetectors. The ID consists of a cylindrical central region (full coverage for $|\eta|<1.5$) arranged around the beam pipe, and two endcaps. Disks in the endcap region are placed perpendicular to the beam axis, covering $1.5<|\eta|<2.5$. Starting from the interaction point, the high-granularity silicon pixel detector segmented in $r$--$\phi$ and $z$ covers the vertex region and typically provides four three-dimensional measurements per track.  The ID includes a new innermost layer, the insertable B-layer \cite{Abbott:2018ikt}, with a mean radius of 33 mm, while the remaining three layers of the pixel system are located at mean radii of 50.5, 88.5, and 122.5 mm respectively.  The coverage in the endcap region is enhanced by three disks on either side of the interaction point.  A semiconductor tracker consisting of modules with two back-to-back silicon microstrip sensors with small-angle stereo readout surrounds the pixel detector, providing typically eight two-dimentional hits translating to four three-dimensional measurements, per track at intermediate radii (275 mm $<r<$ 560 mm). The outermost region of the ID (563 mm $<r<$ 1066 mm) is covered by a transition radiation tracker (TRT) consisting of straw drift tubes filled with a gas mixture consisting of about 70\% Xe, 27\% CO$_{2}$ and 3\% O$_{2}$,\footnote{During part of the 2016 data-taking some TRT layers were filled with argon instead of xenon.}  interleaved with polypropylene/polyethylene transition radiators.  The inner detector allows an accurate reconstruction and transverse momentum measurement of tracks from the primary proton–proton collision region. It also identifies tracks from secondary vertices, permitting the efficient reconstruction of photon conversions up to a radial distance of about 80 cm from the beam-line.

The solenoid is surrounded by a high-granularity lead/liquid-argon (LAr) sampling electromagnetic (EM) calorimeter with an accordion geometry. The EM calorimeter (EMC)  measures the energy and the position of electromagnetic showers with $|\eta|<3.2$. It is divided into a barrel section, covering the pseudorapidity region $|\eta|<1.475$, and two endcap sections, covering the pseudorapidity regions $1.375<|\eta|<3.2$. The transition region between the barrel and the endcaps, $1.37 \leq |\eta| < 1.52$, has a large amount of material upstream of the first active calorimeter layer and is not considered for the precision photon measurements reported in this paper. The EM calorimeter is composed, for $|\eta|<2.5$, of three sampling layers, longitudinal in shower depth. The first layer has a thickness of about 4.4 radiation lengths ($X_0$) at $\eta = 0$.\footnote{The depth of the calorimeter layers varies with $\eta$, generally increasing at higher pseudorapidity.} In the ranges $|\eta|<1.4$ and $1.5<|\eta|<2.4$, the first layer is segmented into high-granularity strips in the $\eta$ direction, with a typical cell size of $0.003 \times 0.0982$ in $\Delta\eta \times \Delta\phi$ in the barrel. For $1.4<|\eta|<1.5$ and $2.4<|\eta|<2.5$ the $\eta$ segmentation of the first layer is coarser, and the cell size is  $\Delta\eta \times \Delta\phi$ = $0.025 \times 0.0982$. The fine $\eta$ granularity of the strips is sufficient to provide, for transverse momenta up to $\mathcal{O}(100\,\GeV)$, an event-by-event discrimination between single-photon showers and two overlapping showers originating from the decays of neutral hadrons, mostly $\pi^0$ and $\eta$ mesons in jets, in the fiducial pseudorapidity region $|\eta|<1.37$ or $1.52 \leq |\eta|<2.37$. The second layer has a thickness of about 16 $X_0$ at $\eta$ = 0, and a granularity of $0.025 \times 0.0245$ in  $\Delta\eta \times \Delta\phi$. It collects most of the energy deposited in the calorimeter by photon and electron showers. The third layer has a granularity of $0.05 \times 0.0245$ in  $\Delta\eta \times \Delta\phi$ and a depth of about 2 $X_0$ at $\eta$ = 0. It is used to correct for leakage of high-energy showers into material beyond the EM calorimeter. In front of the accordion calorimeter, a thin presampler (PS) layer, covering the pseudorapidity interval $|\eta|<1.8$, is used to correct for energy loss upstream of the calorimeter. The PS consists of an active LAr layer with a thickness of 1.1 cm (0.5 cm) in the barrel (endcap) and has a granularity of  $\Delta\eta \times \Delta\phi$ = $0.025 \times 0.0982$. The material upstream of the PS has a thickness of about 2 $X_0$ for $|\eta|<0.6$. In the region $0.6<|\eta|<0.8$ this thickness increases linearly from 2 $X_0$ to 3 $X_0$. For $0.8 < |\eta| < 1.8$ the material thickness is about or slightly larger than 3 $X_0$, with the exception of the transition region between the barrel and the endcaps and the region near $|\eta| = 1.7$, where it reaches 5–6 $X_0$. A sketch of a the EM calorimeter's longitudinal and lateral segmentation around $\eta = 0$ is shown in Figure~\ref{fig:Accordion}.

\begin{figure}
  \centering
  \includegraphics[width=0.6\textwidth]{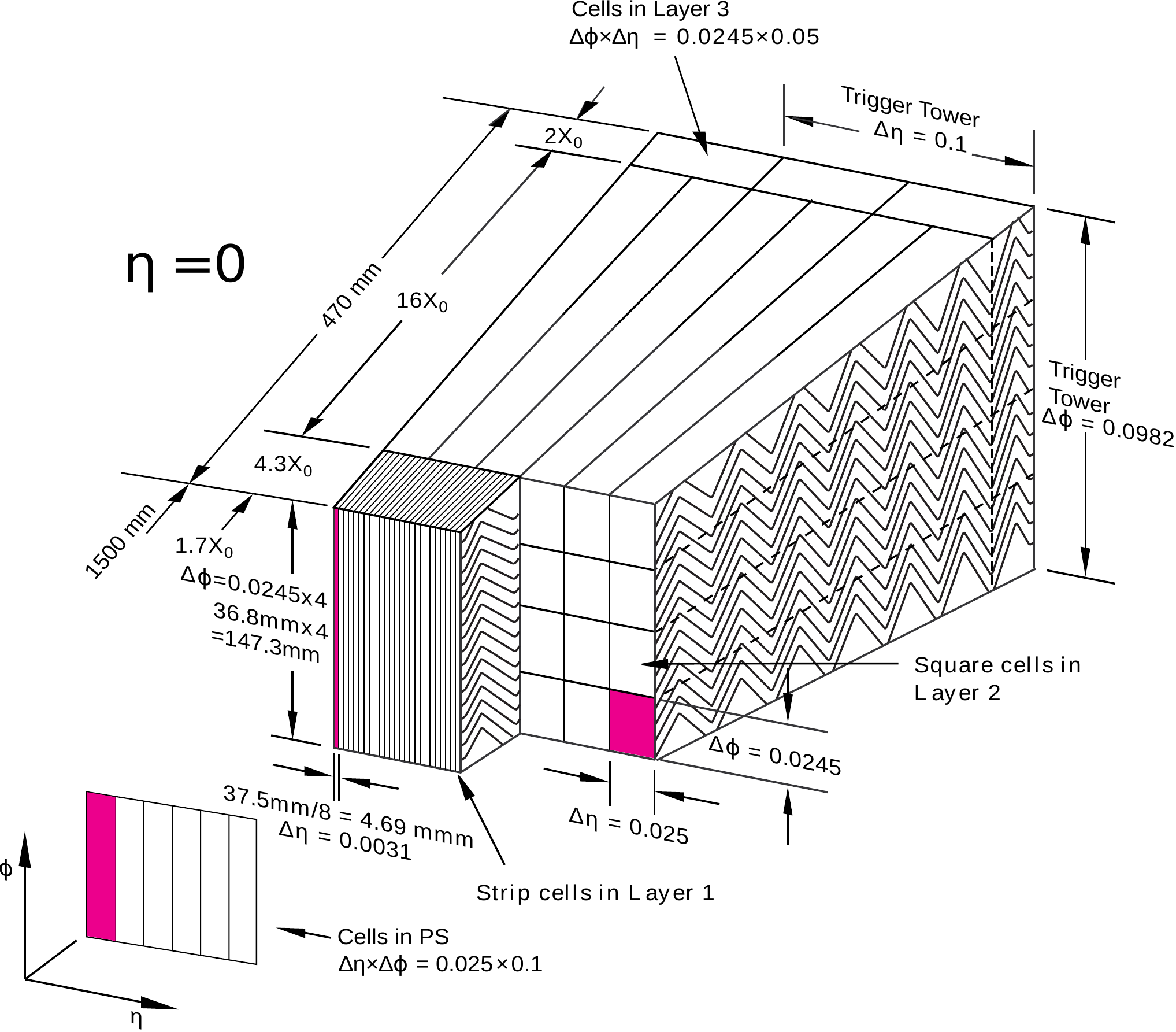}
  \caption{Sketch of the lateral and longitudinal segmentation of the ATLAS electromagnetic calorimeter around $\eta=0$.} 
  \label{fig:Accordion}
\end{figure}

The hadronic calorimeter (HCAL) surrounds the EM calorimeter. It consists of a steel/scintillator tile calorimeter in the central region ($|\eta| < 1.7$), and LAr sampling calorimeters with copper and tungsten absorbers in the endcap ($1.5 < |\eta| < 3.2$) and forward ($3.1 < |\eta| < 4.9$) regions. The muon spectrometer (MS) surrounds the calorimeters. It consists of three large superconducting air-core toroid magnets, each with eight coils, a system of precision tracking chambers ($|\eta| < 2.7$), and fast tracking chambers ($|\eta| < 2.4$) for triggering.

A two-level trigger system, custom hardware followed by a software-based level, is used for online event selection and to reduce the event rate to about 1 kHz for offline reconstruction and storage \cite{TRIG-2016-01}. To reduce the data acquisition rate of low-threshold triggers, used for collecting various control samples, prescale factors $N$ can be applied to each trigger, such that only one in $N$ events passing the trigger causes an event to be accepted at that trigger level.

\section{Photon reconstruction and identification}
\label{sec:PhotonRecoID}

\subsection{Photon reconstruction}
\label{sec:PhotonReco}

The interactions of photons and electrons with the ATLAS EMC produce similar electromagnetic showers, depositing a significant amount of energy in a restricted number of neighbouring calorimeter cells. As photons and electrons have very similar signatures in the EMC, their reconstruction proceeds in parallel. The reconstruction of electron candidates, including a dedicated, cluster-seeded track-finding algorithm to increase the efficiency for the reconstruction of low-momentum electron tracks, is described in Ref.~\cite{ATLAS-CONF-2016-024}. The reconstruction of unconverted and converted photons in Run 2 data collected in 2015 and 2016 is largely unchanged from the reconstruction used Run 1 and described in Ref.~\cite{PERF-2013-04}, and can be summarised as follows:
\begin{itemize}
  
\item A sliding window with a size of 3 $\times$ 5 in units of $\Delta\eta\times\Delta\phi$ = 0.025~$\times$~0.0245, corresponding to the granularity of the EM calorimeter middle layer, is used to search for electromagnetic cluster seeds as longitudinal towers with total cluster transverse energy above 2.5 GeV. The clusters are then formed around the seeds using a clustering algorithm \cite{Lampl:1099735} that allows for removal of duplicates. The cluster kinematics are reconstructed using an extended window depending on the cluster position in the calorimeter. The efficiency of the cluster search in simulation is higher than 99\% for photons with \ET\ > 20 \GeV.
  
\item Tracks reconstructed in the inner detector are loosely matched to seed clusters. Seed clusters that pass loose shower shape requirements in hadronic leakage and energy distribution in $\eta$ are used to create regions-of-interest (ROIs), within which standard track pattern reconstruction \cite{Cornelissen:1020106} is first performed. If the pattern recognition fails for a silicon track seed that is within an ROI, a modified pattern reconstruction algorithm is performed based on a Kalman filter formalism \cite{FRUHWIRTH1987444}, allowing for up to 30\% energy loss at each material intersection. Track candidates are then fitted with the global $\chi^2$ fitter \cite{Cornelissen:2008zza}, allowing for additional energy loss in cases where the standard track fit fails. Tracks with silicon hits loosely matched to EM clusters are re-fitted using a Gaussian-sum filter (GSF) fitter \cite{ATLAS-CONF-2012-047}, a non-linear generalization of the Kalman filter, for improved track parameter estimation.

\item The loosely-matched tracks serve as input to the conversion vertex reconstruction. Tracks with silicon hits (referred to as \emph{Si tracks}) and tracks reconstructed only in the TRT (referred to as \emph{TRT tracks}) are used for the conversion reconstruction. Two-track conversion vertices are reconstructed from two tracks forming a vertex consistent with that of a massless particle, while single-track vertices are built from tracks without hits in the innermost sensitive layers. To increase the converted photon purity, the tracks used to build conversion vertices must generally have a high probability to be electron tracks as determined by the TRT \cite{Aaboud:2017odu}, especially for single-track vertices and conversion vertices constructed from TRT tracks. If there are multiple conversion vertices matched to a cluster, double-track conversions with two silicon tracks are preferred over other double-track conversions, followed by single-track conversions. Within each category, the vertex with the smallest conversion radius is preferred.

\item An arbitration relying on the properties of the tracks and conversion vertices matched to a given electromagnetic cluster is performed, to determine whether an object is reconstructed as an electron, a converted or an unconverted photon, or both as an electron and a photon object in the ambiguous cases: 
clusters to which neither a conversion vertex candidate nor any track has been matched during the electron reconstruction are considered unconverted photon candidates;
clusters matched to a conversion vertex candidate are considered converted photon candidates;
converted photon candidates that are also reconstructed as electrons, the electron track is evaluated against the properties of the track(s) originating from the conversion vertex candidate matched to the same cluster;
unconverted photon candidates are recovered from reconstructed electron candidates depending on the track hits, momentum and $E/p$ properties.
This procedure is discussed in details in Ref.~\cite{PERF-2013-04}.
  
\end{itemize}

Since the analysis reported in Ref.~\cite{PERF-2013-04}, the reconstruction of converted photon candidates has undergone a few changes to improve both reconstruction efficiency and rejection of fake converted photons. Improvements are made especially in track reconstruction and conversion vertex building for TRT tracks:
\begin{itemize}
\item the reconstruction of tracks using the outside-in tracking algorithm~\cite{Cornelissen:1020106} is restricted to ROIs defined by electromagnetic clusters;
\item the efficiency for the reconstruction of double-track TRT conversions is improved by allowing the reconstruction of TRT tracks which share up to 70\% of hits;
\item the fraction of unconverted photons reconstructed as double- or single-track TRT conversions is reduced by tightening the requirements on the TRT tracks: the tracks are required to have at least 25\% precision hits (a precision hit is defined as a hit with a track-to-wire distance within 2.5 times the drift circle uncertainty \cite{Aaboud:2017odu});
\item the determination of the probability of a track to be an electron using high-threshold hit information from the TRT is improved, taking into account the TRT occupancy as a measure of the pile-up level of an event \cite{TRT20152016}.
\end{itemize}
With this improved reconstruction of converted photons, the efficiency to reconstruct a true converted photon\footnote{A true converted photon is defined as a photon undergoing a conversion into an electron--positron pair within a distance $r<80$ cm from the interaction point.} is higher than 70\% for simulated photons with true $\ET > 20\,\GeV$. This efficiency is higher at lower $\mu$ values, being greater than 75\% at $\mu \sim 0$ and decreasing to about 65\% at $\mu \sim 60$. The fraction of true unconverted photons in simulation that are erroneously reconstructed as converted photons is below 9\% for $\mu$ = 60, and decreases with $\mu$\
to become smaller than 1\% for $\mu$ < 24.

The photon energy measurement is performed using information from the calorimeter. The photon energy calibration, which accounts for upstream energy loss and both lateral and longitudinal leakage, is based on the same procedure developed in Run~1 \cite{calib_paper}, but specifically tuned to the Run 2 detector configuration \cite{Aaboud:2018ugz} in order to account for a different amount of material upstream of the EMC, due to the presence of the insertable B-layer. The energy of the electromagnetic clusters associated with the photon candidates is corrected in subsequent steps using a combination of simulation-based and data-driven correction factors, with the calibration regression being separately optimised for converted and unconverted photons. The uniformity corrections and the intercalibration of the longitudinal calorimeter layers are unchanged from to those determined in Run 1.

In the following the photon \ET\ is computed from the photon cluster’s calibrated energy $E$ and the pseudorapidity $\eta$ of the barycentre of the cluster in the second layer of the EMC as $\ET = E / \cosh(\eta)$. 

\subsection{Photon identification}
\label{sec:PhotonID}

The identification of photon candidates in ATLAS relies on rectangular cuts using calorimetric variables which deliver good separation between prompt-photons and fake signatures from non-prompt-photons originating from the decay of neutral hadrons in jets, or QCD jets depositing a large energy fraction in the EMC. Such variables, listed in Table~\ref{tab:IsEM-DV} and depicted in Figure~\ref{fig:IsEM-DV} with their respective definitions, characterize the lateral and longitudinal electromagnetic shower development in the EMC and the shower leakage fraction in the HCAL.\footnote{The $\Rhadone$ variable was initially used by ATLAS along the whole $\eta$ acceptance range~\cite{CSC_PhotonReco}; however, the use of the normalised total hadronic energy $\Rhad$ is found to be more effective in discriminating hadronic showers in the region 0.8 $< |\eta| <$ 1.37~\cite{ATL-PHYS-PUB-2011-007}.} Prompt-photons typically produce narrower energy deposits in the EMC and have smaller leakage to the HCAL compared to background photons from jets. Additionally, background candidates from $\pi^0\to\gamma\gamma$ decays are often characterized by two separate local energy maxima in the finely segmented strips of the EMC first layer.

Two reference sets of cuts – \emph{loose} and \emph{tight} – are specifically defined for the $pp$ data collected at $\sqrt{s}$ = 13 \TeV\ in 2015 and 2016. While the same set of discriminating variables employed by the photon identification in Run 1 \cite{PERF-2013-04} are used, the selection cuts are tuned to reduce the dependency of the identification efficiency on pile-up, in order to cope with the harsher Run 2 conditions. This mostly results in looser selections for converted photons, where broader electromagnetic showers tend to be more affected by the larger number of interactions per beam bunch crossing.

The loose selection is based on shower shapes in the second layer of the electromagnetic calorimeter and on the energy deposited in the hadronic calorimeter. The tight selections add information from the finely segmented strip layer of the calorimeter, and are separately optimised for unconverted and converted photons, to account for the generally broader lateral shower profile of the latter. The thresholds of the selection criteria are different in seven intervals of the reconstructed photon $|\eta|$ ($0.0$--$0.6$, $0.6$--$0.8$, $0.8$--$1.15$, $1.15$--$1.37$, $1.52$--$1.81$, $1.81$--$2.01$, $2.01$--$2.37$) to account for the calorimeter geometry, and for different effects on the shower shapes from the material upstream of the calorimeter.

The distributions of the discriminating variables for both the prompt and background photons are affected by additional soft $pp$ interactions that may accompany the hard-scattering collision, referred to as in-time pile-up, as well as by out-of-time pile-up arising from bunches before or after the bunch where the event of interest was triggered. Pile-up collisions result in the presence of low-\et\ activity in the detector, including energy deposits in the EMC. A greater number of superimposed $pp$ events, $\mu$, would generally broaden the photon shower shapes because of these additional energy deposits in the calorimeter, thus resulting in a lower identification efficiency for larger $\mu$ values, as discussed in Section~\ref{sec:PileUp}.

\begin{figure}
  \includegraphics[width=1.\textwidth]{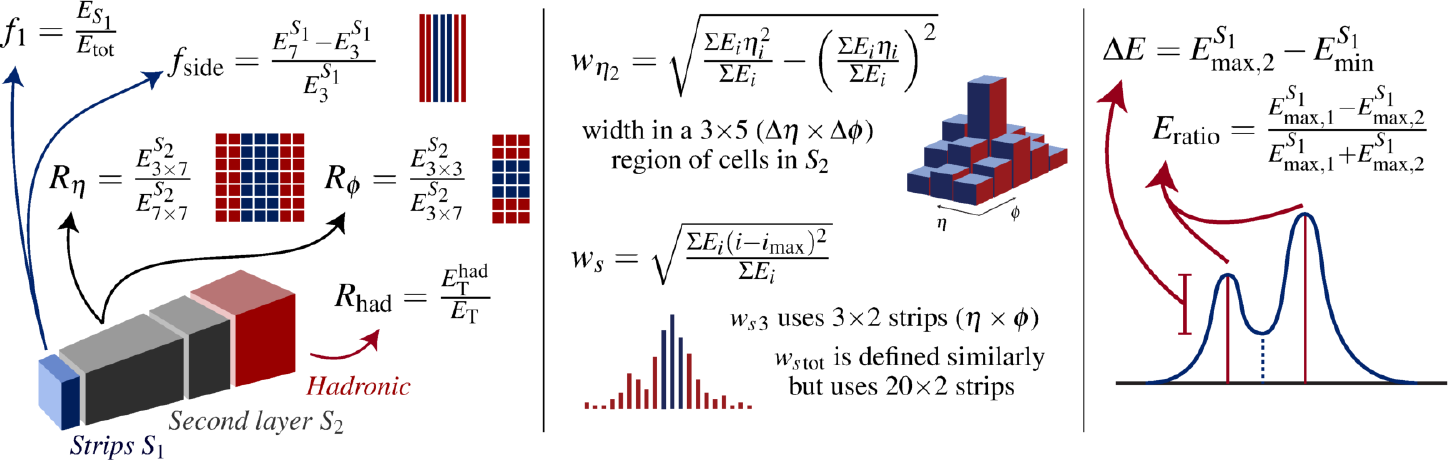}
  \caption{Schematic representation of the photon identification discriminating variables, from Ref.~\cite{Saxon:1746004}. $E^{S_N}_C$ identify the electromagnetic energy collected in the $N$-th longitudinal layer of the electromagnetic calorimeter in a cluster of properties $C$, identifying the number and/or properties of selected cells. $E_i$ is the energy in the $i$-th cell, $\eta_i$ the pseudorapidity centre of that cell.}
  \label{fig:IsEM-DV}
\end{figure}

\begin{table*}[tp]
  \centering
  \caption{Discriminating variables used for \textit{loose} and \textit{tight} photon identification.}
  \def\arraystretch{1.5}
  \small
  \begin{tabular}{
    l
    >{\RaggedRight}p{0.55\textwidth}
    l|cc}
  \hline \hline
  Category & Description & Name & \textit{loose} & \textit{tight} \\ 
  \hline \hline
    Acceptance
           & $|\eta|<2.37$, with $1.37 \leq |\eta|< 1.52$ excluded  & -- & $\checkmark$ & $\checkmark$ \\
    Hadronic leakage
           & Ratio of $\et$ in the first sampling layer of the hadronic calorimeter to $\et$ of the 
             EM cluster (used over the range $|\eta| < 0.8$ or $|\eta| > 1.52$)  & $\Rhadone$ & $\checkmark$ & $\checkmark$ \\ 
           & Ratio of $\et$ in the hadronic calorimeter to $\et$ of the EM cluster 
             (used over the range $0.8 < |\eta| < 1.37$) & $\Rhad$ & $\checkmark$ & $\checkmark$ \\
    EM middle layer
           & Ratio of the energy in 3 $\times$ 7 $\eta\times\phi$ cells over the energy in  7~$\times$~7 cells centered around the photon cluster position & $\Reta$ & $\checkmark$ & $\checkmark$ \\
           & Lateral shower width, 
             $\sqrt{(\Sigma E_i \eta_i^2)/(\Sigma E_i) - ((\Sigma E_i \eta_i)/(\Sigma E_i))^2}$, 
             where $E_i$ is the energy and $\eta_i$ is the pseudorapidity of cell $i$ and the sum is calculated within a window of  3~$\times$~5 cells & $\wetatwo$ & $\checkmark$ & $\checkmark$ \\
           & Ratio of the energy in 3~$\times$~3 $\eta\times\phi$ cells over the energy of 3~$\times$~7 cells centered around the photon cluster position & $\Rphi$  & & $\checkmark$ \\
    EM strip layer
           & Lateral shower width,
             $\sqrt{(\Sigma E_i (i - i_{\textrm{max}})^2)/(\Sigma E_i)}$,
             where $i$ runs over all strips in a window of 3~$\times$~2 $\eta\times\phi$ strips, and $i_{\textrm{max}}$ is the index of the highest-energy strip
             calculated from three strips around the strip with maximum energy deposit  & $\wthree$ & & $\checkmark$ \\
           & Total lateral shower width
             $\sqrt{(\Sigma E_i (i - i_{\textrm{max}})^2)/(\Sigma E_i)}$,
             where $i$ runs over all strips in a window of 20~$\times$~2 $\eta\times\phi$ strips, and $i_{\textrm{max}}$ is the index of the highest-energy strip
             measured in the strip layer & $\wtot$  & & $\checkmark$ \\
           & Energy outside the core of the three central strips but within seven strips divided by energy within the three central strips & $\Fside$  & & $\checkmark$ \\ 
           & Difference between the energy associated with the second maximum in the strip layer
             and the energy reconstructed in the strip with the minimum value found between the first and second maxima  & $\DeltaE$ & & $\checkmark$ \\
           & Ratio of the energy difference between the maximum energy deposit and the energy deposit in the secondary maximum in the cluster to the sum of these energies & $\Eratio$  & & $\checkmark$\\
           & Ratio of the energy in the first layer to the to the total energy of the EM cluster & $\fone$ & & $\checkmark$\\
      \hline \hline
  \end{tabular}
  \label{tab:IsEM-DV}
\end{table*}

\subsection{Photon isolation}
\label{sec:PhotonIso}

The identification efficiencies presented in this paper are measured for photon candidates passing an isolation requirement, similar to those applied to reduce hadronic background in prompt-photon cross-section measurements \cite{STDM-2016-08}, $H\to\gamma\gamma$ measurements \cite{Aaboud:2018xdt,ATLAS-CONF-2017-046}, or searches for exotic processes with photons \cite{Aaboud:2017yyg,SUSY-2016-04,EXOT-2015-12}.  The choice of a specific isolation criterion is determined by the actual physics analysis, since it depends on the different background sources, the signal-to-background ratio, and the background rejection needs. On the other hand, it is shown in Section~\ref{sec:PIDvsIso} that the photon identification efficiency does not show a significant dependence on the chosen isolation criterion. Additionally, it is shown in Section~\ref{sec:measurements:comb} that the corrections meant to address the mismodelling by simulation of the photon identification efficiency measured in data (\emph{scale factors}) do not depend, within uncertainties, on the physics process used to measure it and the isolation criterion of choice.

The definition of photon isolation in ATLAS is based on the transverse energy in a cone with angular size $\Delta R$ around the direction of the photon candidate. This transverse energy is characterized by two quantities, the calorimeter isolation and the track isolation. The calorimeter isolation \Etiso\ is obtained from the sum of transverse energies of topological clusters~\cite{Lampl:1099735} in the calorimeters, after subtracting on an event-by-event basis the energy deposited by the photon candidate and the contribution from the underlying event and pile-up. This uses the method described in Refs.~\cite{Cacciari:area, Cacciari:UE, Cacciari:PU} and is discussed in more detail in Ref.~\cite{PERF-2013-04}. The track isolation \ptiso\ is obtained by summing the transverse momenta of all the tracks with transverse momentum above 1~\GeV\ and having a distance of closest approach to the primary vertex~\cite{PERF-2015-01} along the beam axis $|z_{0}\sin{\theta}| < 3\,\mathrm{mm}$, and excluding the tracks associated with photon conversions.

ATLAS analyses selecting final-state photons use a variety of isolation selection criteria. The most commonly adopted are a \emph{loose} isolation requirement, based on both the calorimeter isolation and the track isolation, in both cases computed in a cone with $\Delta R=0.2$:
\begin{equation*}
  \left.\Etiso\right|_{\Delta R<0.2}<0.065\cdot\ET \,\text{and}\, \left.\ptiso\right|_{\Delta R<0.2}<0.05\cdot\ET \,;
  \label{eq:FixedCutLoose}
\end{equation*}
a \emph{tight} isolation requirement, based on the calorimeter isolation computed in a cone with $\Delta R = 0.4$, and the track isolation computed in a cone with $\Delta R=0.2$: 
\begin{equation*}
  \left.\Etiso\right|_{\Delta R<0.4} <0.022\cdot\ET + 2.45\,\GeV \,\text{and}\, \left.\ptiso\right|_{\Delta R<0.2}<0.05\cdot\ET \,;
  \label{eq:FixedCutTight}
\end{equation*}
an alternative version of the tight isolation requirement (\emph{calorimeter-only tight}), based only on the calorimeter isolation:
\begin{equation*}
  \left.\Etiso\right|_{\Delta R<0.4} <0.022\cdot\ET + 2.45\,\GeV \,;
  \label{eq:FixedCutTightCaloOnly}
\end{equation*}
and a legacy isolation requirement, requiring a fixed selection on the calorimeter isolation:
\begin{equation*}
  \left.\Etiso\right|_{\Delta R<0.4} < 4\,\GeV \,.
  \label{eq:FixedCutLegacy}
\end{equation*}
The data/MC corrections to the electromagnetic shower shape variables discussed in Section~\ref{sec:DataMC} are computed using photon candidates satisfying the calorimeter-only tight isolation criterion. The measurements of photon identification efficiency reported in Section~\ref{sec:measurements} are performed for isolated photon candidates meeting the loose criterion, apart from the measurement using radiative $Z$ decays, which is nominally performed for the tight criterion, and repeated using the loose isolation and the calorimeter-only tight isolation criteria in order to evaluate the potential dependency of the identification efficiency on the photon isolation (Section~\ref{sec:PIDvsIso}). The measurement of the electron-to-photon fake rate discussed in Section~\ref{sec:EGammaFakeRate} is performed for isolated photon candidates satisfying the loose criterion, and its dependency on the isolation selection verified on candidates meeting the tight and calorimeter-only tight criteria. The conversion reconstruction efficiency reported in Section~\ref{sec:ConvRecoEff} is performed for isolated photon candidates satisfying the legacy criterion.

\section{Collision and simulated data samples}
\label{sec:DataMC}

The measurements presented in this paper use proton--proton ($pp$) collisions at $\sqrt{s}$ = 13\,\TeV\ recorded by the ATLAS detector in 2015 and 2016 during the LHC Run 2. The data are required to pass good quality requirements on the detector performance and object reconstruction, leading to 36.1~\ifb\ of integrated luminosity. The inclusive photon measurement discussed in Section~\ref{sec:measurements:MM} relaxes the requirement on the performance of the ATLAS muon spectrometer, and uses 36.7~\ifb\ of integrated luminosity. In these datasets the mean number of interactions per bunch crossing 
is 13.5 in 2015 data and 24.9 in 2016 data.

Two of the methods used to measure the photon identification efficiency described in Section~\ref{sec:measurements} rely on the use of $Z$ boson decays into electron--positron pairs $Z\to e^+e^-$ and on $Z$ boson radiative decays $Z\to\ell\ell\gamma$ ($\ell=e,\mu$): these events are selected in data collected with the lowest-threshold unprescaled lepton triggers. The single-electron trigger has a transverse momentum threshold of 24\,\GeV\ in 2015 and in most of 2016, and of 26\,\GeV\ in the last data-taking period of 2016; the single-muon trigger uses a transverse momentum threshold increased from 20\,\GeV\ to 26\,\GeV\ in 2015 depending on the instantaneous luminosity, and of 24\,\GeV\ in 2016. The dielectron trigger has a transverse momentum threshold of 9\,\GeV\ in 2015 and in most of 2016, and of 10\,\GeV\ in the last data-taking period of 2016; the dimuon trigger has a transverse momentum threshold of 8\,\GeV\ in both 2015 and 2016. The third method for measuring the photon identification efficiency uses a sample of single-photon candidates, selected in data from events collected with single-photon triggers with loose identification requirements and large prescale factors, thus exploiting only a fraction of the total luminosity. The lowest transverse momentum threshold of these single-photon triggers is 10\,\GeV\ in both 2015 and 2016.

Simulated MC samples of prompt-photon production were generated with \textsc{Pythia8}~\cite{Sjostrand:2007gs,Sjostrand:2006za}. 
Such samples include the leading-order $\gamma$ + jet events from $q g \rightarrow q \gamma$ and $q \bar{q} \rightarrow g \gamma$ hard scattering, as well as prompt-photons from quark fragmentation in QCD dijet events. Samples of background photons in jets were produced by generating with \textsc{Pythia8} all tree-level 2$\rightarrow$2 QCD processes, removing $\gamma$ + jet events from quark fragmentation. Simulated samples of $Z\to \ell\ell\gamma$ ($\ell=e,\mu$) events were generated with \textsc{Sherpa}~\cite{sherpa2} or with \textsc{Powheg-Box}~\cite{Frixione:2007vw,Alioli:2010xd} interfaced to \textsc{Photos}~\cite{Golonka:2005pn} for the modelling of QED final-state radiation and to \textsc{Pythia8} for showering, hadronization and modelling of the underlying event.  $Z(\to\ell\ell)$+jet MC events were generated for both $\ell=e$ and $\ell=\mu$ with \textsc{Sherpa}.  All MC samples were processed through a full simulation of the ATLAS detector response~\cite{ATLASsimulation} using \textsc{Geant4}~\cite{GEANT4}.  Pile-up $pp$ interactions in the same and nearby bunch crossings are included in the simulation.  MC samples were reweighted to reproduce the distribution of $\mu$ observed in data.

\subsection{Data-driven corrections to shower shapes in simulated data}
\label{sec:DataMC:FF}

The distributions of the photon transverse shower shapes in the ATLAS MC simulation do not perfectly describe those observed in data. While these distributions in simulation are rather similar in shape to those found in the data, small systematic differences in their average values are observed, pointing to a mismodelling in MC simulation of the lateral profile development of the electromagnetic showers, while, overall, the longitudinal electromagnetic shower profiles are well described. These differences between data and MC distributions are measured and parameterised as simple shifts to be applied to the MC-simulated values to align with the distributions observed in data. The shifts are calculated by minimizing the $\chi^2$ between the data and the shifted MC distributions of photon candidates satisfying the \textit{tight} identification criteria and the calorimeter isolation requirement described in the previous section. The shifts are computed in intervals of the reconstructed photon pseudorapidity and transverse momentum. The pseudorapidity intervals are the same as those used to define the photon selection criteria, while the \ET\ bin boundaries are 8, 15, 20, 25, 30, 40, 50, 60, 80, 100, 250, and 1000\,\GeV. Photon candidates from $Z\to\ell\ell\gamma$ events are used for $\ET~<~50~\GeV$, while candidates from single-photon events are used for $\ET~>~50~\GeV$. The correction factors are measured from both types of events in the overlapping \ET\ region around 50\,\GeV\ and found to be compatible.

The typical size of the correction is 10\% of the root-mean-square of the distribution of the corresponding variable in data. The corresponding correction to the photon efficiency predicted by simulation varies with pseudorapidity between --10\% and --5\% for photon transverse momenta close to 10\,\GeV, and approaches zero for transverse momenta above 50\,\GeV.
Examples of the simulated discriminating variable distributions before and after corrections, for converted and unconverted photon candidates originating from $Z$ boson radiative decays, are shown in Figure~\ref{fig:FF_examples}. For comparison, the distributions observed in data for candidates passing the $Z$ boson radiative decay selection illustrated in Section~\ref{sec:measurements:RadZ}, are also shown. Improved agreement between the shower shape distributions in data and simulation after applying such corrections is clearly visible. Residual discrepancies are observed in the tail of the distributions. Their effect on the MC description of the photon identification efficiency is addressed with data/MC scale factors.
Similarly, while there are correlations between variables and this might not be perfectly described by simulation, these simple corrections do not attempt to address such a potential mismodelling, whose impact would instead be collectively taken into account with the same data/MC scale factors mentioned above.

\begin{figure}
   \centering
   \includegraphics[width=0.49\textwidth]{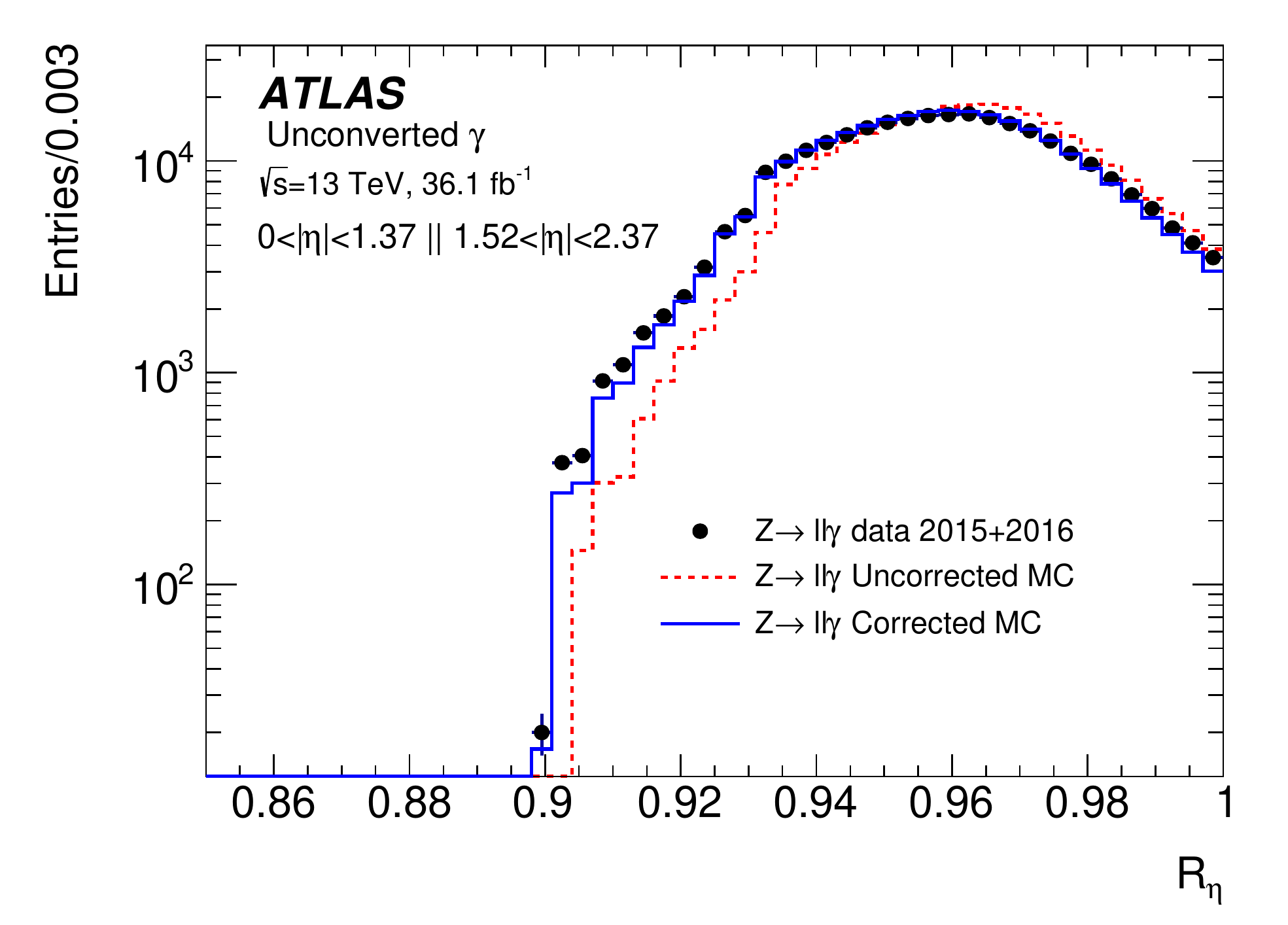}
   \includegraphics[width=0.49\textwidth]{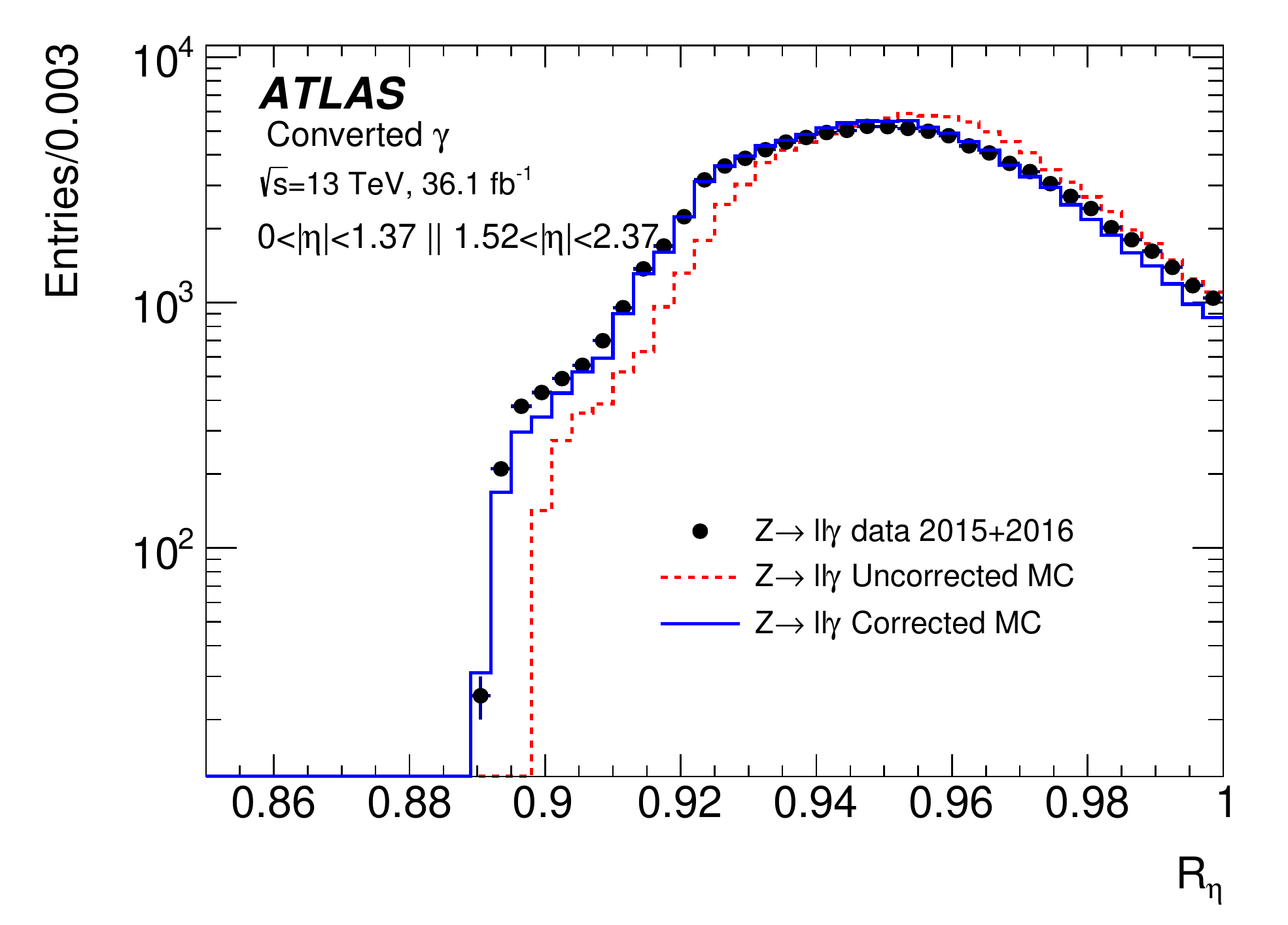} \\
   \includegraphics[width=0.49\textwidth]{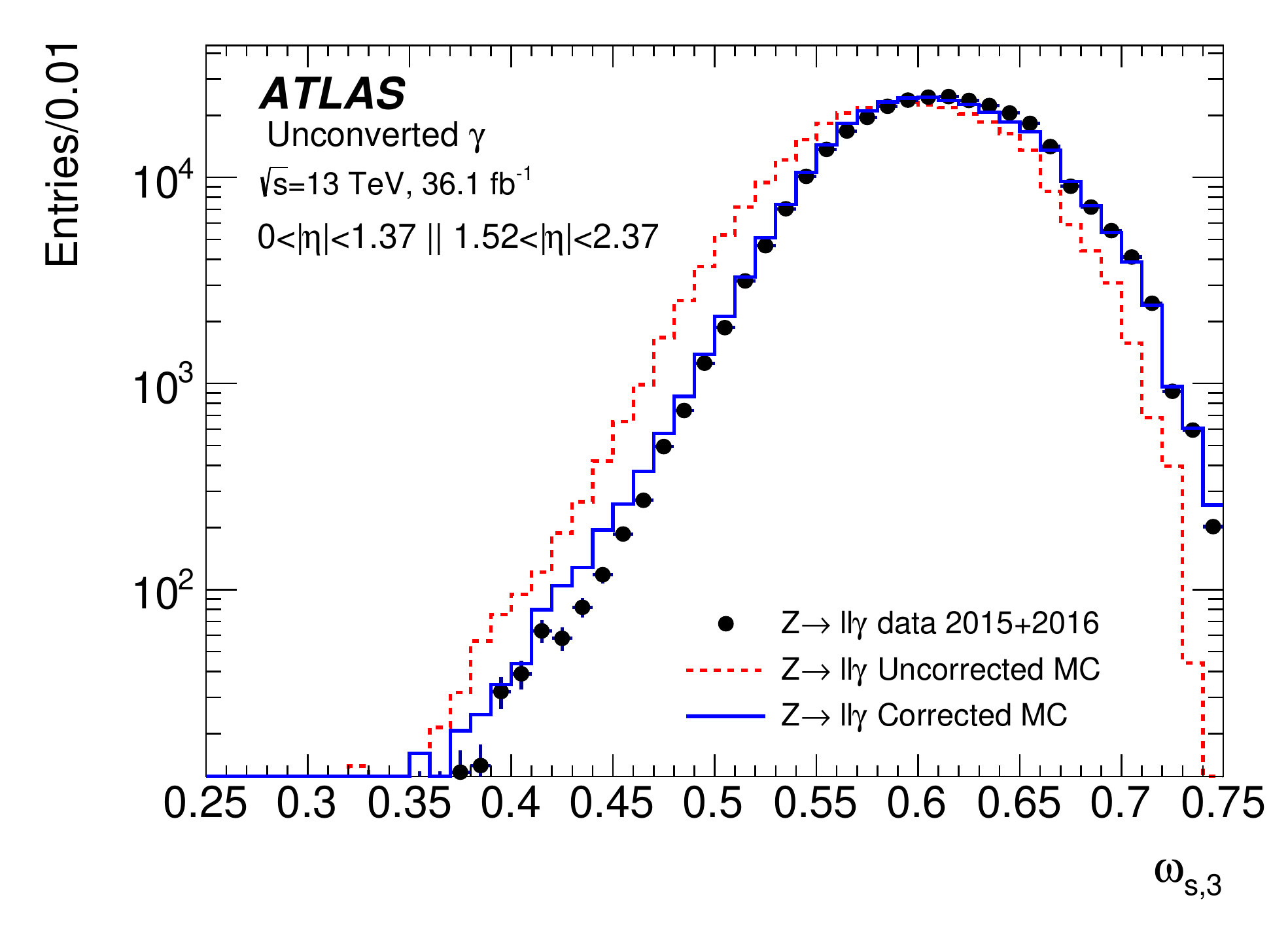}
   \includegraphics[width=0.49\textwidth]{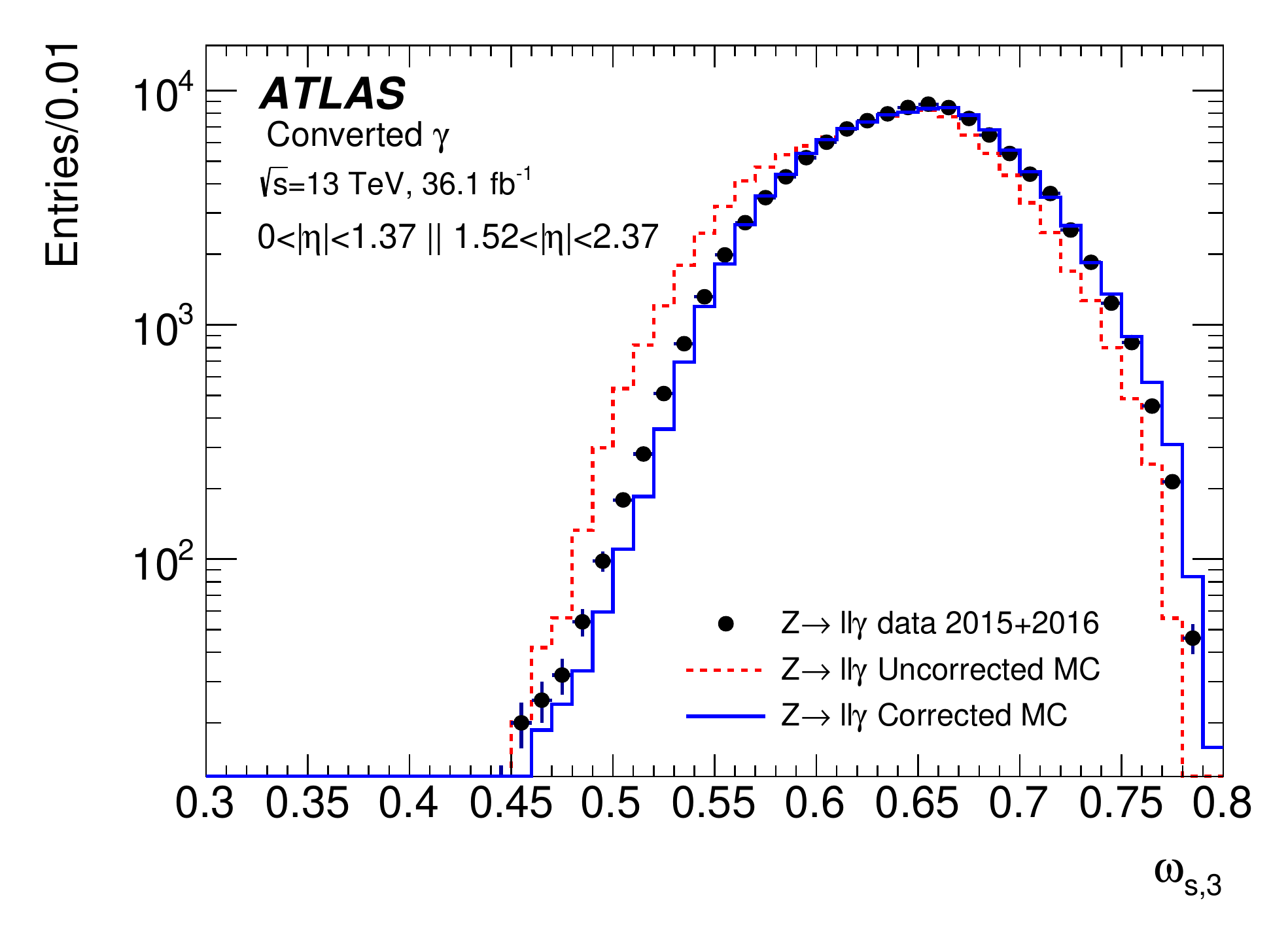}   
   \caption{Distributions of the calorimetric discriminating variables $\Reta$ and $\wthree$ for converted and unconverted photon candidates with $\ET \in [10,50]~\GeV$ and $|\eta| <2.37$ (excluding $1.37\leq|\eta|<1.52$) selected from $\ell^+\ell^-\gamma$ events (black dots). The distributions for true photons from simulated $Z\to\ell\ell\gamma$ events are also shown.  for the uncorrected simulation (dashed red line) and the simulation corrected by the average shift between data and simulation distributions (solid blue line). The definition of the plotted variables is given in Table~\ref{tab:IsEM-DV} and Figure~\ref{fig:IsEM-DV}.}
   \label{fig:FF_examples} 
\end{figure}

In all the analyses described in Section~\ref{sec:measurements} the reference simulated samples are modified with the correction factors described above, while in Section~\ref{sec:measurements:comb} the measured photon identification efficiencies are compared with the values in both the uncorrected and corrected MC samples.

\section{Measurements of the photon identification efficiency}
\label{sec:measurements}

The efficiency \effID\ of the tight photon identification criterion described in Section~\ref{sec:PhotonID} is measured in data using three methods:
\begin{itemize}
\item \textit{Radiative Z decays}: this method uses a clean sample of low-energy photons obtained from $Z \rightarrow \ell \ell \gamma$ decays ($\ell = e,\mu$). This allows measurements of \effID\ from $\ET = 10\,\GeV$, below which photons are not reconstructed, to $\ET \sim 100\,\GeV$, beyond which event yields are insufficient. The method is described in detail in Section~\ref{sec:measurements:RadZ} below.
\item \textit{Electron extrapolation}: this method uses a sample of electromagnetic showers from electrons originating from $Z \rightarrow ee$ decays, identified using a tag-and-probe method. These showers are modified so that their shape information matches the properties of photon showers, and used to measure \effID\ in the region $25 < \ET < 150\,\gev$ where sufficient numbers of $Z \rightarrow ee$ electron candidates are available. The method is described in detail in Section~\ref{sec:measurements:EE} below.
\item \textit{Inclusive photons}:\footnote{Called \emph{Matrix Method} in Ref.~\cite{PERF-2013-04}.}
this method uses an inclusive photon sample collected using single-photon triggers. The efficiency of a tight track-based isolation criterion is used to obtain the fraction of prompt-photons in the full sample and in the subsample satisfying the tight identification criterion, from which a measurement of \effID\ can be derived. The measurement is performed over a wide kinematic range spanning $25\,\gev < \ET < 1.5\,\tev$. At low transverse energy this is limited by the prescaling of single-photon triggers below $\ET = 140\,\gev$, and at high transverse energy by limited event yields. The method is described in detail in Section~\ref{sec:measurements:MM} below.
\end{itemize}
The efficiencies are reported in each case for converted and unconverted photons separately, since their distributions for the discriminating variables listed in Table~\ref{tab:IsEM-DV} typically differ due to differences in electromagnetic shower development. Efficiencies are measured in a two-dimensional interval grid in photon $\ET$ and $|\etacal|$ with boundaries at 10, 15, 20, 25, 30, 35, 40, 45, 50, 60, 80, 100, 125, 150, 175, 250 and $1500\,\gev$ and 0, 0.6, 1.37, 1.81 and 2.37 respectively; each method covers only a subspace of this region, as described above.

The various methods provide measurements covering overlapping kinematic ranges, where the measured values of the photon identification efficiency can be compared. Because of the different compositions of the prompt-photon samples used to measure the efficiencies (i.e.\ the varying fraction of photons originating from fragmentation processes), the three methods are not necessarily expected to provide the same efficiency values for the same photon \et\ and $\eta$. On the other hand, one expects any residual mismodelling of the photon identification efficiency by MC simulation to be independent of the physics process used to measure the efficiency, since it would mostly be due to an imperfect modelling of the detector response. For this reason, the corrections for this mismodelling are expected to be universal, and the correction values obtained from the various methods are therefore combined for increased precision (see Section~\ref{sec:measurements:comb}).

\subsection{Photons from $Z$ boson radiative decays}
\label{sec:measurements:RadZ}

Radiative $Z$ decays are selected by requiring the presence of a photon candidate and an opposite-charge pair of electron or muon candidates. Photon candidates are required to have a transverse momentum $\ET > 10\,\gev$ and a pseudorapidity in the range $|\etacal| < 1.37$ or $1.52 \leq |\etacal| < 2.37$. The loose isolation selection described in Section~\ref{sec:PhotonIso} is applied for the nominal results, while the effect of applying alternative isolation criteria is studied in Section~\ref{sec:PIDvsIso}. No other selection is applied to the photon, in order to avoid biases due to the photon selection in the efficiency measurement.

The events are selected using unprescaled single-lepton and dilepton triggers with the lowest transverse momentum threshold.

Muons candidates are required to be reconstructed from hits in both the MS and the ID~\cite{PERF-2015-10}. They must fulfil the conditions $\ET > 10\,\gev$ and $|\etacal|<2.5$, and the impact parameters of their track must be compatible with originating from the primary event vertex. They must fulfil the \textit{medium} identification criterion~\cite{PERF-2015-10}, which is based on the overall quality of the track fit and the compatibility of track parameters measured in the ID and the MS, with selection cuts chosen to be 99\% efficient. They must also satisfy the \textit{loose} isolation criterion~\cite{PERF-2015-10}, defined similarly to those described for photons in Section~\ref{sec:PhotonIso}.

Electron candidates are required to have $\ET > 10\,\gev$ and $|\etacal|<2.47$, excluding the barrel--endcap transition region, and their track must fulfil loose impact parameter selections. They are required to satisfy the \textit{medium} identification criterion~\cite{ATLAS-CONF-2016-024}, which relies mainly on information about the shape of the associated cluster in the electromagnetic calorimeter and transition radiation emission in the TRT. Electrons are also required to meet the \textit{loose} isolation criterion~\cite{ATLAS-CONF-2016-024} similar to the one described in Section~\ref{sec:PhotonIso}.

Radiative decays are selected by requiring $40 < m_{\ell \ell} < 83\,\gev$ and $80 < m_{\ell \ell \gamma} < 100\,\gev$, were $m_{\ell \ell}$ is the invariant mass of the dilepton system and $m_{\ell \ell \gamma}$ that of the two leptons and the photon.  These selections are meant to isolate the radiative decays from events where the photon originates from initial state radiation.  Distributions of the $ m_{\ell \ell}$ and $ m_{\ell \ell \gamma}$ quantities for the electron channel are shown in the left panel of Figure~\ref{fig:RadZ_illustration}. Separations $\Delta R > 0.2$ and $\Delta R > 0.4$ are required between the photon and the closest muon and electron candidate respectively, in order to avoid biases in the photon shower shape and isolation variables.  About $170\,000$ unconverted photons and $60\,000$ converted photons are found to pass all selections in the \mmg\ channel, and about $90\,000$ and $30\,000$ respectively in the \eeg\ channel.

A small background contamination occurs due to $Z \rightarrow \ell\ell$ decays accompanied by a jet which is misidentified as a photon, particularly if the photon candidate has low transverse momentum. The size of these contributions is estimated using a fit of the $m_{\ell \ell \gamma}$ shape in data over the range $65 < m_{\ell \ell \gamma} < 105\,\gev$. The model uses signal and background shapes obtained in simulated samples of $Z\to\ell^+\ell^-\gamma$ and $Z\to\ell^+\ell^-$ + jet production respectively, described in Section~\ref{sec:DataMC}. The signal and background yields are determined by the fit, an example of which is shown in the right panel of Figure~\ref{fig:RadZ_illustration} for the electron channel. In the region $10 < \ET < 25\,\gev$, the \textit{purity} of the selection, defined as the ratio of the $Z \rightarrow \ell \ell \gamma$ yield to the total sample size, is measured to be about 82\% in the \mmg\ channel and 86\% in the \eeg\ channel. After applying photon identification cuts, this purity becomes about 94\% in the \mmg\ channel and 97\% in the \eeg\ channel.  The photon identification efficiency is then computed as $\effID = (P\pass N\data\pass)/(P\total N\data\total)$, where $N\data\total$ ($P\total$) and $N\data\pass$ ($P\pass$) are the numbers of events (purities) in the full sample and the subset in which the photon passes the identification cuts respectively. The computation is performed separately for the \mmg\ and \eeg\ channels in $(\ET, |\etacal|)$ intervals ranging up to $\ET = 100\,\gev$. For $\ET > 25\,\gev$, the purity in the full sample is above 96\% and no correction is applied, but a systematic uncertainty is included to account for the residual background level. The \mmg\ and \eeg\ results are combined as discussed in Section~\ref{sec:measurements:comb}.

The following sources of systematic uncertainties are considered:
\begin{itemize}

\item A closure test is made by performing the measurement on a sample consisting of known fractions of simulated signal and background events. This test is only performed for $\ET < 25\,\gev$, because of the the limited number of MC events at higher transverse momenta. Deviations from the true identification efficiency are included as systematic uncertainties. Their value is below 1\% in all regions.
  
\item An uncertainty in the level of background contamination is assessed by computing \effID\ with and without accounting for the background component, and using the differences between the two results in each region as a systematic uncertainty. Its values are less than 2.5\%, except in the region $\ET < 15\,\gev$ where they are as large as 8\%.

\item An uncertainty in the description of the detector in simulation is assessed by using an alternative geometry with additional inactive material in front of the calorimeter when obtaining the simulated signal distribution. The amount of additional material is chosen to be compatible with the measurements performed using Run 1 data~\cite{calib_paper}. The determination of \effID\ is repeated with this configuration, and the relative changes in the results for each region are counted as systematic uncertainties. Their values are typically below 2\%, but up to 5\% in the endcap and negligible for $\ET > 25\,\gev$.
\item Similarly to the above, the generator used in the signal simulation is changed from \textsc{Sherpa} to \textsc{Powheg-Box}. The impact of this change on the computed \effID\ is typically 3\% or less except for $\ET < 15\,\gev$ where it is as large as 10\%, and is included as a systematic uncertainty.
\end{itemize}
The statistical uncertainty is obtained from the $m_{\ell \ell \gamma}$ shape fit. It remains typically below 1\% for $\ET<40\,\gev$ but rises to about 5\% at $80\,\gev$. The total uncertainty reaches 5--15\% for $\ET < 15\,\gev$, about 5\% for $15 < \ET < 25\,\gev$, 1\% for $25 < \ET < 40\,\gev$ and then follows a rise driven by the statistical uncertainty. 
Results are shown in Figures~\ref{fig:eff_unconv} and~\ref{fig:eff_conv}.

\begin{figure}
  \centering
  \includegraphics[width=0.49\textwidth]{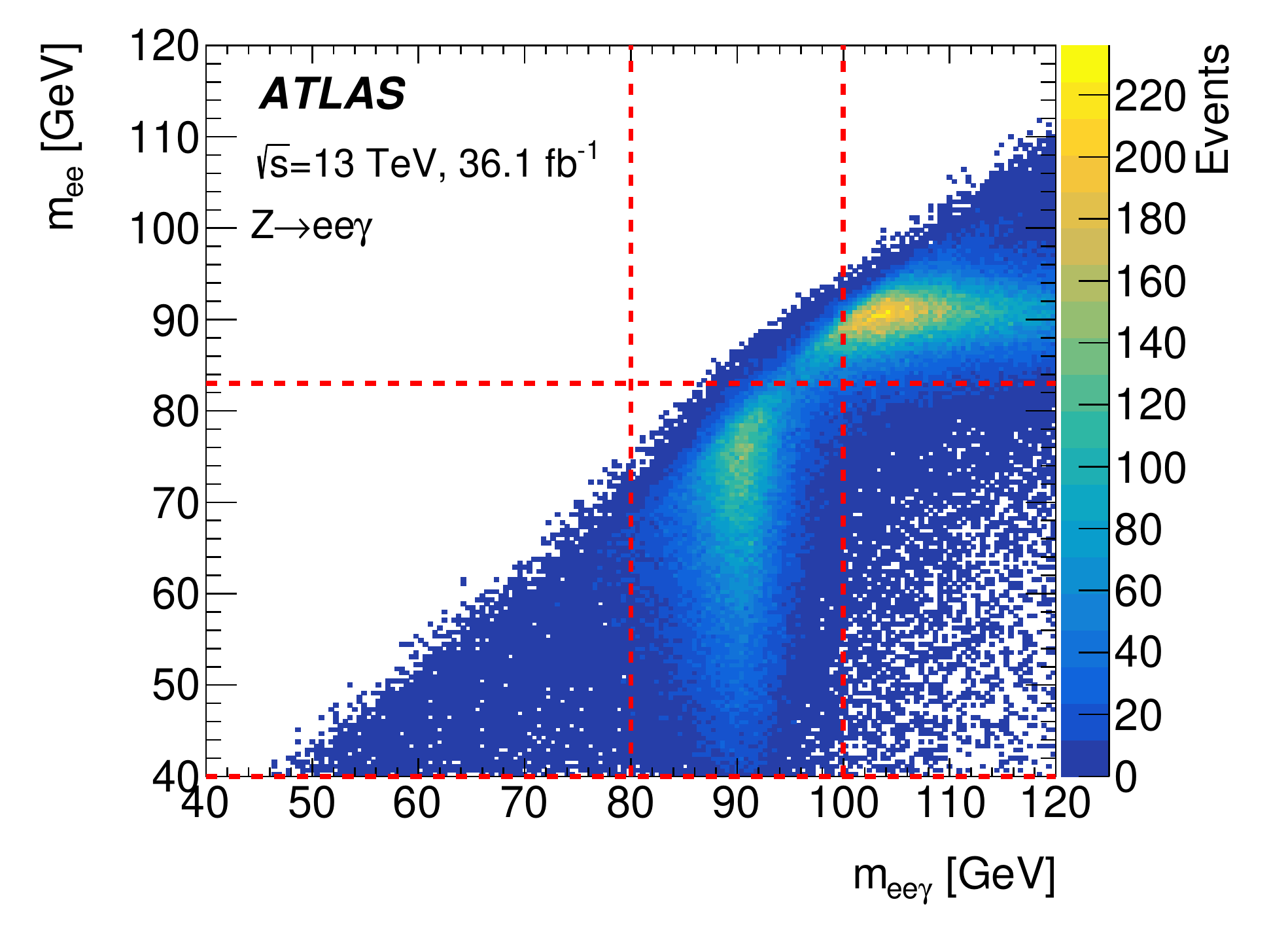}
  \includegraphics[width=0.49\textwidth,angle=0]{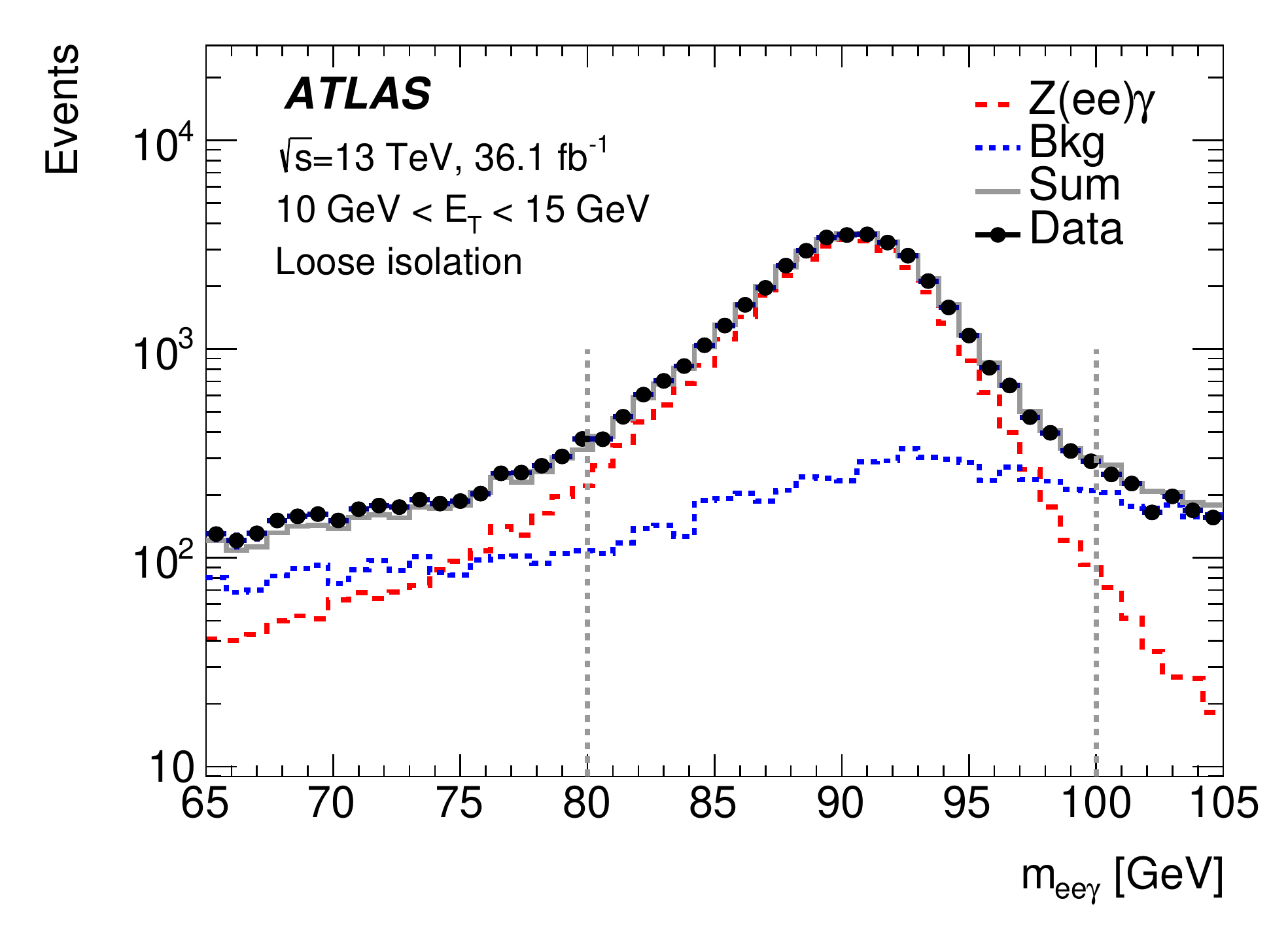}
  \caption{Left: distribution of $m_{ee\gamma}$ vs. $m_{ee}$ in events satisfying all $Z\to e e \gamma$ selection criteria except those for $m_{e e \gamma}$ and $m_{e e}$; the horizontal and vertical dashed lines show the selections used to isolate radiative decays, $40 < m_{e e} < 83\,\gev$ and $80 < m_{e e \gamma} < 100\,\gev$.
    Right: distribution of the invariant mass $m_{ee\gamma}$ for events meeting all selection criteria except the $m_{ee\gamma}$ selection, and in which the photon has $10<\ET<15$ GeV (black dots). The solid gray line represents the result of the fit to the data of the distribution to the sum of invariant mass templates for signal (dashed red line) and background (dotted blue line), both obtained from \textsc{Sherpa} MC simulation. The vertical dashed lines define the mass window used in the measurement.}
  \label{fig:RadZ_illustration}
\end{figure}

\subsection{Electron extrapolation}
\label{sec:measurements:EE}

The electron extrapolation method uses a clean sample of electron candidates from $Z\rightarrow ee$ decays. The distributions of the shower shapes associated with electron candidates are then modified with a Smirnov transform, estimated from simulation and discussed below, to reproduce those associated with photon candidates.

Electrons are selected using a tag-and-probe method, in order to avoid selection biases in the electron shower shape distributions: most of the selections are applied to one of the electrons (the \emph{tag}), while only a loose selection is applied to the other electron (the \emph{probe}) from which shower shape distributions are then obtained. The events are required to pass a single-electron trigger selection, and the trigger object must match the tag electron. Both electrons are required to have $\ET > 25\,\gev$ and $|\etacal|<2.37$, excluding the transition region between barrel and endcap calorimeters. The tag electron is required to pass the \textit{tight} identification requirement~\cite{ATLAS-CONF-2016-024}, while the probe electron is only required to have a track with at least seven track hits in the semiconductor tracker and at least one hit in the pixel detector.  However, in order to match the photon selection described below, the probe electron is required to pass the loose isolation requirement as described in Section~\ref{sec:PhotonIso}.  There must also be no more than one jet (the one reconstructed from the energy deposited in the calorimeter by the electron) with $\ET > 20\,\gev$ and within $\Delta R = 0.4$ of the probe electron. The two electrons are required to have opposite charges and an invariant mass in the range $70 < m_{ee} < 110\,\gev$.

\begin{figure}
\centering
\includegraphics[width=.45\textwidth]{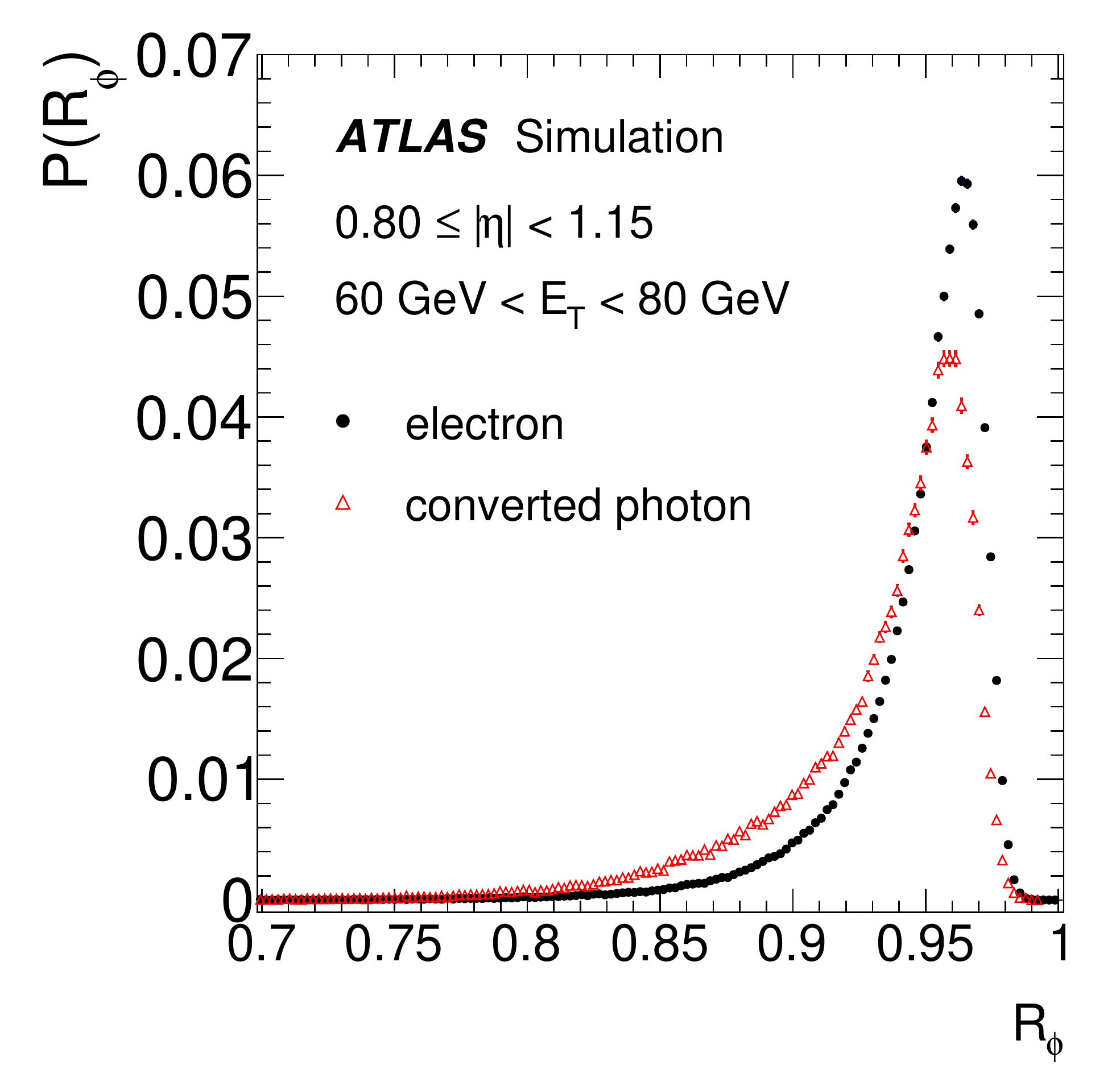}
\includegraphics[width=.45\textwidth]{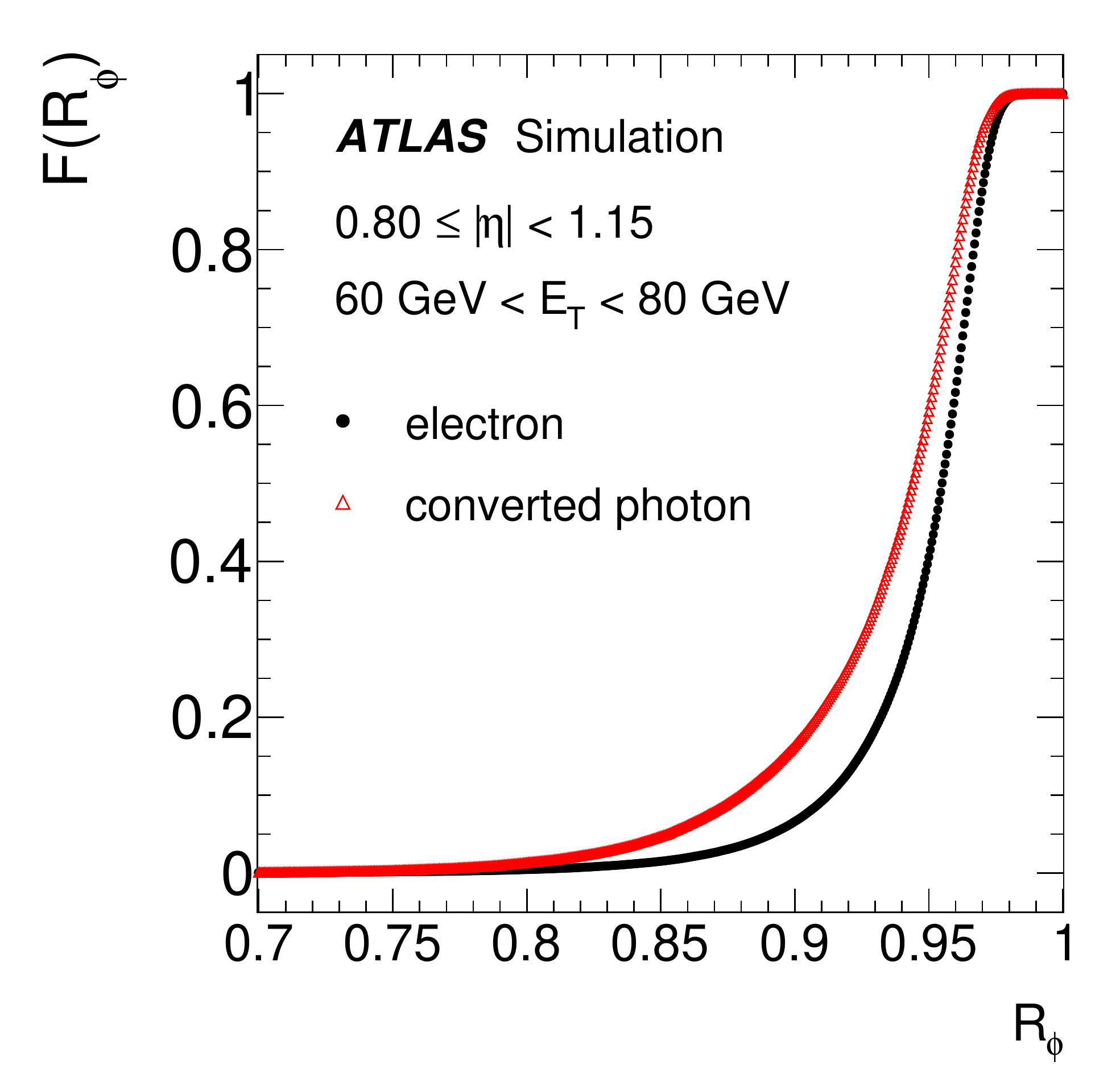} \\
\includegraphics[width=.45\textwidth]{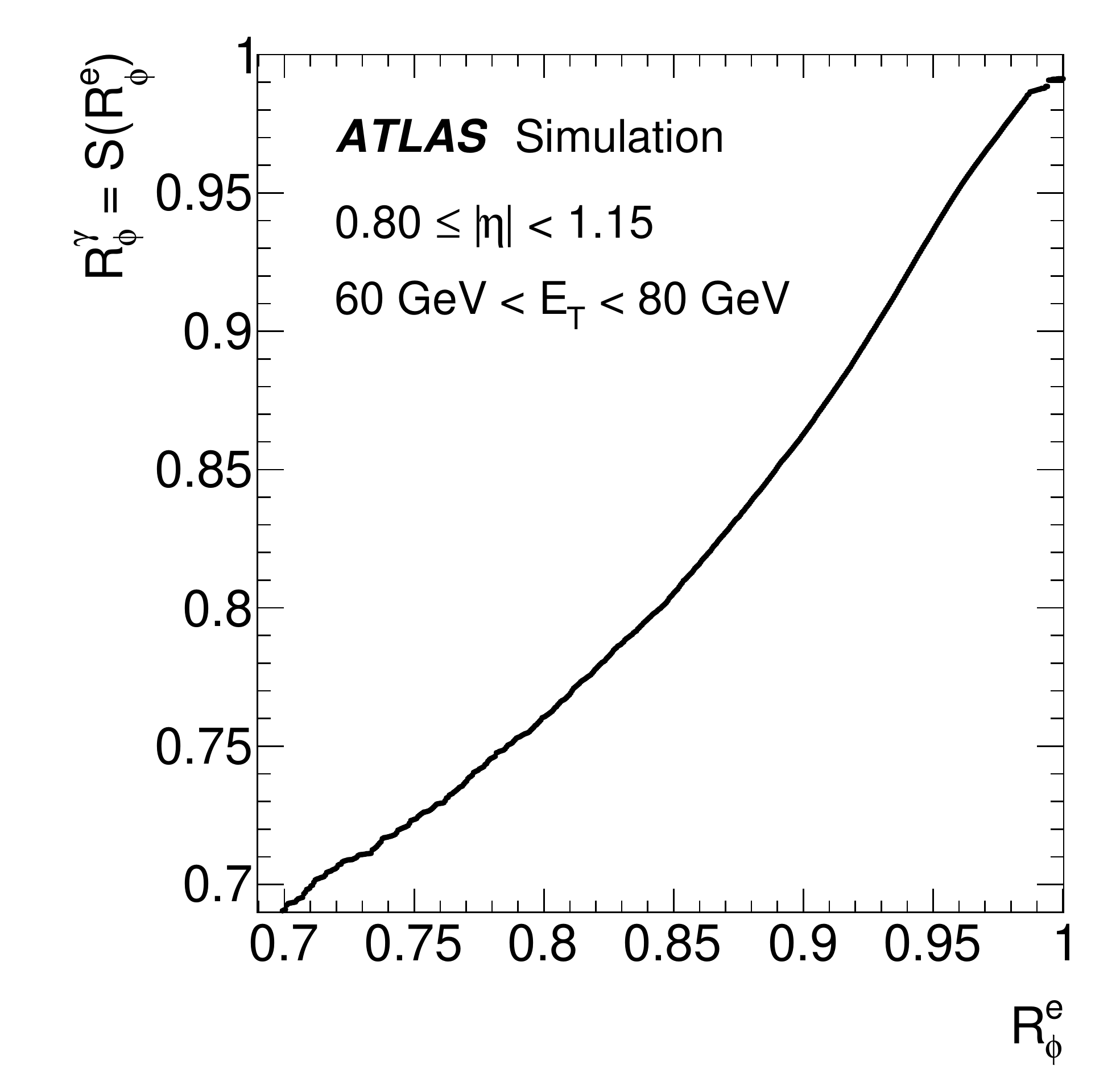}
\includegraphics[width=.45\textwidth]{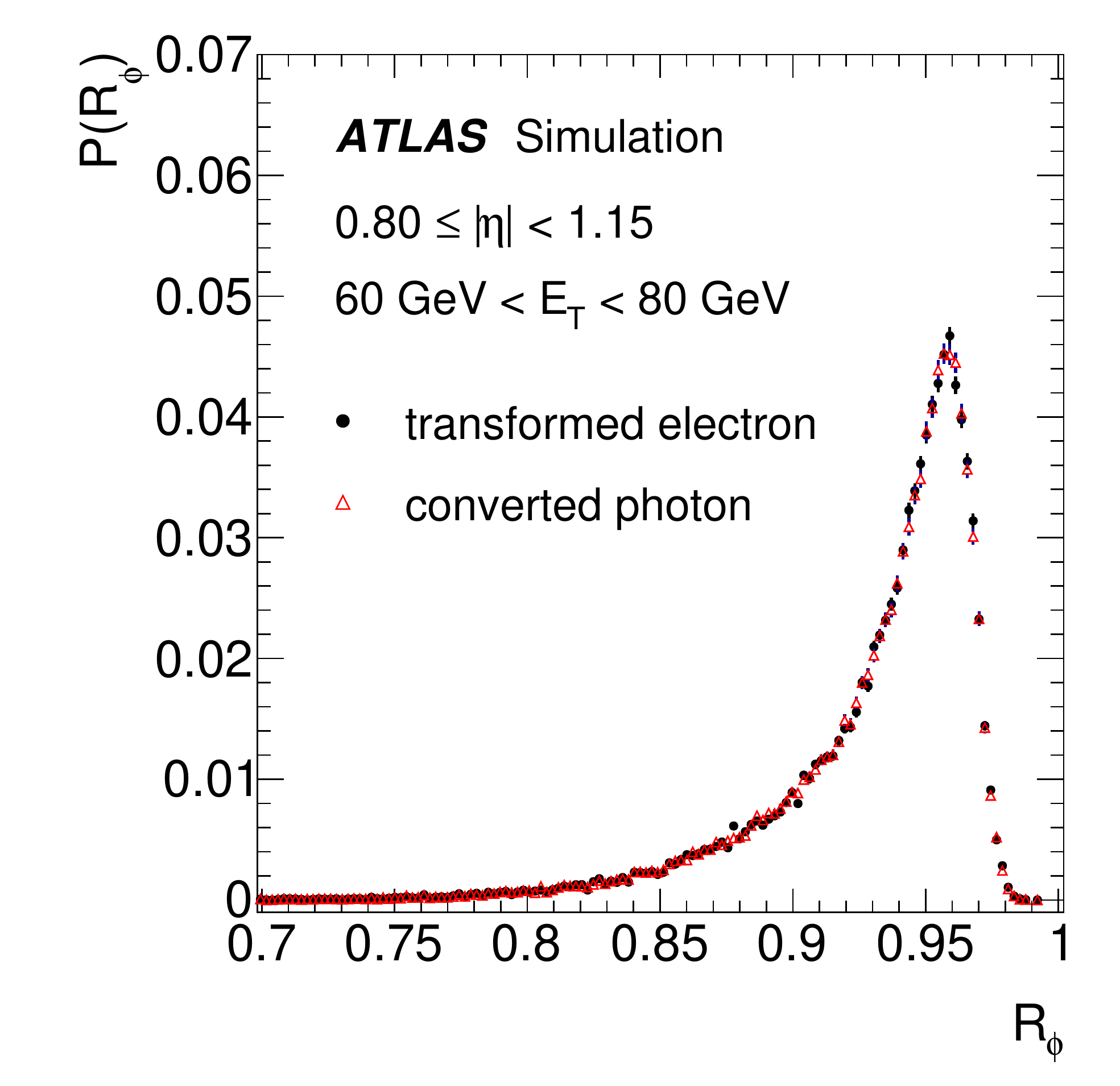}
\caption{Illustration of the Smirnov transform technique applied to
  photon and electron shower shapes. $R_{\phi}$ is chosen as an
  example of a shower shape which differs notably between
  electrons and photons. The $R_{\phi}$ distribution in each sample
  (top left) is used to calculate the corresponding cumulative
  distributions (top right). From the two cumulative distributions, a
  Smirnov transformation can be derived (bottom left). Applying the
  transformation leads to an $R_{\phi}$ distribution which closely matches the photon
  distribution.
  The definition of the $\Rphi$ is given in Table~\ref{tab:IsEM-DV} and Figure~\ref{fig:IsEM-DV}.
}
\label{fig:EE_illustration}
\end{figure}

The shower-shape variables listed in Table~\ref{tab:IsEM-DV} are obtained from the electromagnetic clusters of the electron in the same way as for photons. Differences between the distributions of photon and electron shower shapes are corrected using simulation. A set of simulated probe electrons is selected by applying the selection above to $Z\rightarrow ee$ simulated signal events. A set of simulated photons is selected in single-photon simulated samples by applying the same $\ET$, $|\etacal|$ and isolation selections as described above for the probe electrons, and requiring the photon candidate to be matched to a true photon object. 

In each case, the distribution of each shower-shape variable $x_i$ is then obtained, with shifts applied to the photon distributions as described in Section~\ref{sec:DataMC:FF}, and a similar procedure applied to the electron distributions. Smirnov transforms\footnote{The Smirnov transform is also known as the \emph{inverse probability integral transform} \cite{Devroye1986}.}
$S_i$ are defined by the relations:
\begin{equation*}
  x_{\gamma,i} = S_i\left(x_{e,i}\right) \equiv F_{\gamma,i}^{-1}F_{e,i}(x_{e,i}) \, ,
\end{equation*}
where $F_{\gamma,i}$ and $F_{e,i}$ are the cumulative distributions of $x_i$ for simulated photons and electrons respectively~\cite{Devroye1986}. The transformations are such that for an input $x_{e,i}$ following the electron distribution, the output $x_{\gamma,i}$ follows the photon distribution. They are therefore applied to the shower shape properties of data electrons in order to match the expected photon profiles. Transformations
are separately computed for converted and unconverted photons for all discriminating variables. An example of the procedure is shown in Figure~\ref{fig:EE_illustration} for the $R_{\phi}$ variable for converted photons.

The transformed variables are then used to apply the photon identification selection to the electrons in the same way as for photon candidates. The ratio of the number of transformed electron candidates passing the photon selections to the total number of electron candidates is used to estimate \effID\ separately in each $(\ET, |\etacal|)$ bin, and separately for converted and unconverted photons.

The data sample includes a small contamination from processes where the probe is a fake electron, mainly from multijet and $W$+jets production. The size of this contamination is estimated in each $\ET$ and $|\etacal|$ bin using a shape fit of the $m_{ee}$ variable over the range $70 < m_{ee} < 110\,\gev$ with a signal and a background component. The shape of the signal component is obtained from the simulated electron sample. The background shape is obtained from the data by requiring that the probe electron fail at least two of the \textit{loose} electron identification selections of the cut-based selection defined in Ref.~\cite{ATLAS-CONF-2016-024}, as well as the calorimeter-based requirement of the loose isolation selection. 
The background contribution is estimated separately in each analysis bin and subtracted from both the numerator and denominator in the computation of \effID.

The following sources of systematic uncertainties are considered:
\begin{itemize}
\item A closure test of the Smirnov transform procedure is performed by comparing photon identification efficiencies for transformed MC electrons and MC photons. This check accounts in particular for a difference in the correlations of shower-shape variables between photons and electrons, which are not modified by the per-variable Smirnov transforms. The effect is found to be at most 1\% for converted photons and 2\% for unconverted photons. 
\item An uncertainty is assigned to the background subtraction technique by repeating the measurement while using the range $80 < m_{ee} < 100\,\gev$ for the template fit. The difference between this result and the nominal result is used as a systematic uncertainty.
\item An uncertainty is assigned because of the difference in the fraction of converted photons between data and simulation, which impacts the simulated shower shapes used to derive the Smirnov transforms. The fraction of true converted photons in the simulated photon sample is varied by $\pm 10\%$, an amplitude which covers the differences between data and simulation reported in Section~\ref{sec:ConvRecoEff}; the resulting change in \effID\ is used to estimate the uncertainty. The effect is of the order of 0.2\% or less, and up to about 1\% in the first endcap $|\etacal|$ bin.
\item As described in Section~\ref{sec:DataMC:FF}, shifts are applied to simulated shower shape distributions to align them with those in data. These do not, however, capture the full difference between data and simulation if the shapes cannot be reconciled by simple shifts. The impact of the residual differences is accounted for by defining for each variable a range of shift values such that, for any value of the variable, the data distribution can be locally matched to the simulated distribution by a shift belonging to the range of allowed shift values. The measurement is then repeated with the endpoints of the range replacing the nominal value of the shift for each variable. The sum in quadrature of the maximum changes relative to the nominal measurement for each variable is used as an uncertainty. The uncertainties are typically below 1\% at low \ET. However, the relatively tight cut on the $\Fside$ variable in the $1.81 \leq |\etacal| < 2.37$ bin leads to uncertainties of about 5\% for unconverted photons and 2\% for converted photons.
\item An uncertainty is assigned to the fraction of photons originating from fragmentation processes in the simulation. These photons are less isolated than direct photons and have broader showers, which affects the Smirnov transforms. The uncertainty is computed as the variation in \effID\ when the number of fragmentation photons is varied by $\pm$~50\% in simulation. The uncertainty is typically 0.3\% or less, rising to 1\% at high \ET.
\item Finally, an uncertainty is assigned to account for statistical uncertainties in the simulation sample. The uncertainty is computed by iteratively resampling the simulated samples, recomputing \effID\ for each iteration, and the uncertainty is extracted as the width of the resulting distribution. The uncertainties are typically 0.3\%, and up to 0.6\% at high \ET.
\end{itemize}
The statistical uncertainty is computed by iteratively resampling the data as described above for the simulated samples. It remains below 0.1\% over the range $25 < \ET < 150\,\gev$ covered by this measurement.
Overall, the total uncertainty reaches about 2\% at low \ET, and is typically below 1\% for $\ET > 40\,\gev$. However, values of up to 5\% are reached for unconverted photons in the bin $1.81 \leq |\etacal| < 2.37$ due to the data--MC differences  noted above in the $\Fside$ variable.
Results are shown in Figures~\ref{fig:eff_unconv} and~\ref{fig:eff_conv}.

\subsection{Inclusive photon method}
\label{sec:measurements:MM}

This method is based on an inclusive photon sample collected by single-photon triggers. These triggers have thresholds ranging from $10\,\gev$ to $140\,\gev$ and require loose photon identification selections. They are prescaled except at the $140\,\gev$ threshold, but provide large photon datasets at high $\ET$, with a few hundred events in each bin for $\ET > 250 \,\gev$, allowing efficiency measurements to be performed up to $\ET \sim 1.5\,\tev$.

In addition to the trigger requirements, the photons are required to have $\ET > 25\,\gev$ and $|\etacal|<2.37$, excluding the region $1.37\leq|\etacal|< 1.52$, and to pass the \textit{loose} isolation requirements described in Section~\ref{sec:PhotonIso}. However, the \textit{purity} of the sample, defined as the fraction of true photon candidates, is low, especially at low photon $\ET$, both with and without the identification cuts applied. The identification efficiency can be estimated as
\begin{equation*}
\effID = \frac{P\pass N\pass}{P\total N\total}
\end{equation*}
where $P\total$ and $P\pass$ are the purities in the full sample and the subset passing the tight photon identification selection respectively, and $N\total$ and $N\pass$ are the total number of photon candidates in each case. As described below, the purities are estimated using a tight isolation criterion which requires that no track with $\pT > 1\,\gev$ is within $0.1 < \Delta R < 0.4$ of the photon cluster, the lower bound in $\Delta R$ being introduced to avoid selecting conversion tracks.

Purities are obtained by comparing the efficiency for this selection in data before (after) tight photon identification cuts are applied, $\epsilon\total$ ($\epsilon\pass$), with reference efficiency values for the true photon component, $\epsilon\total\sig$ ($\epsilon\pass\sig$), and background component, $\epsilon\total\bkg$ ($\epsilon\pass\bkg$), as 
\begin{equation}
P\passtot = \frac{\epsilon\passtot - \epsilon\passtot\bkg}{\epsilon\passtot\sig - \epsilon\passtot\bkg}.
\label{eq:P}
\end{equation}
The $\epsilon\sig$ are estimated from simulation, and the $\epsilon\bkg$ from data. The efficiency $\epsilon\total\bkg$ in the full sample is measured in the subset which fails the tight photon identification selection, in order to reduce the contamination from true photons. The efficiency $\epsilon\pass\bkg$ after tight photon identification cuts, is similarly evaluated by inverting some photon identification cuts to reduce the contamination from true photons. The inverted cuts are chosen to be the criteria for the $\wthree$, $\Fside$, $\DeltaE$ and $\Eratio$ quantities, which are measured in the finely segmented first layer of the calorimeter and thus expected to be largely uncorrelated with isolation. In both cases, the residual contamination from true photons is subtracted using identification and track isolation efficiencies obtained from simulation and a data-driven overall normalisation. The evolution of the tight isolation efficiencies and the sample purities as a function of photon $\ET$ is shown in Figure~\ref{fig:MM_illustration} for unconverted photons in $|\etacal|<0.6$. The isolation efficiency for prompt-photons is nearly constant in $\ET$, while for fake photons the efficiency decreases with $\ET$ since higher-energy fake photons are typically associated with higher-energy jets, which are more likely to contain tracks.
For unconverted photons, the purities $P\total$ before photon identification selections are found to range from about 30\% at low $\ET$ to about 85\% at high \ET. After applying the tight photon identification selection, this rises to about 50\% at low \ET\ and about 90\% at high \ET. For converted photons the purities are lower, ranging from 20\% to 60\% before the identification requirement and 40\% to 80\% after it is applied.

\begin{figure}
  \centering
  \includegraphics[width=0.49\textwidth]{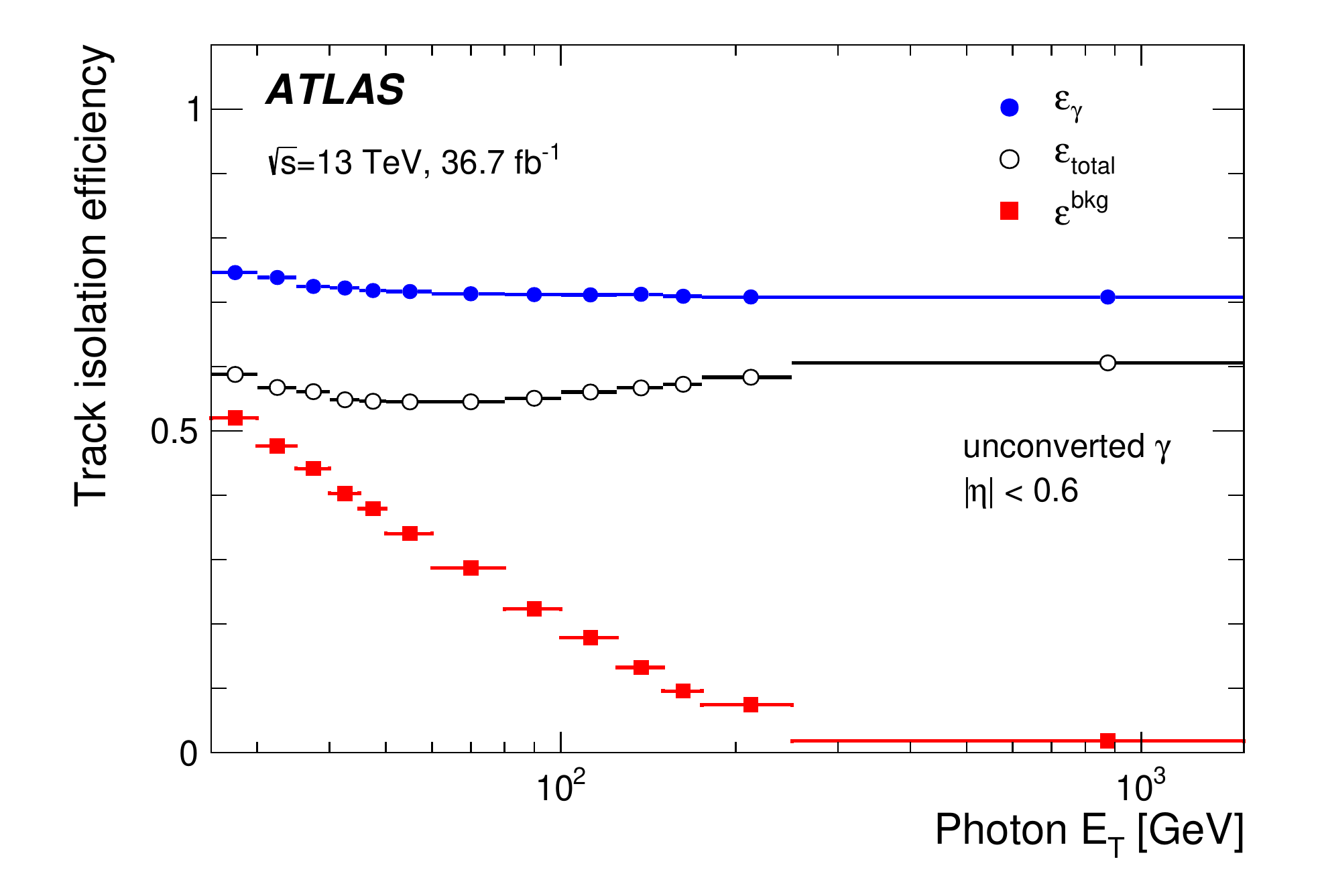}
  \includegraphics[width=0.49\textwidth]{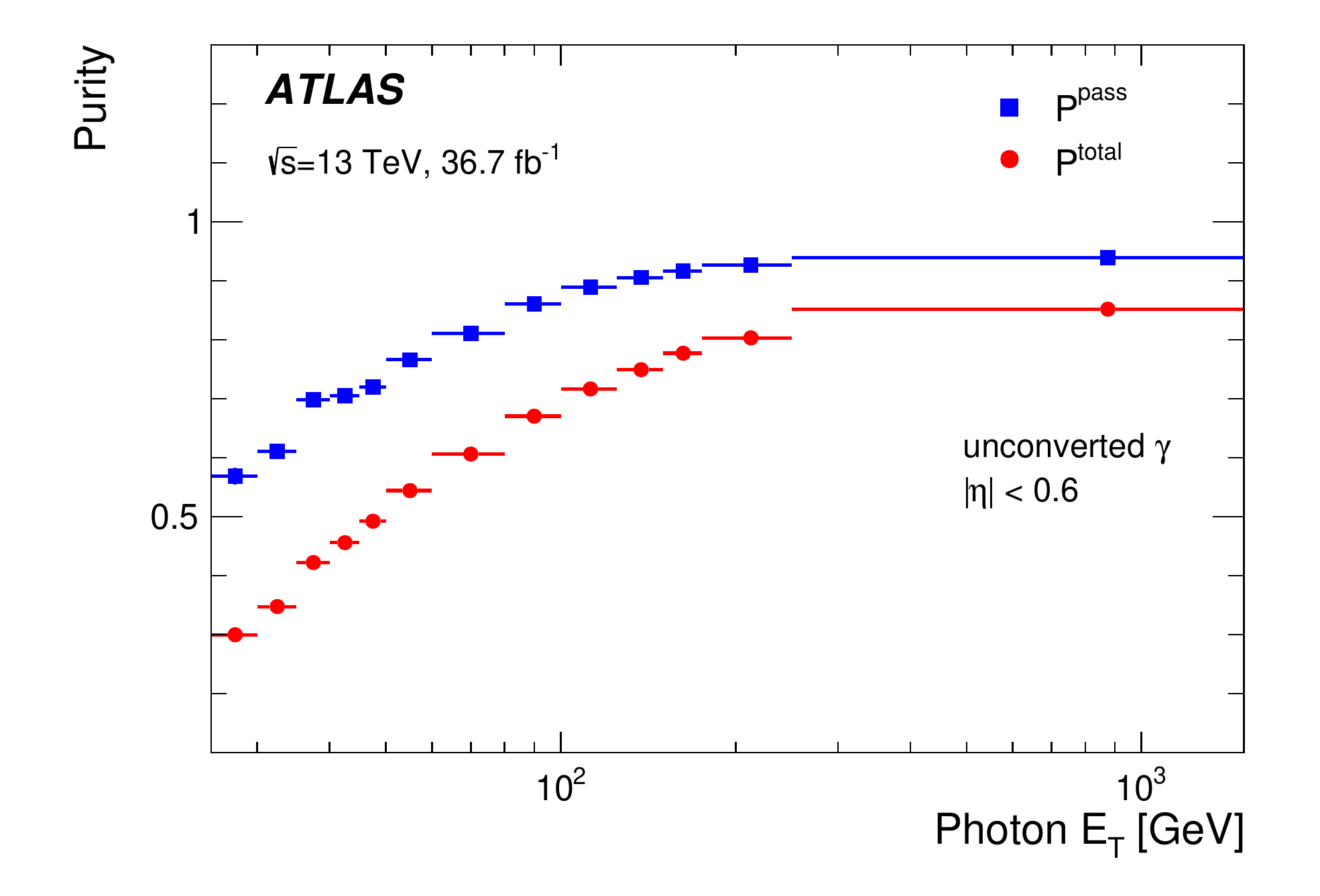}
  \caption{Left: Track isolation efficiency in the inclusive sample separately for prompt (blue), fake (red) and all (black) photons for unconverted photons in the region $|\eta|<0.6$. Right: signal purity $P\pass$ for unconverted photons satisfying the tight identification criteria (blue) and signal purity $P\total$ in the inclusive sample (red) for the region $|\eta|<0.6$. The $P\total$ curve on the right plot is obtained from the $\epsilon\total$ curve relative to the $\epsilon\sig$ and $\epsilon\bkg$ ones in the left plot, following Eq.~(\ref{eq:P}). In both plots statistical uncertainties are shown as error bars but are generally not visible.}
  \label{fig:MM_illustration}
\end{figure}

Finally, since the efficiency $\epsilon$ is computed using a sample of photons which pass the \textit{loose} identification selection, a correction obtained from prompt-photon simulation is applied to measure \effID.

The following sources of systematic uncertainties are considered:
\begin{itemize}
\item A closure check on the computation of the background efficiencies $\epsilon\total\bkg$ and $\epsilon\pass\bkg$ is performed by determining their values as described above in a sample of simulated multijet production (see Section~\ref{sec:DataMC}). The relative differences between these values and the true value is used as a systematic uncertainty. The uncertainty reaches 18\% at low \ET\ but is about 3\%  at 50~\gev\ and below 1\% at high \ET.
\item An uncertainty due to the tight isolation requirement is evaluated by changing the size of the isolation cone from $0.4$ to $0.2$. The uncertainty reaches 8\% at low \ET, but is less than 1\% above 50~\gev.
\item An uncertainty due to the description of the detector in simulated samples is derived in the same way as for the method using radiative $Z$ decays, using samples with variations in the amount of inactive material in front of the calorimeter. The size of the uncertainty is typically 1\% at low \ET\ and at the per-mil level at high \ET, except for the unconverted photons in the first endcap bin where uncertainties reach 4\%.
\item The statistical uncertainties in the simulation samples
are accounted for using the electron
extrapolation method by iteratively resampling the corresponding datasets, and are typically 0.5\% or less.
\end{itemize}
The statistical uncertainty is computed as the width of the distribution of results obtained when repeating the measurement on pseudo-datasets obtained by resampling the data and reach 1--2\% for $\ET < 50\,\GeV$ and typically 0.5\% at higher \ET.
The total uncertainty reaches to 7--18\% at $\ET=25\,\gev$, but 2--3\% at 40~\gev\ and 1\% or less above 100~\gev\ except for unconverted photons for $\ET > 250\,\gev$ and $1.52 \leq |\etacal| < 1.81$ where it reaches 4\% as noted above.
Results are shown in Figures~\ref{fig:eff_unconv} and~\ref{fig:eff_conv}.

\subsection{Results and Combination}
\label{sec:measurements:comb}

\begin{figure}
  \centering
  \subfloat[]{\label{fig:dd_eff_unconv_a}\includegraphics[width=.5\textwidth]{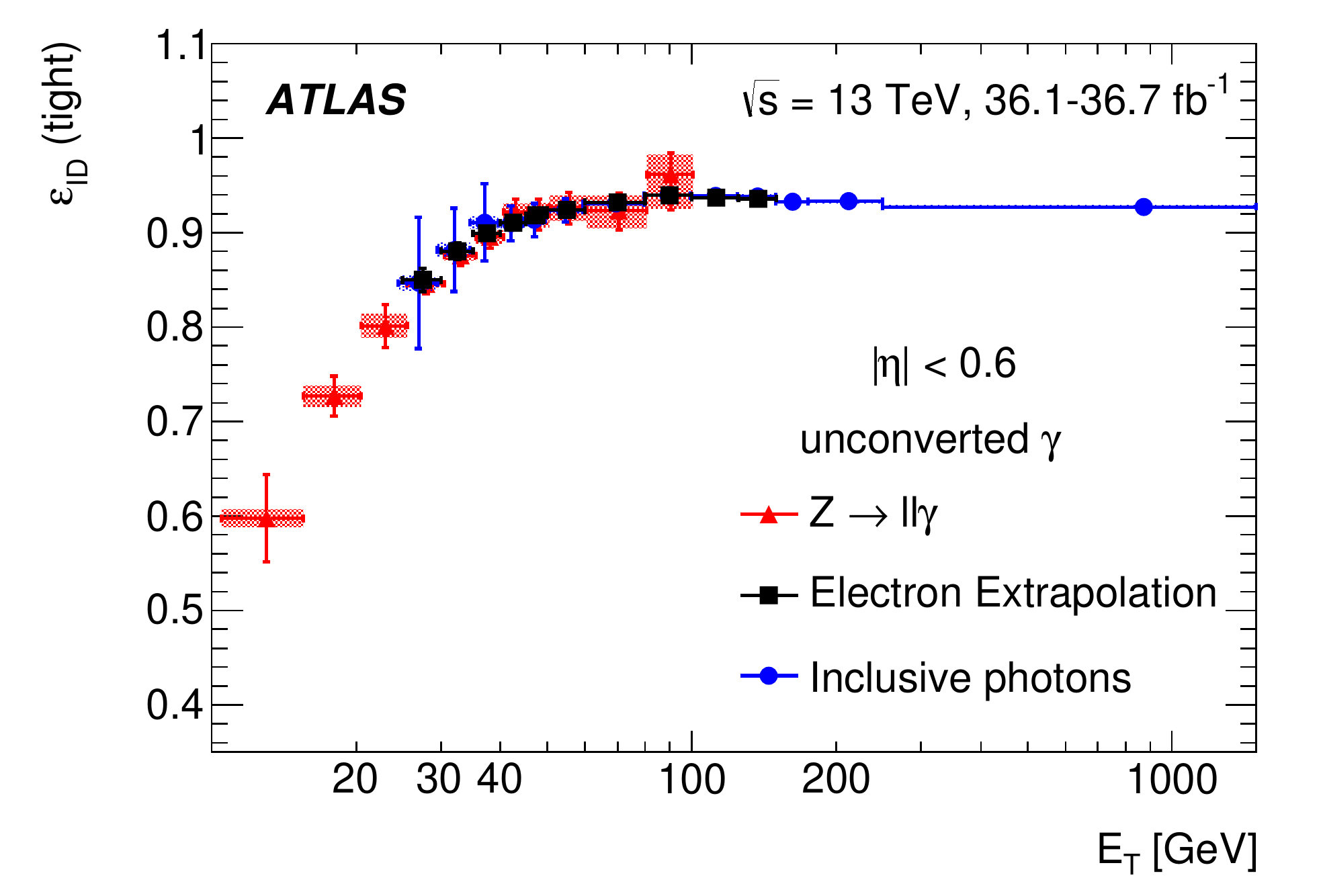}}
  \subfloat[]{\label{fig:dd_eff_unconv_b}\includegraphics[width=.5\textwidth]{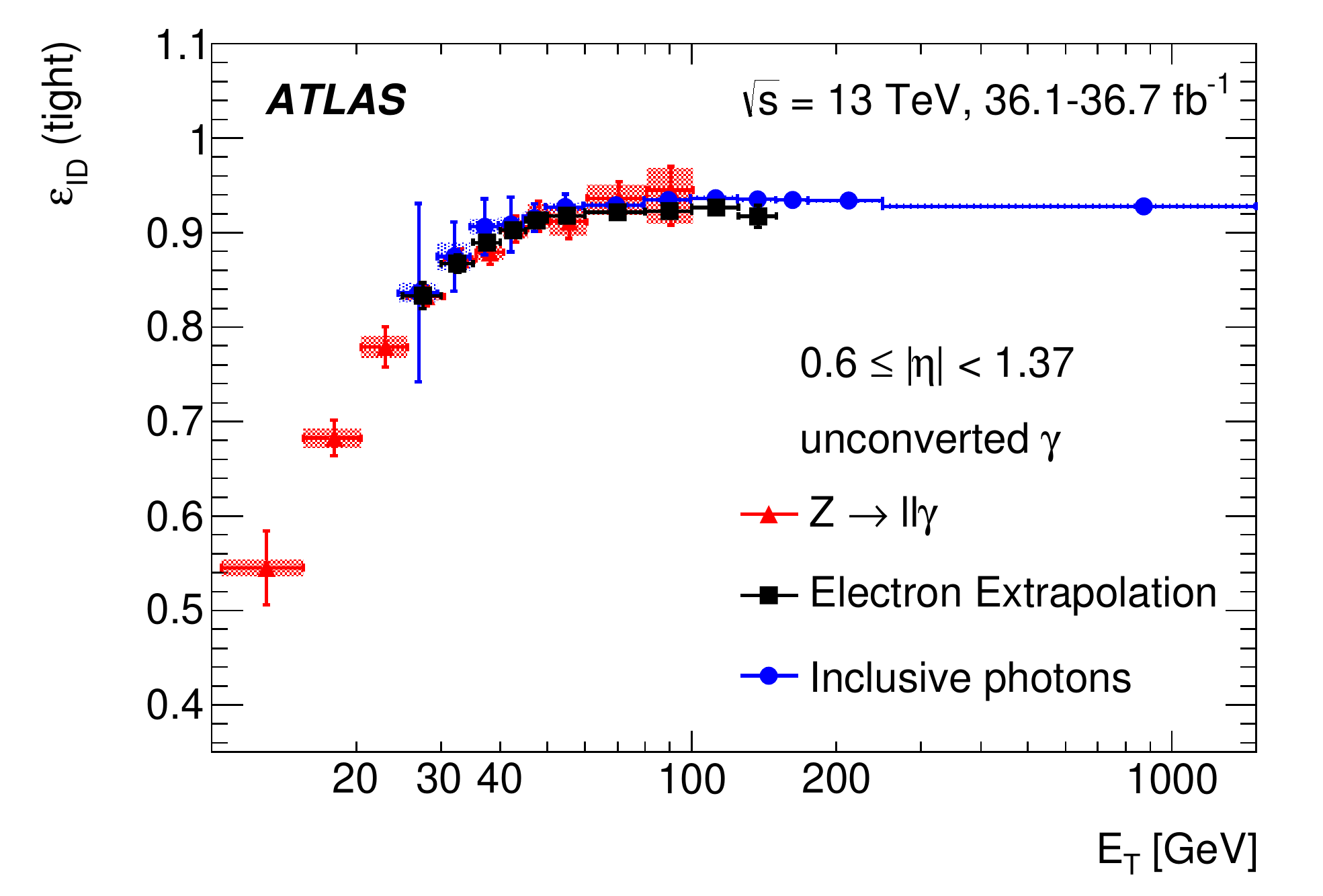}}\\
  \subfloat[]{\label{fig:dd_eff_unconv_c}\includegraphics[width=.5\textwidth]{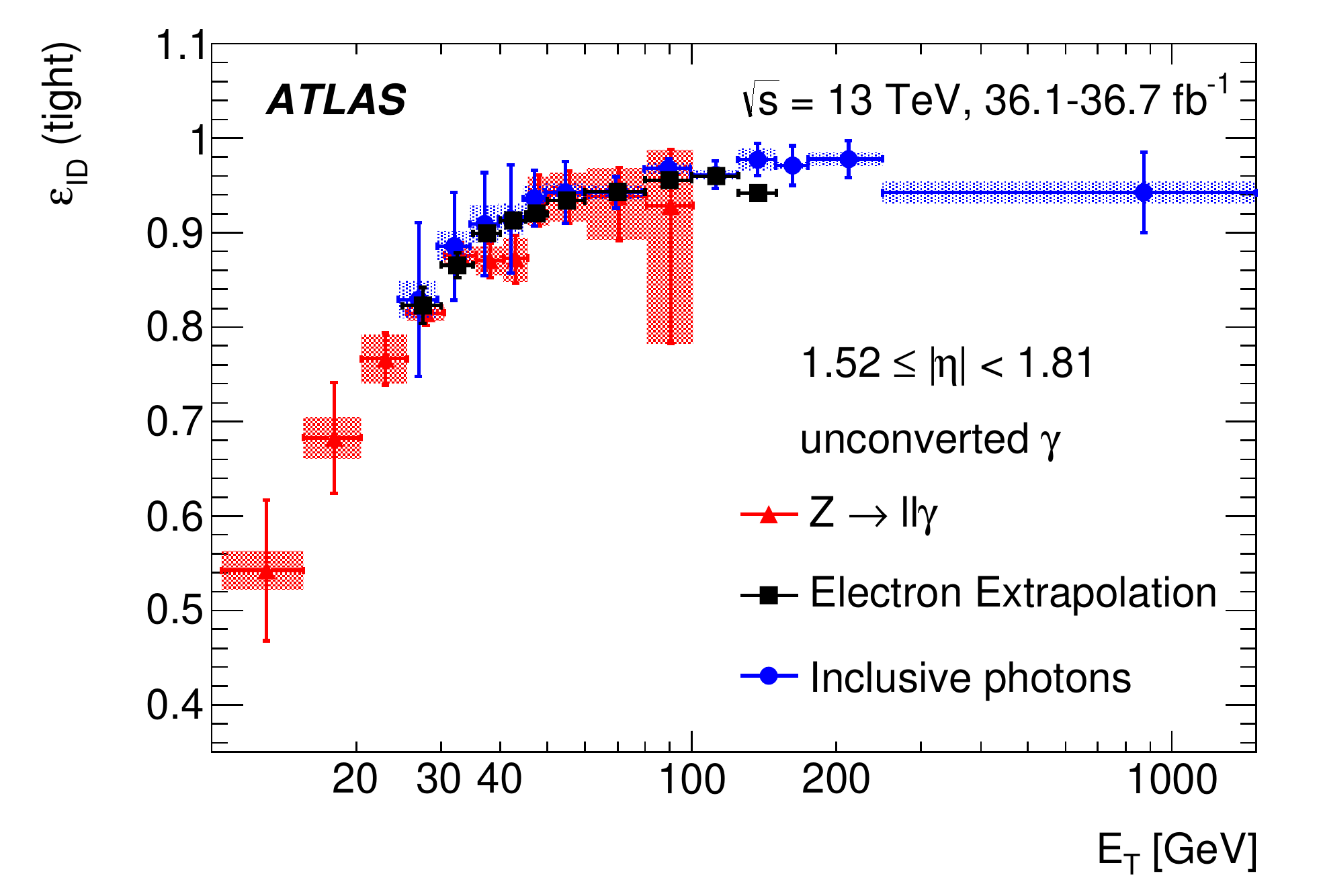}}
  \subfloat[]{\label{fig:dd_eff_unconv_d}\includegraphics[width=.5\textwidth]{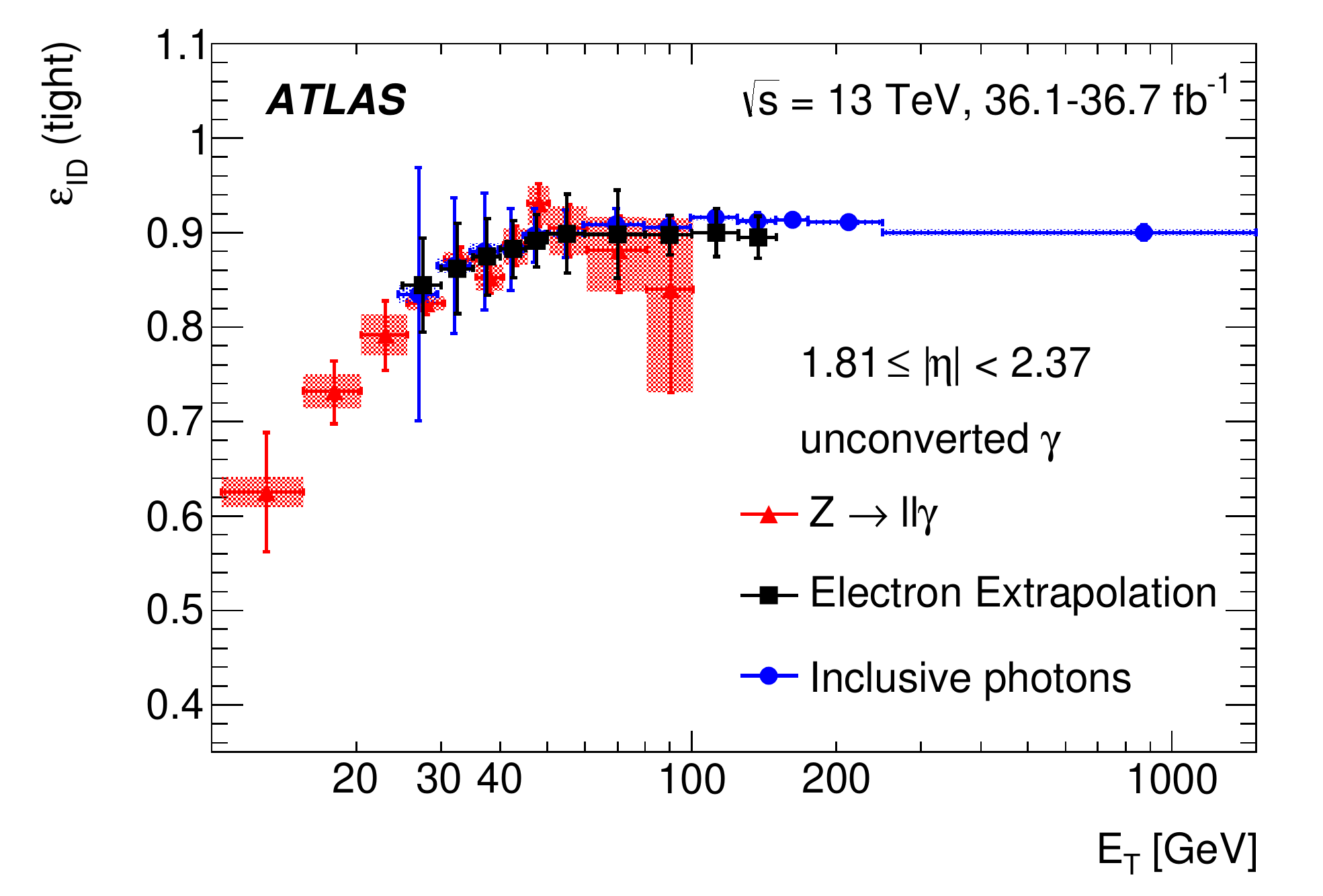}}
  \caption{Comparison of the data-driven measurements of the identification efficiency for unconverted photons as a function of $\ET$, for the four pseudorapidity intervals (a) $|\etacal|<0.6$, (b) $0.6\leq|\etacal|<1.37$, (c) $1.52 \leq |\etacal|<1.81$, and (d) $1.81\leq|\etacal|<2.37$.
    The error bars represent the sum in quadrature of the statistical and systematic uncertainties estimated in each method.
    The shaded areas correspond to the statistical uncertainties.
    The last bin extends to 1.5 \TeV.
  } 
  \label{fig:eff_unconv} 
\end{figure}

\begin{figure}
  \centering
  \subfloat[]{\label{fig:dd_eff_conv_a}\includegraphics[width=.5\textwidth]{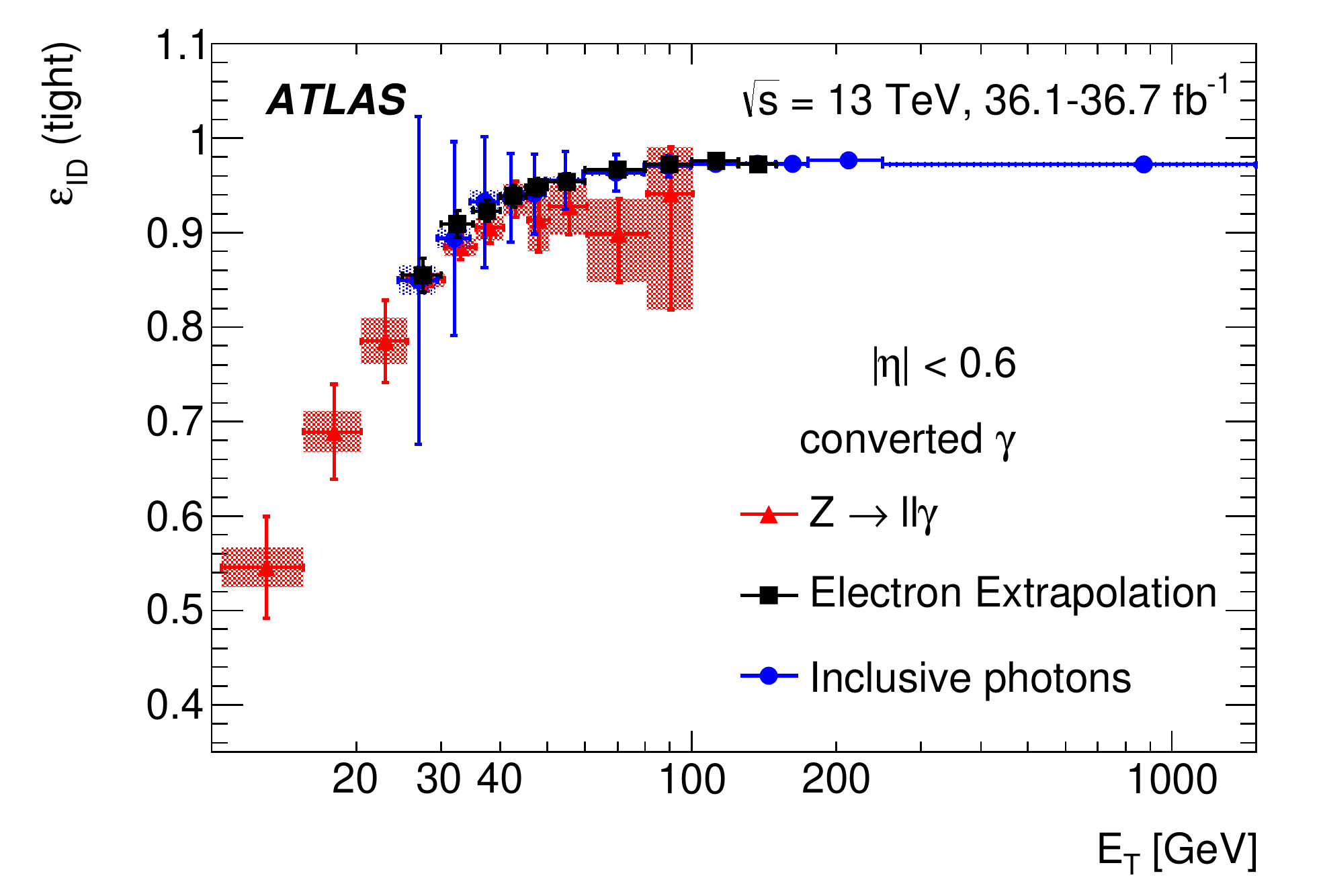}}
  \subfloat[]{\label{fig:dd_eff_conv_b}\includegraphics[width=.5\textwidth]{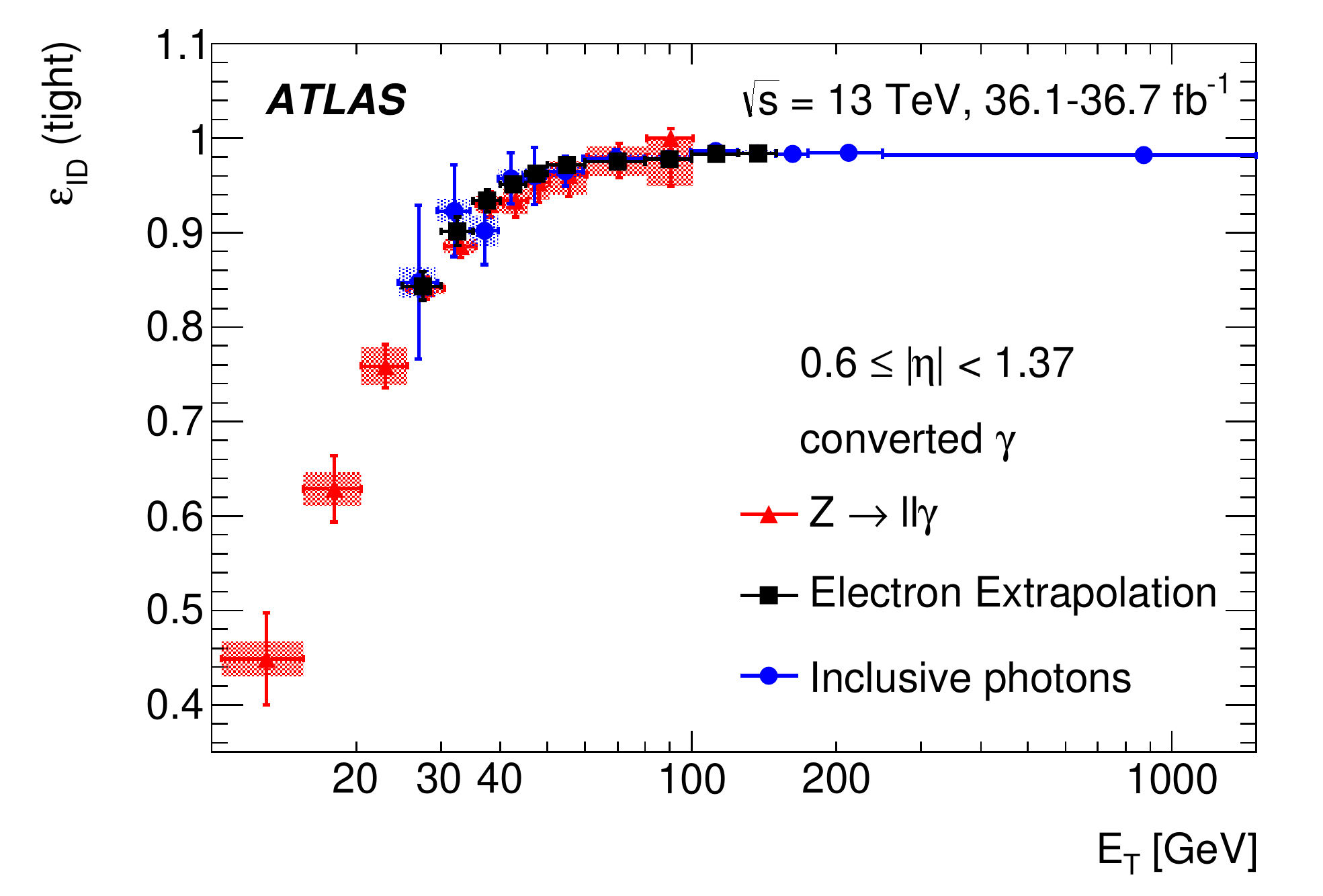}}\\
  \subfloat[]{\label{fig:dd_eff_conv_c}\includegraphics[width=.5\textwidth]{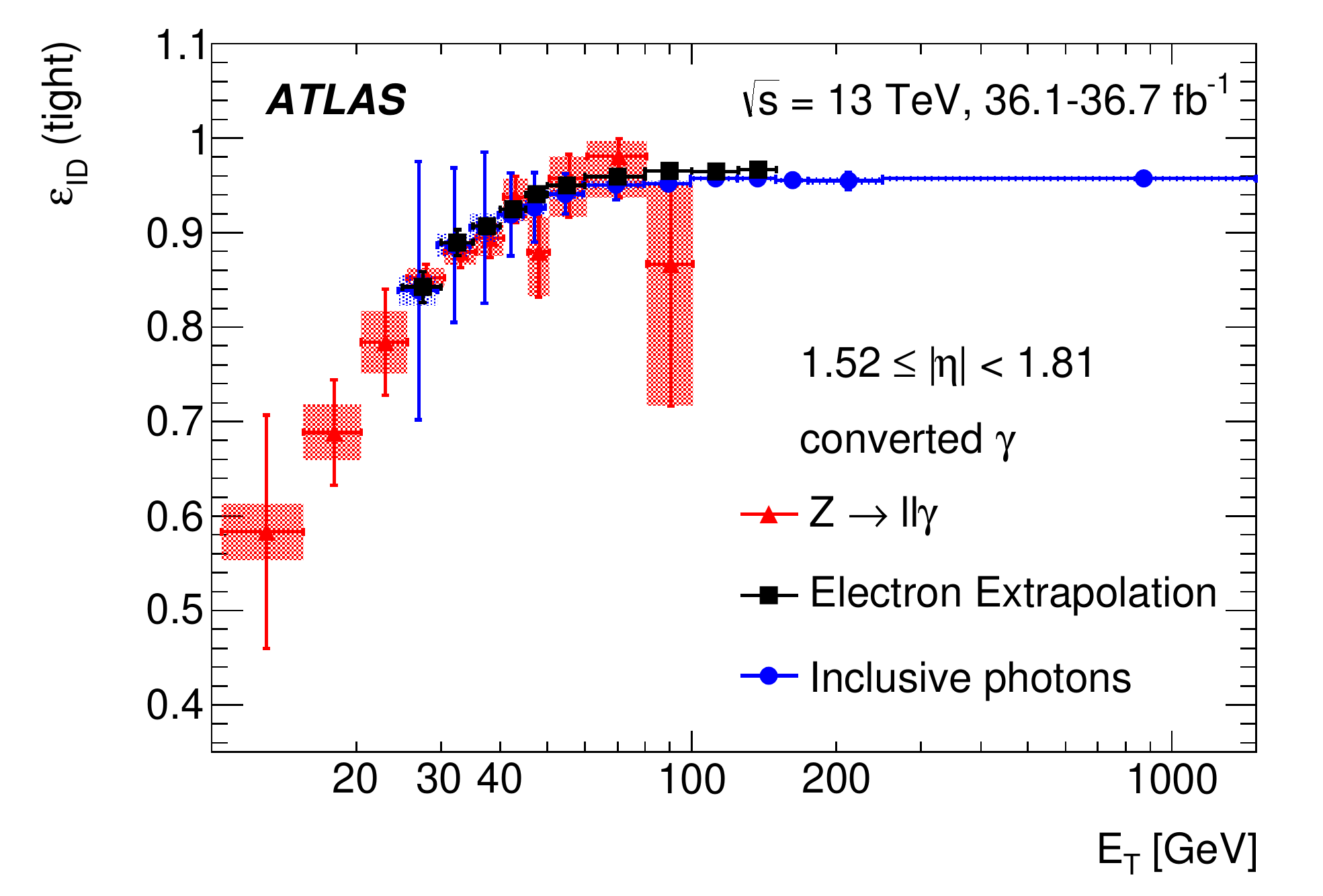}}
  \subfloat[]{\label{fig:dd_eff_conv_d}\includegraphics[width=.5\textwidth]{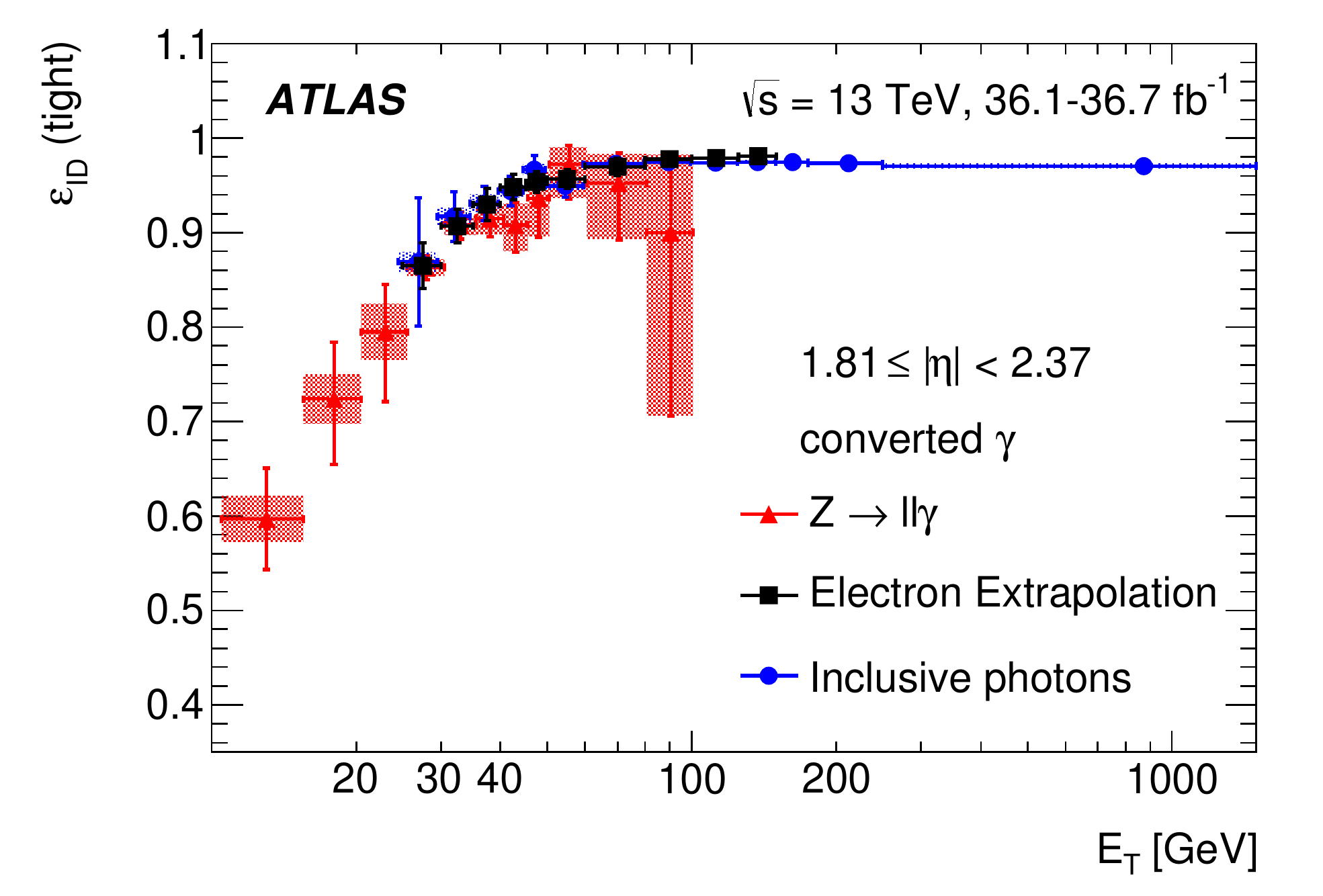}}
  \caption{Comparison of the data-driven measurements of the identification efficiency for converted photons as a function of $\ET$, for the four pseudorapidity intervals (a) $|\etacal|<0.6$, (b) $0.6\leq|\etacal|<1.37$, (c) $1.52 \leq |\etacal|<1.81$, and (d) $1.81\leq|\etacal|<2.37$.
    The error bars represent the sum in quadrature of the statistical and systematic uncertainties estimated in each method.
    The shaded areas correspond to the statistical uncertainties.
    The last bin extends to 1.5 \TeV.
  } 
  \label{fig:eff_conv} 
\end{figure}

The efficiencies \effID\ measured in data for each method are shown in Figure~\ref{fig:eff_unconv} for unconverted photons, and in Figure~\ref{fig:eff_conv} for converted photons. The results obtained in Section~\ref{sec:measurements:RadZ} in the $Z \rightarrow ee\gamma$ and $Z \rightarrow \mu\mu\gamma$ channels are found to be in good agreement and are therefore combined and shown as $Z \rightarrow \ell\ell\gamma$. The combination is performed following the method presented below.

\begin{figure}
   \centering
   \subfloat[]{\label{fig:radz_eff_unconv_barrel_a}\includegraphics[width=.5\textwidth]{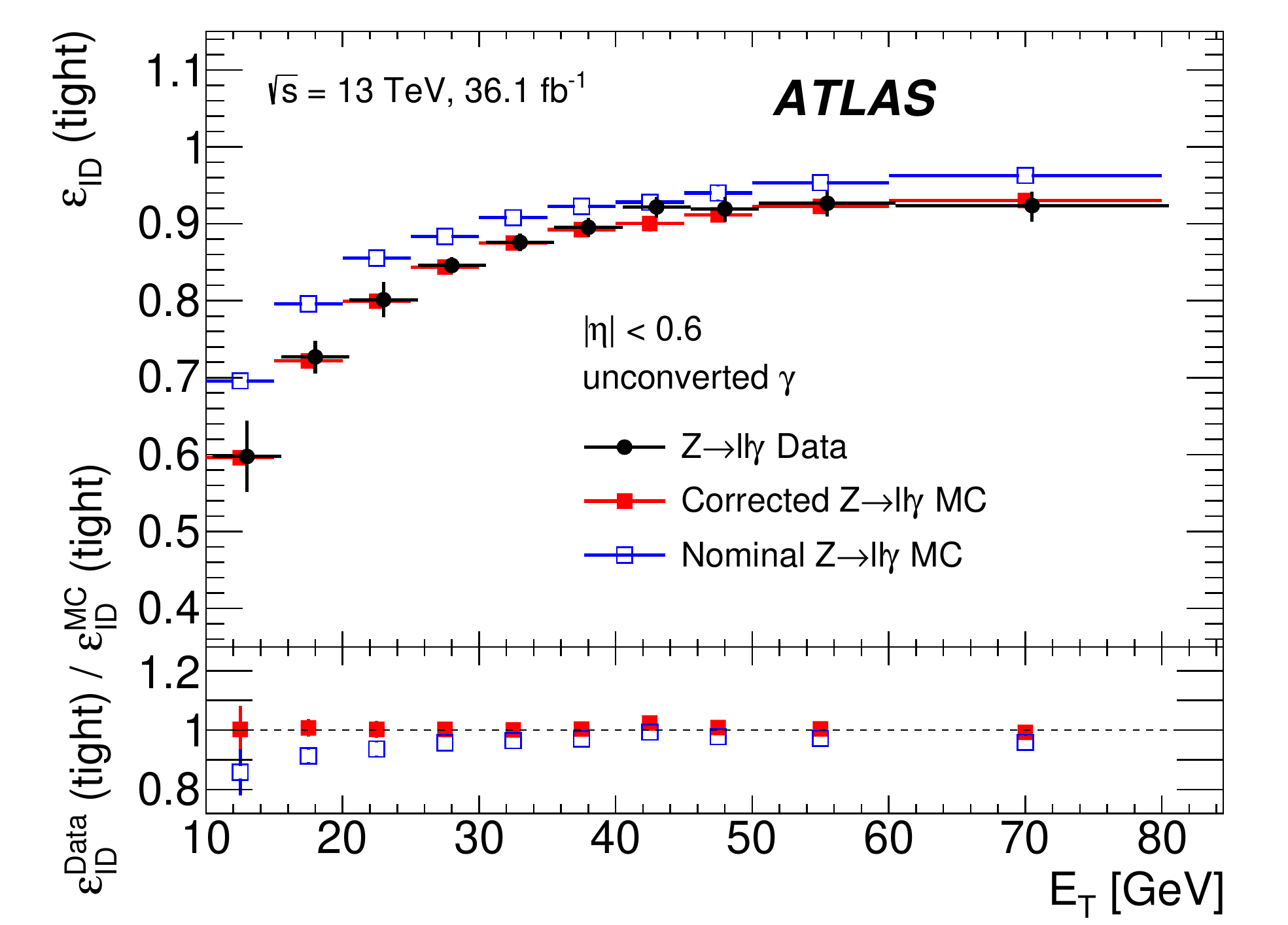}}
   \subfloat[]{\label{fig:radz_eff_unconv_barrel_b}\includegraphics[width=.5\textwidth]{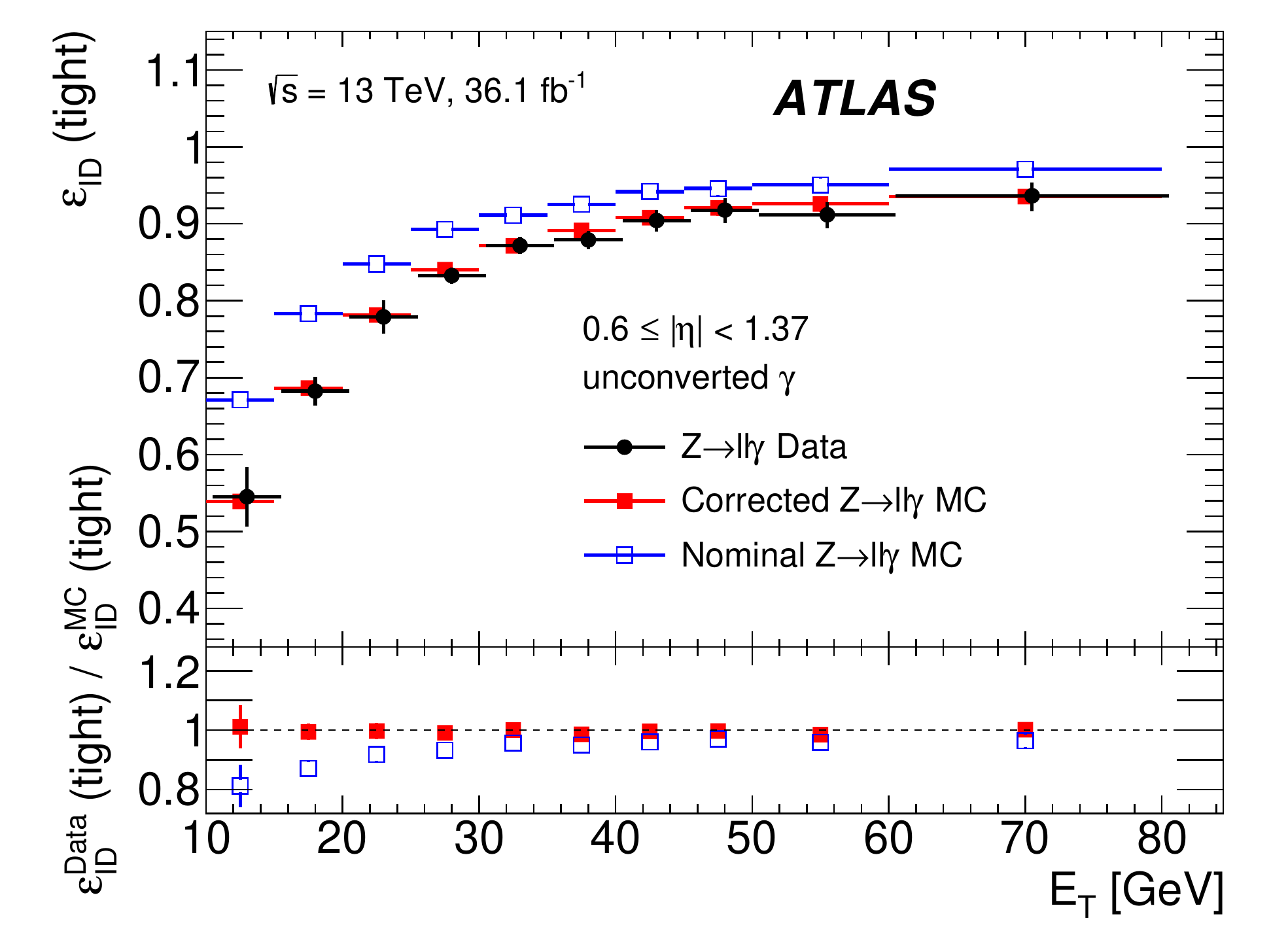}}\\
   \subfloat[]{\label{fig:radz_eff_unconv_endcap_c}\includegraphics[width=.5\textwidth]{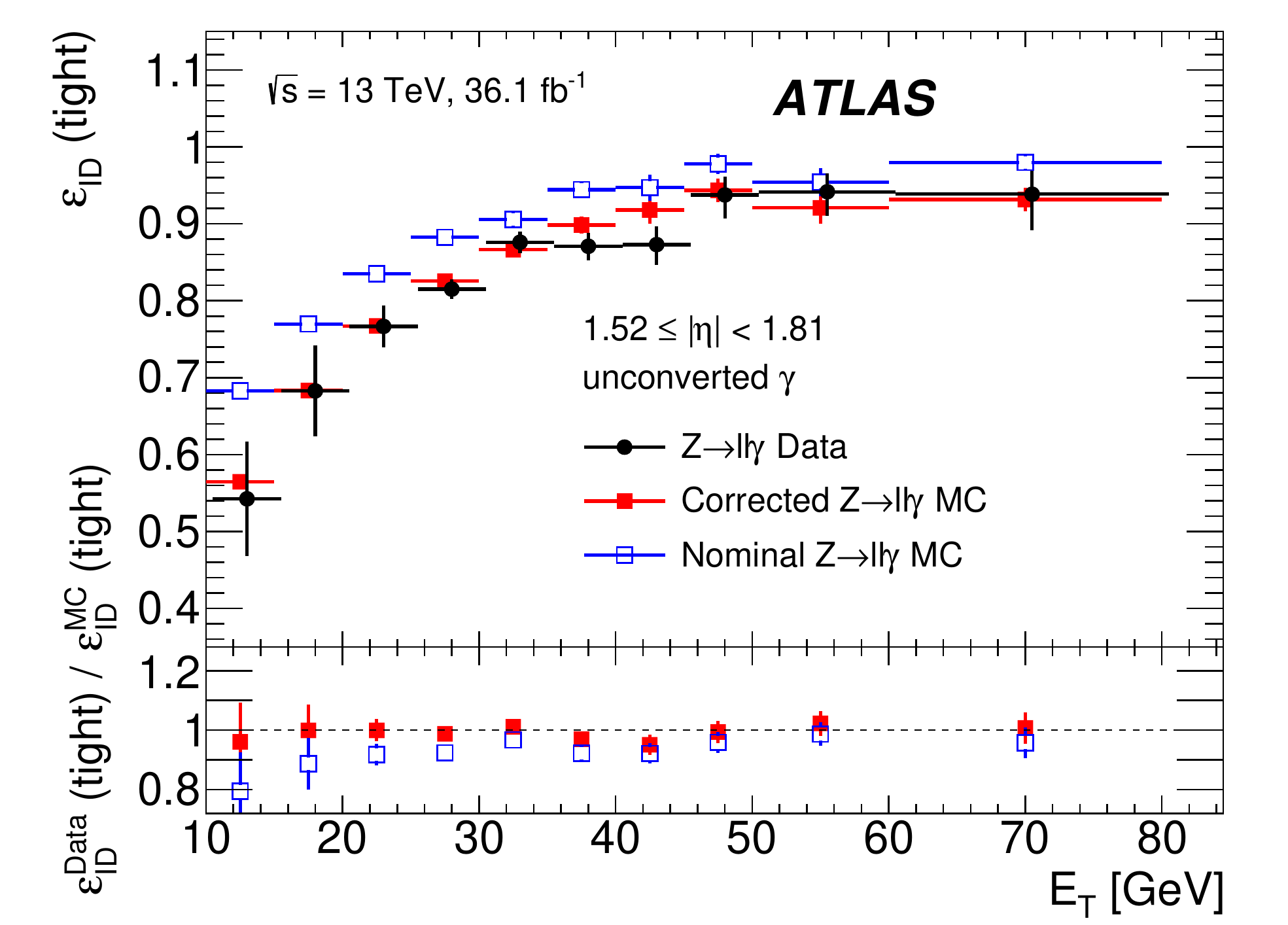}}
   \subfloat[]{\label{fig:radz_eff_unconv_endcap_d}\includegraphics[width=.5\textwidth]{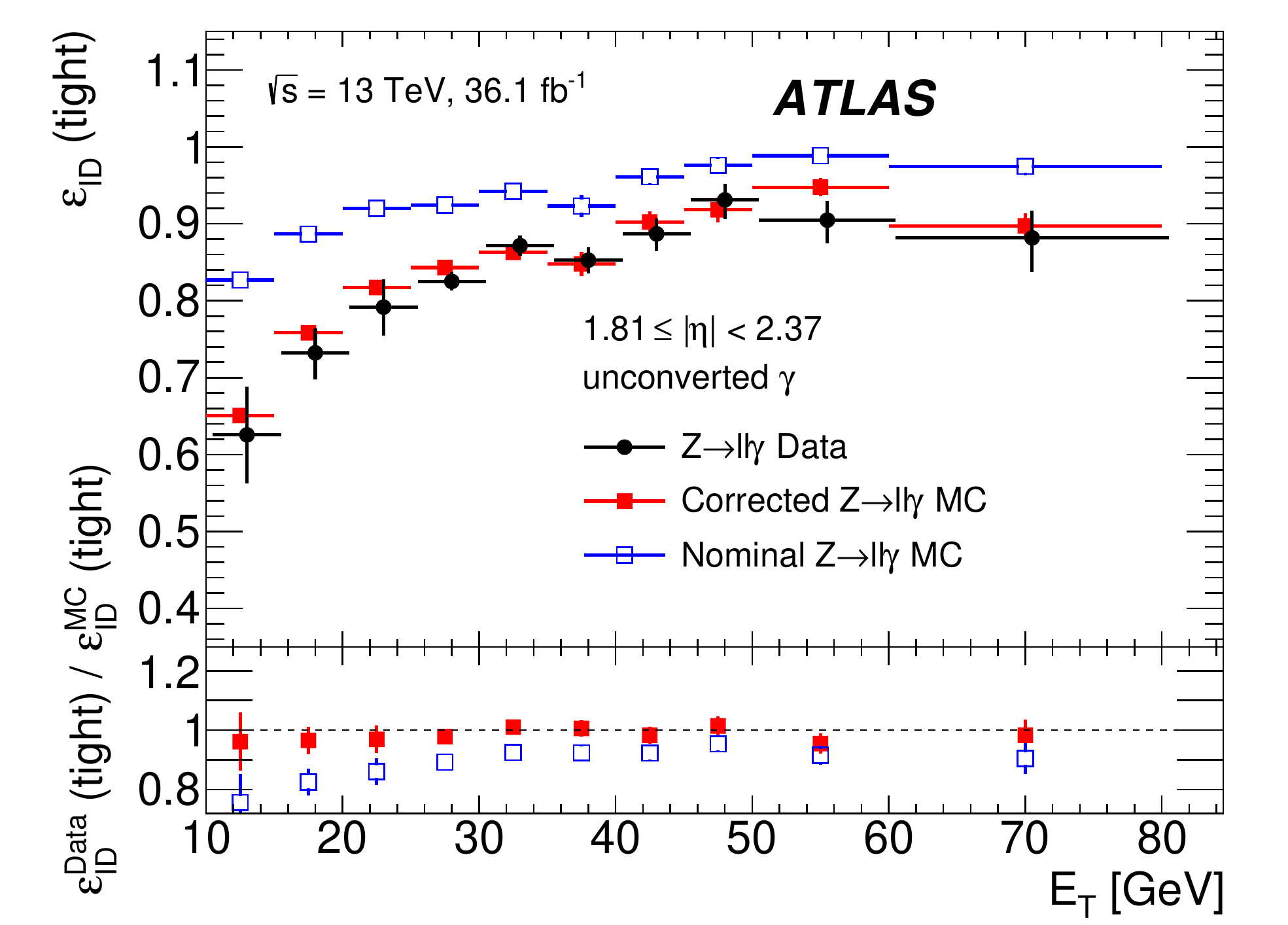}}  
   \caption{Comparison of the measurements of the data-driven identification efficiency for unconverted photons measurements obtained using the radiative $Z$ method with
   the predictions from $Z\to\ell\ell\gamma$ simulation as a function of photon $\ET$, for the four
     pseudorapidity intervals (a) $|\etacal|<0.6$, (b) $0.6\leq|\etacal|<1.37$,
     (c) $1.52 \leq |\etacal|<1.81$, and (d) $1.81\leq|\etacal|<2.37$.
     Predictions are shown for both the nominal simulation and with the corrections described in Section~\ref{sec:DataMC:FF}.
     The bottom panels show the ratio of the data-driven values to the
     MC predictions (referred to as efficiency corrections in the text).} 
   \label{fig:radZ_eff_unconv_vs_mc} 
\end{figure}
\begin{figure}
   \centering
   \subfloat[]{\label{fig:radz_eff_conv_a}\includegraphics[width=.5\textwidth]{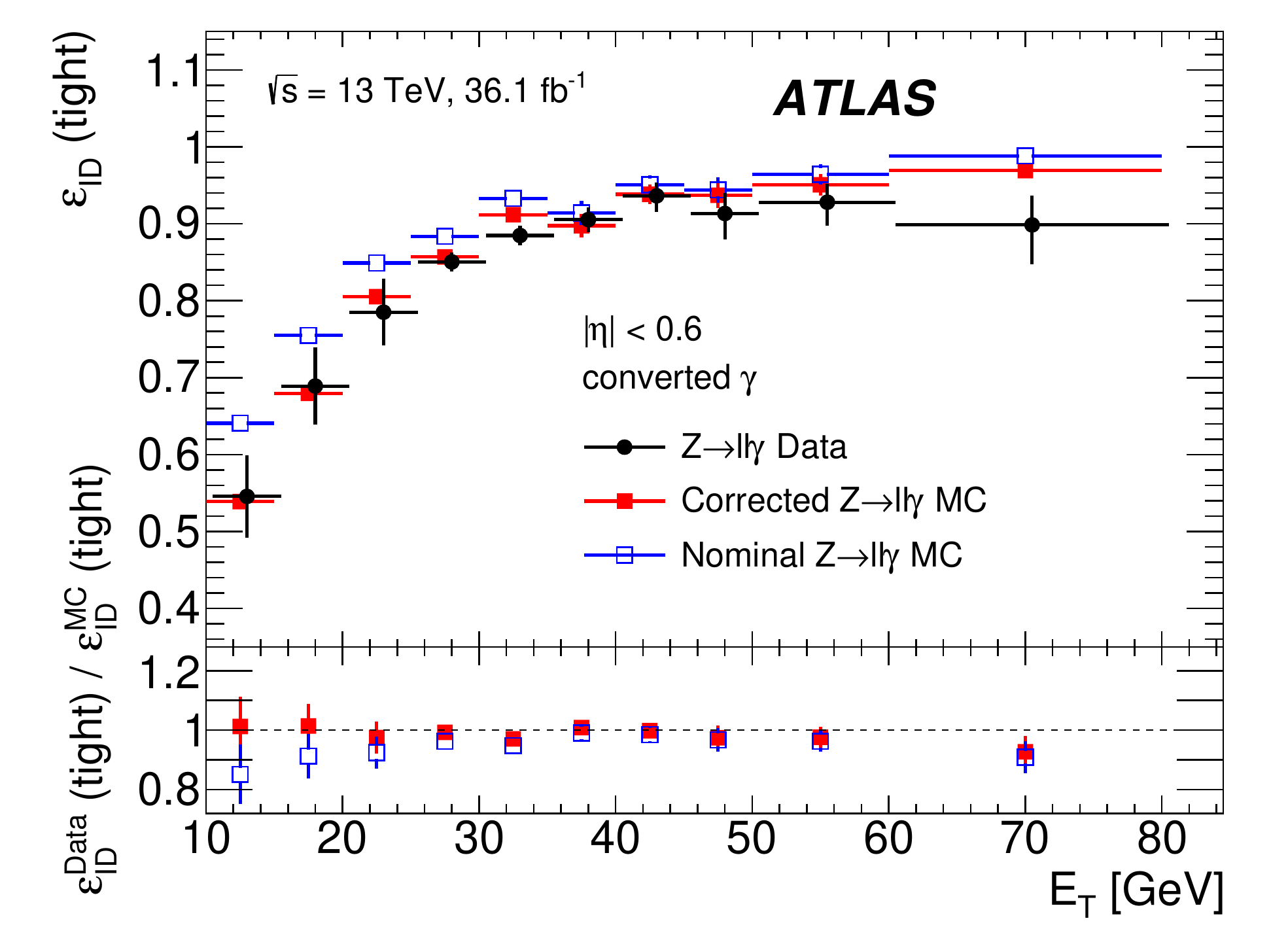}}
   \subfloat[]{\label{fig:radz_eff_conv_b}\includegraphics[width=.5\textwidth]{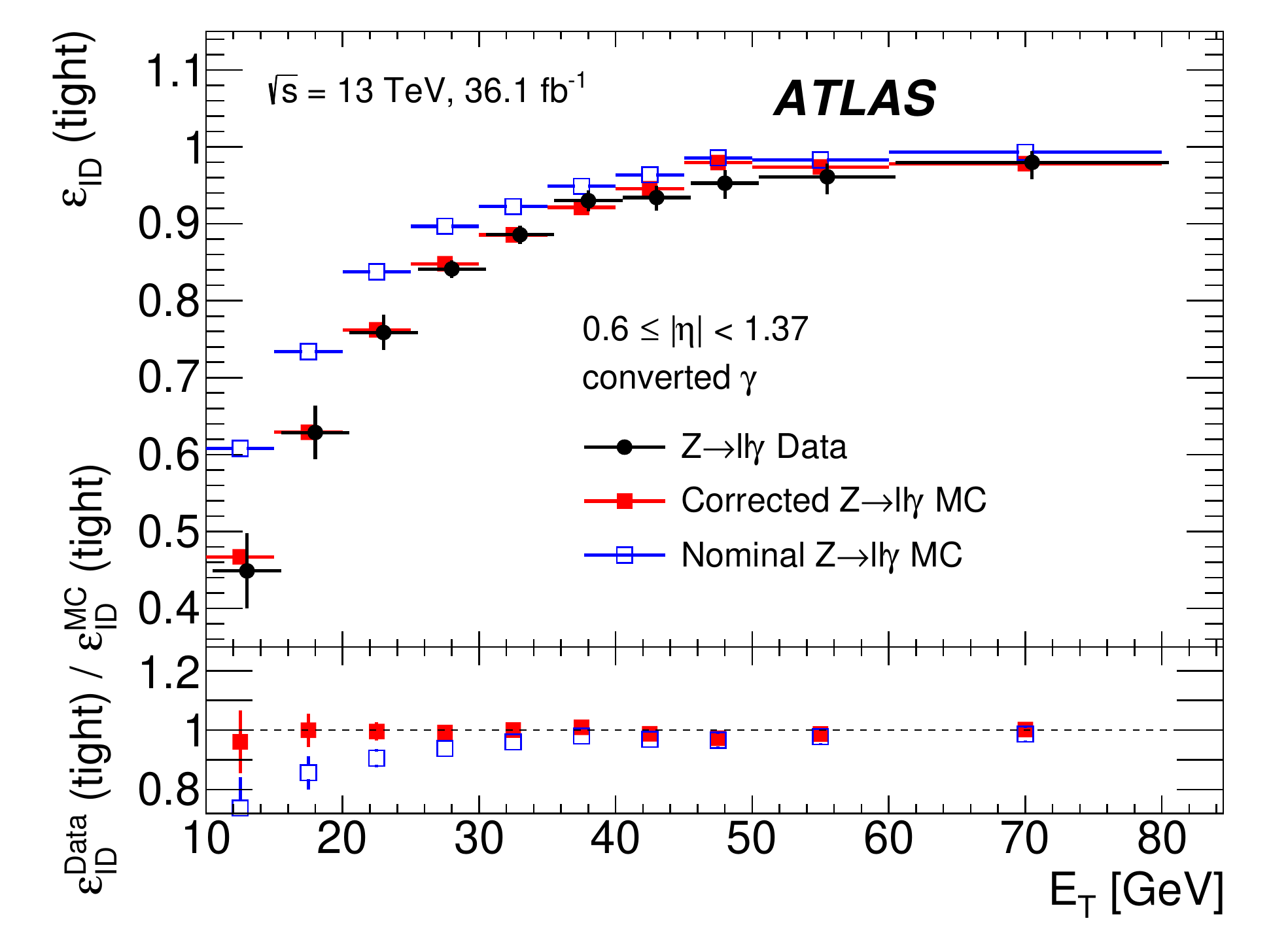}}\\
   \subfloat[]{\label{fig:radz_eff_conv__c}\includegraphics[width=.5\textwidth]{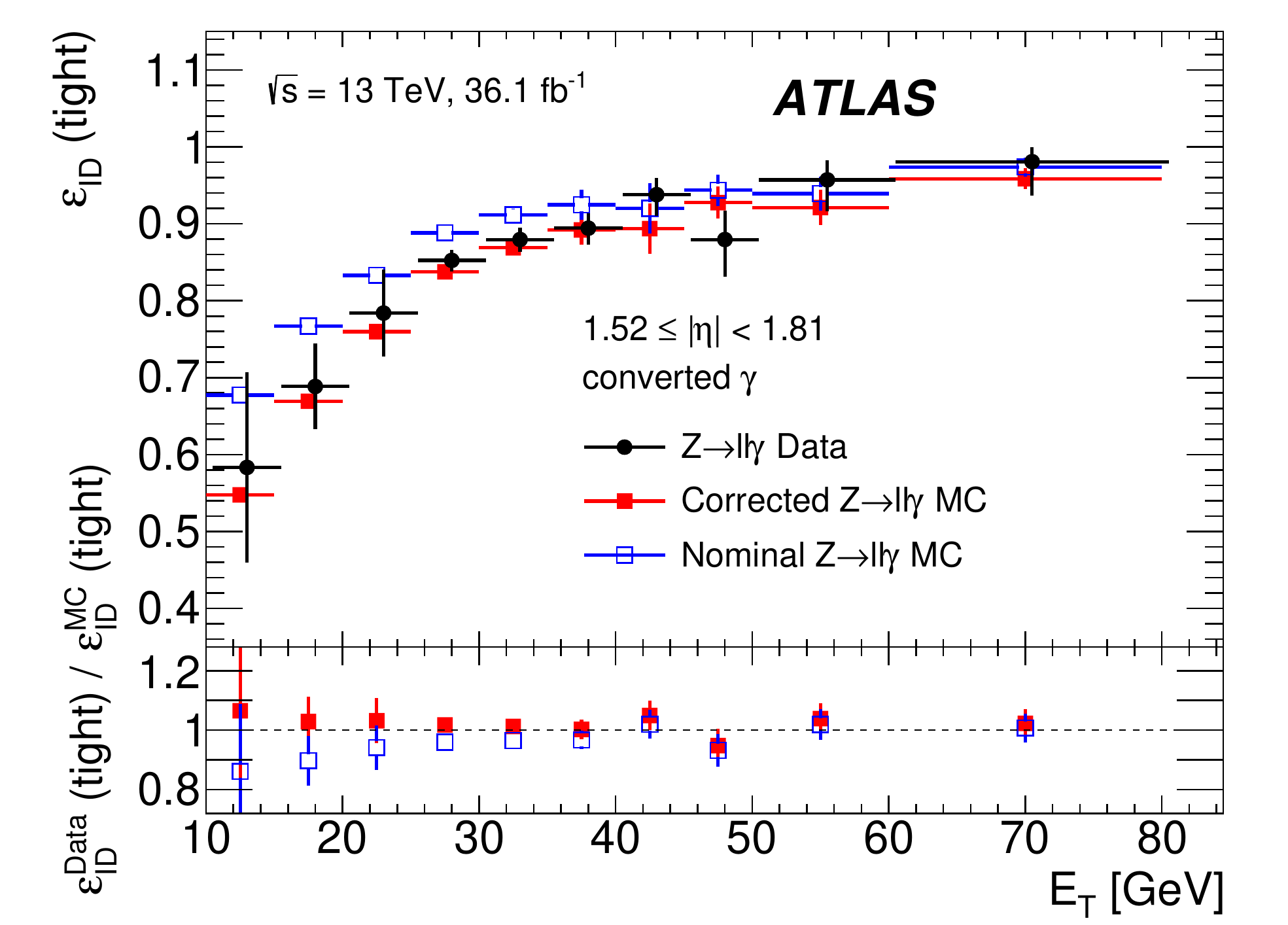}}
   \subfloat[]{\label{fig:radz_eff_conv__d}\includegraphics[width=.5\textwidth]{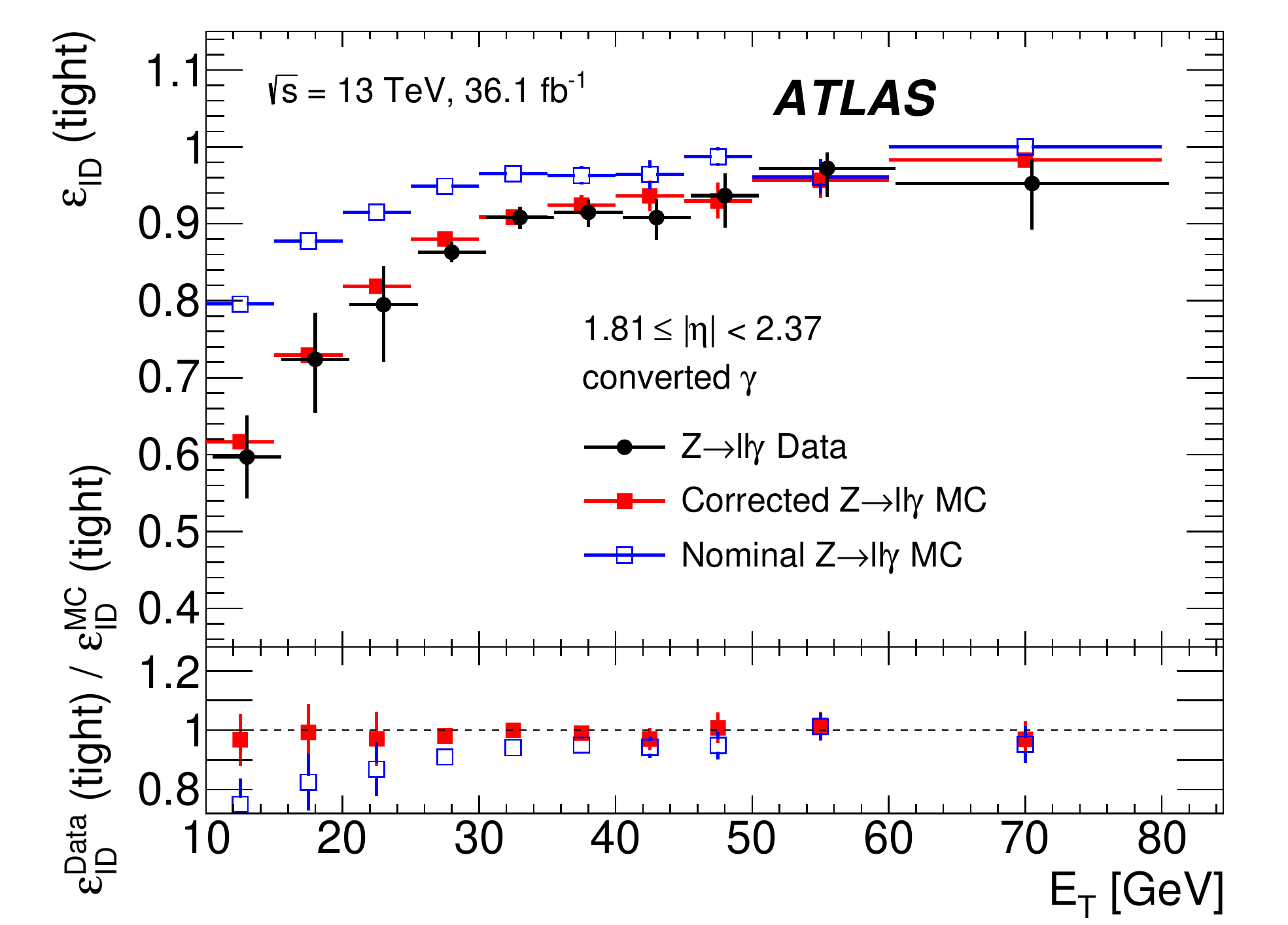}}
   \caption{Comparison of the measurements of the data-driven identification efficiency for converted photons measurements obtained using the radiative $Z$ method with
   the predictions from $Z\to\ell\ell\gamma$ simulation as a function of photon $\ET$, for the four
     pseudorapidity intervals (a) $|\etacal|<0.6$, (b) $0.6\leq|\etacal|<1.37$,
     (c) $1.52 \leq |\etacal|<1.81$, and (d) $1.81\leq|\etacal|<2.37$.
     Predictions are shown for both the nominal simulation and with the corrections described in Section~\ref{sec:DataMC:FF}.
     The bottom panels show the ratio of the data-driven values to the
     MC predictions (referred to as scale factors in the text).} 
   \label{fig:radZ_eff_conv_vs_mc} 
\end{figure}
%
\begin{figure}
   \centering 
   \subfloat[]{\label{fig:eemm_eff_unconv_a}\includegraphics[width=.5\textwidth]{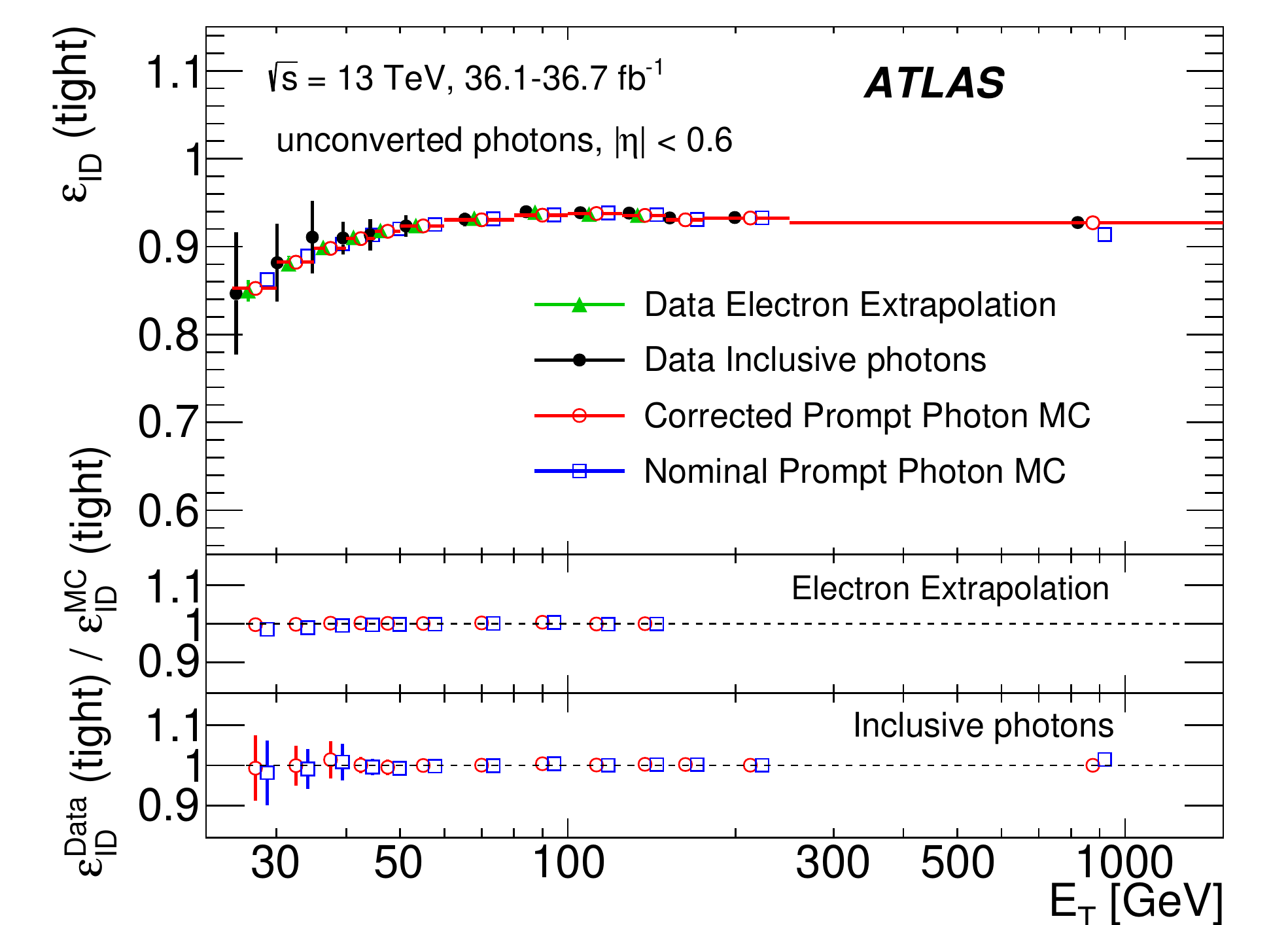}} 
   \subfloat[]{\label{fig:eemm_eff_unconv_b}\includegraphics[width=.5\textwidth]{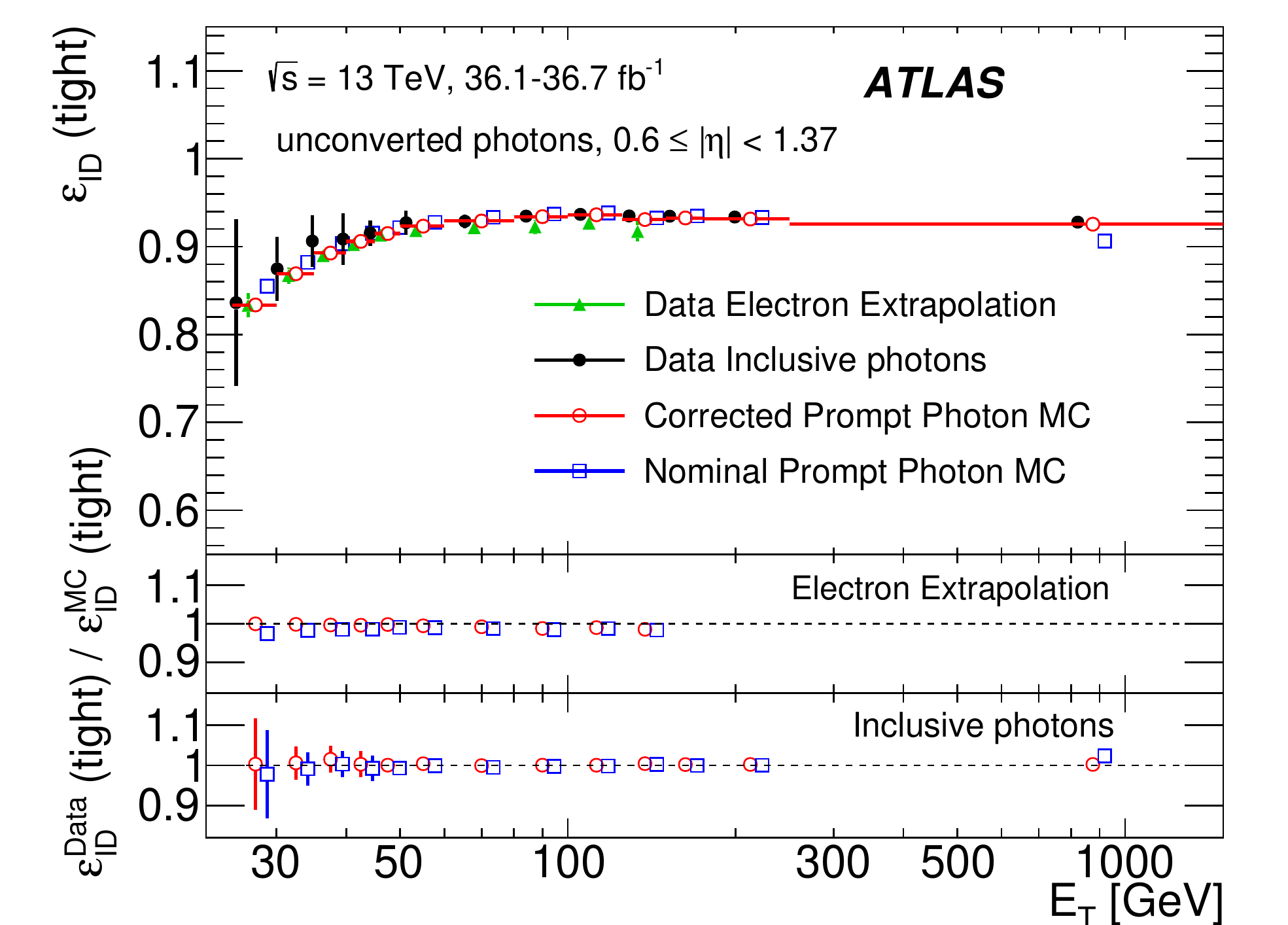}}\\
   \subfloat[]{\label{fig:eemm_eff_unconv_c}\includegraphics[width=.5\textwidth]{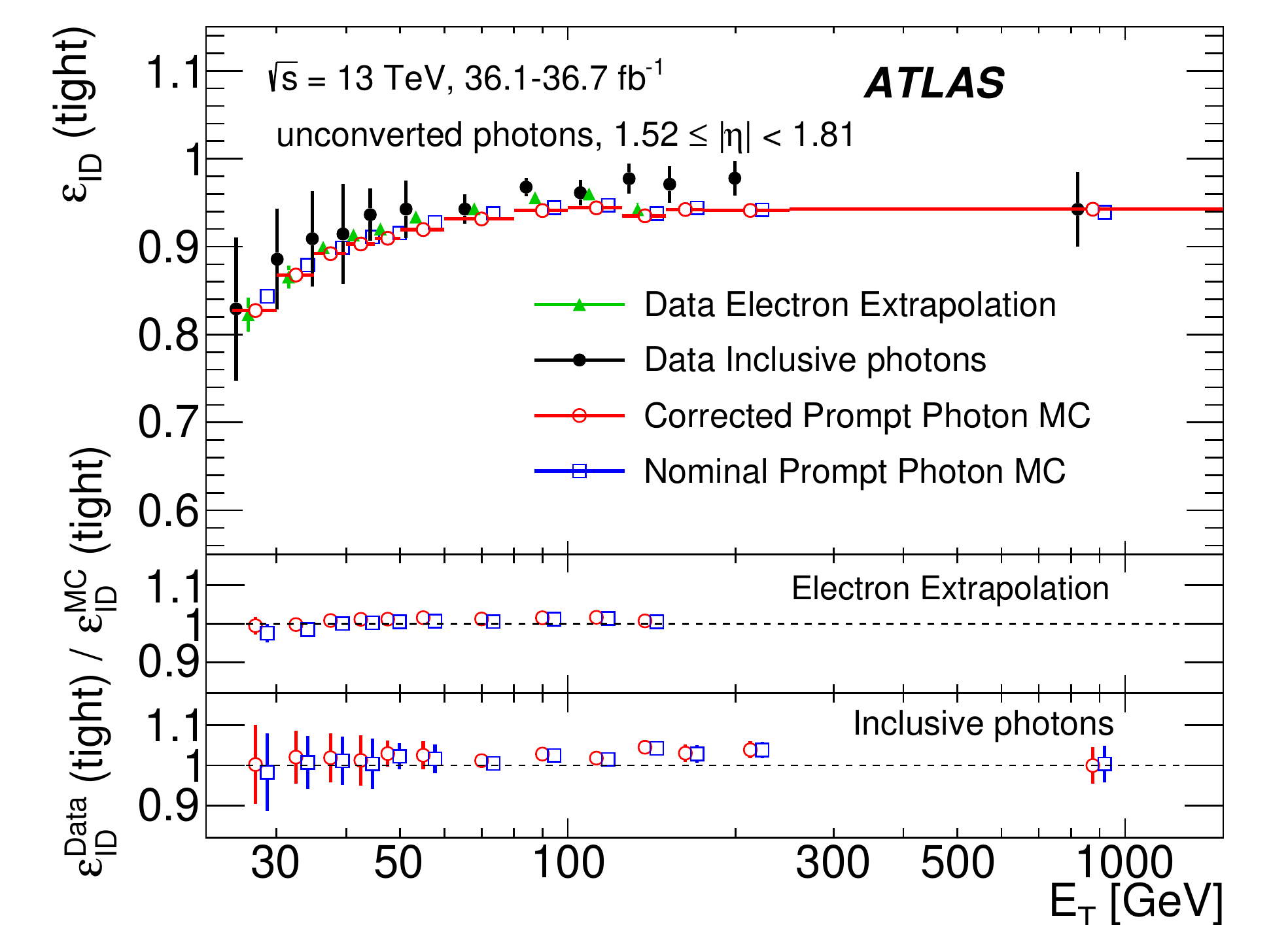}}
   \subfloat[]{\label{fig:eemm_eff_unconv_d}\includegraphics[width=.5\textwidth]{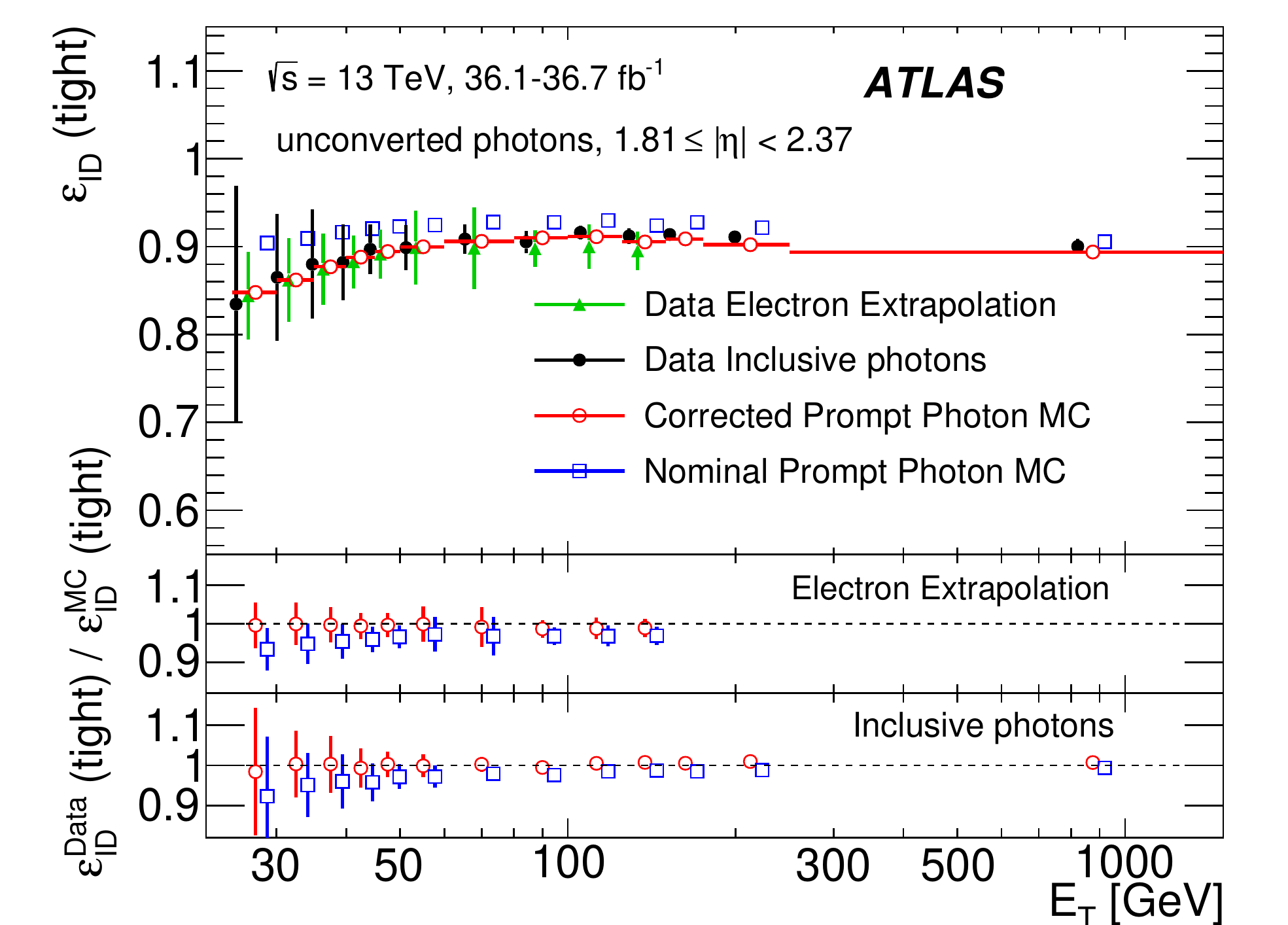}}
   \caption{Comparison of the measurements of the data-driven identification efficiency for unconverted photons obtained using the electron extrapolation and inclusive photon methods with
   the predictions from prompt-photon+jet simulation as a function of photon $\ET$, for the four
     pseudorapidity intervals (a) $|\etacal|<0.6$, (b) $0.6\leq|\etacal|<1.37$,
     (c) $1.52 \leq |\etacal|<1.81$, and (d) $1.81\leq|\etacal|<2.37$.
     Predictions are shown for both the nominal simulation and with the corrections described in Section~\ref{sec:DataMC:FF}.
     The bottom panels show the ratio of the data-driven values to the
     MC predictions (referred to as scale factors in the text).
     The last bin extends to 1.5 \TeV.
   } 
   \label{fig:eemm_eff_unconv_vs_mc} 
\end{figure}
\begin{figure}
   \centering
   \subfloat[]{\label{fig:eemm_eff_conv_a}\includegraphics[width=.5\textwidth]{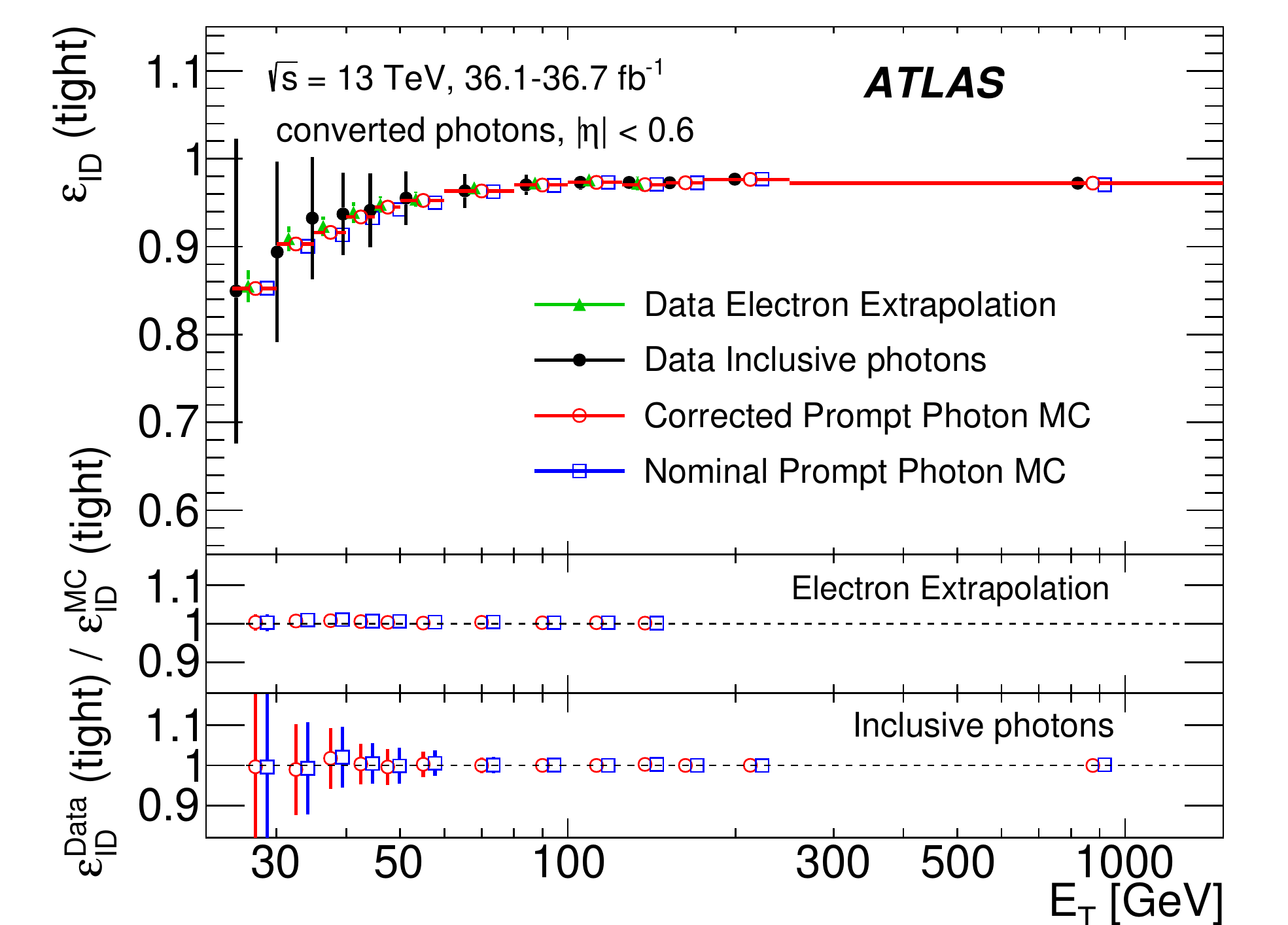}}
   \subfloat[]{\label{fig:eemm_eff_conv_b}\includegraphics[width=.5\textwidth]{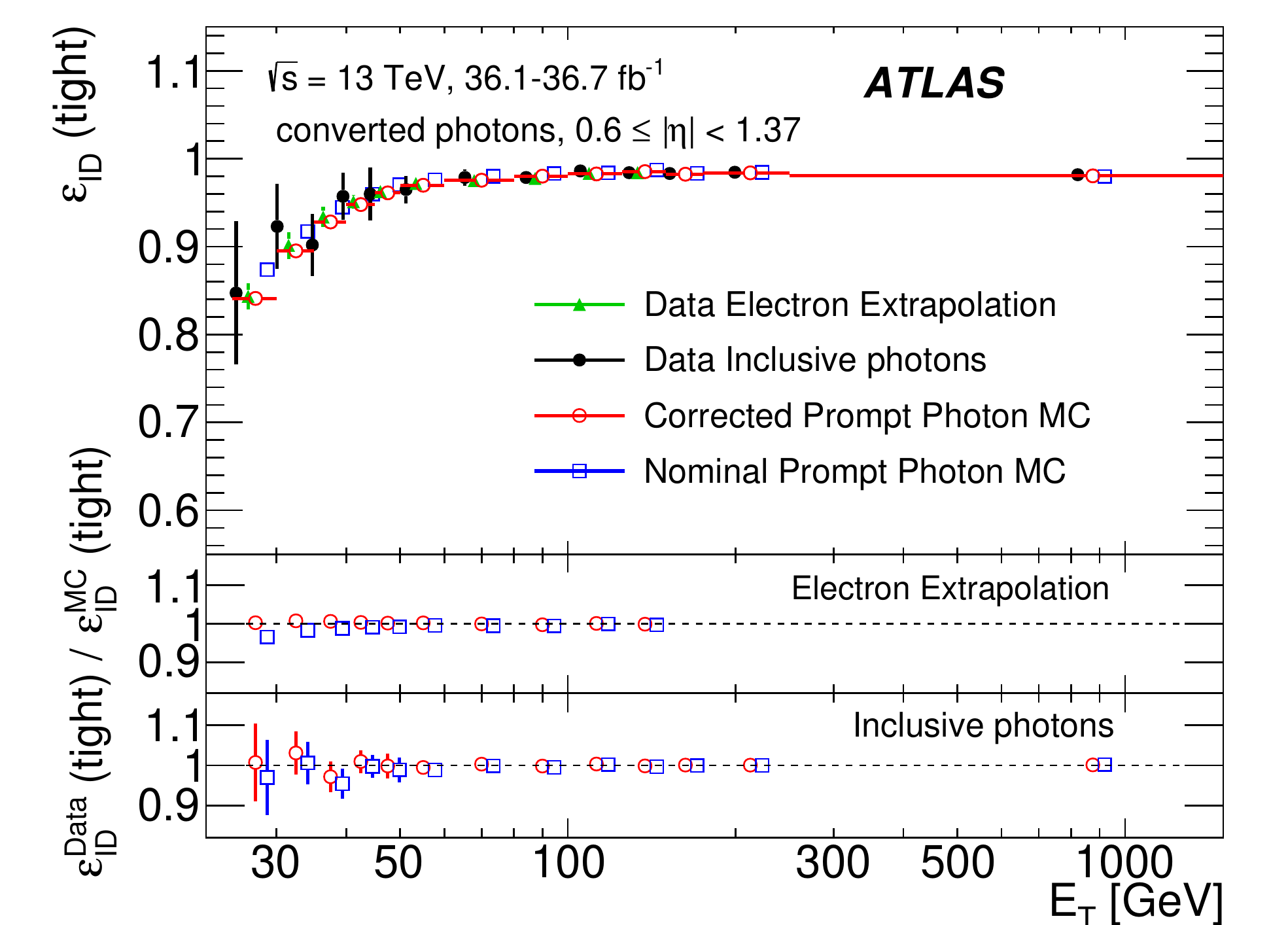}}\\
   \subfloat[]{\label{fig:eemm_eff_conv_c}\includegraphics[width=.5\textwidth]{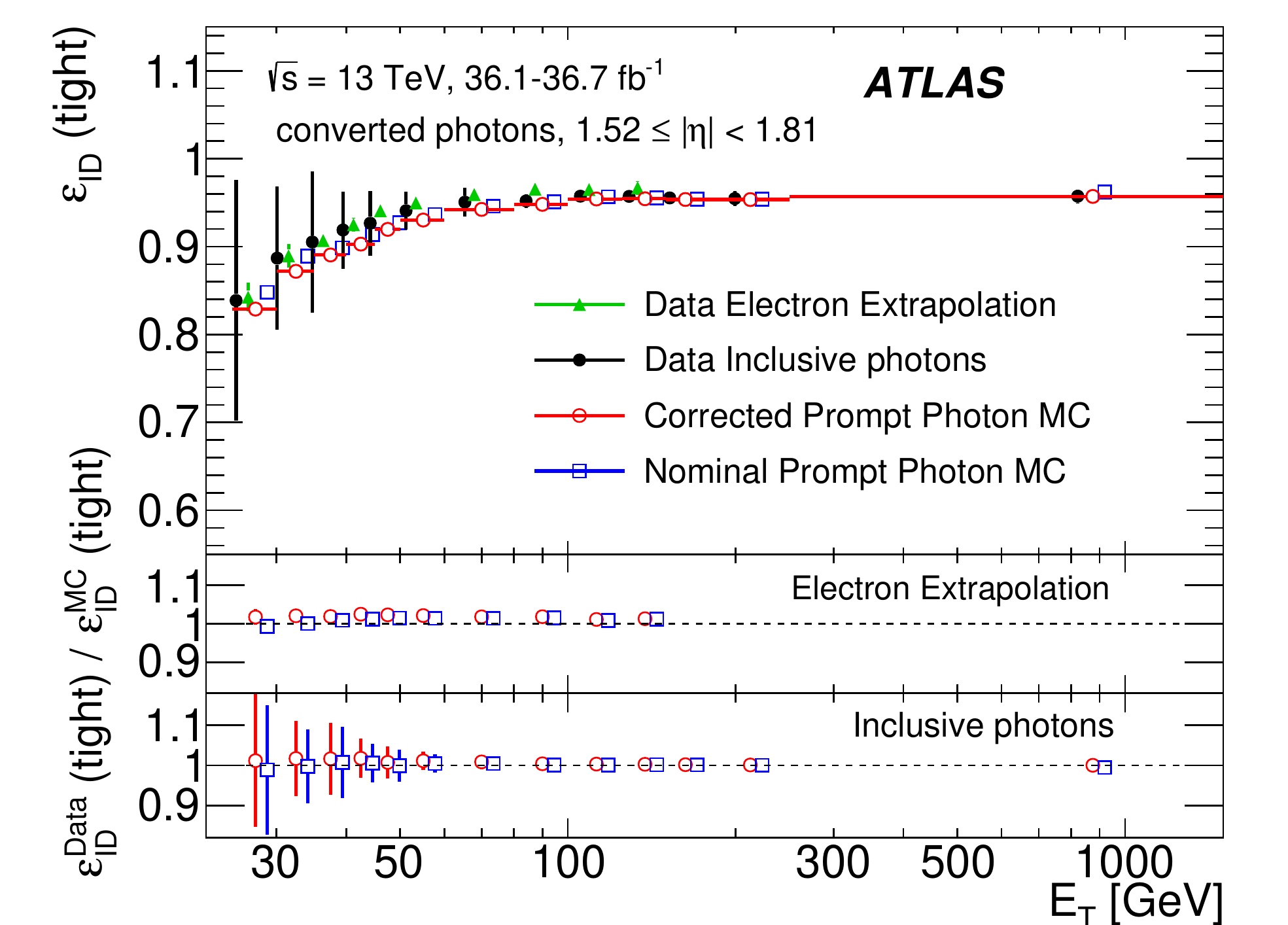}}
   \subfloat[]{\label{fig:eemm_eff_conv_d}\includegraphics[width=.5\textwidth]{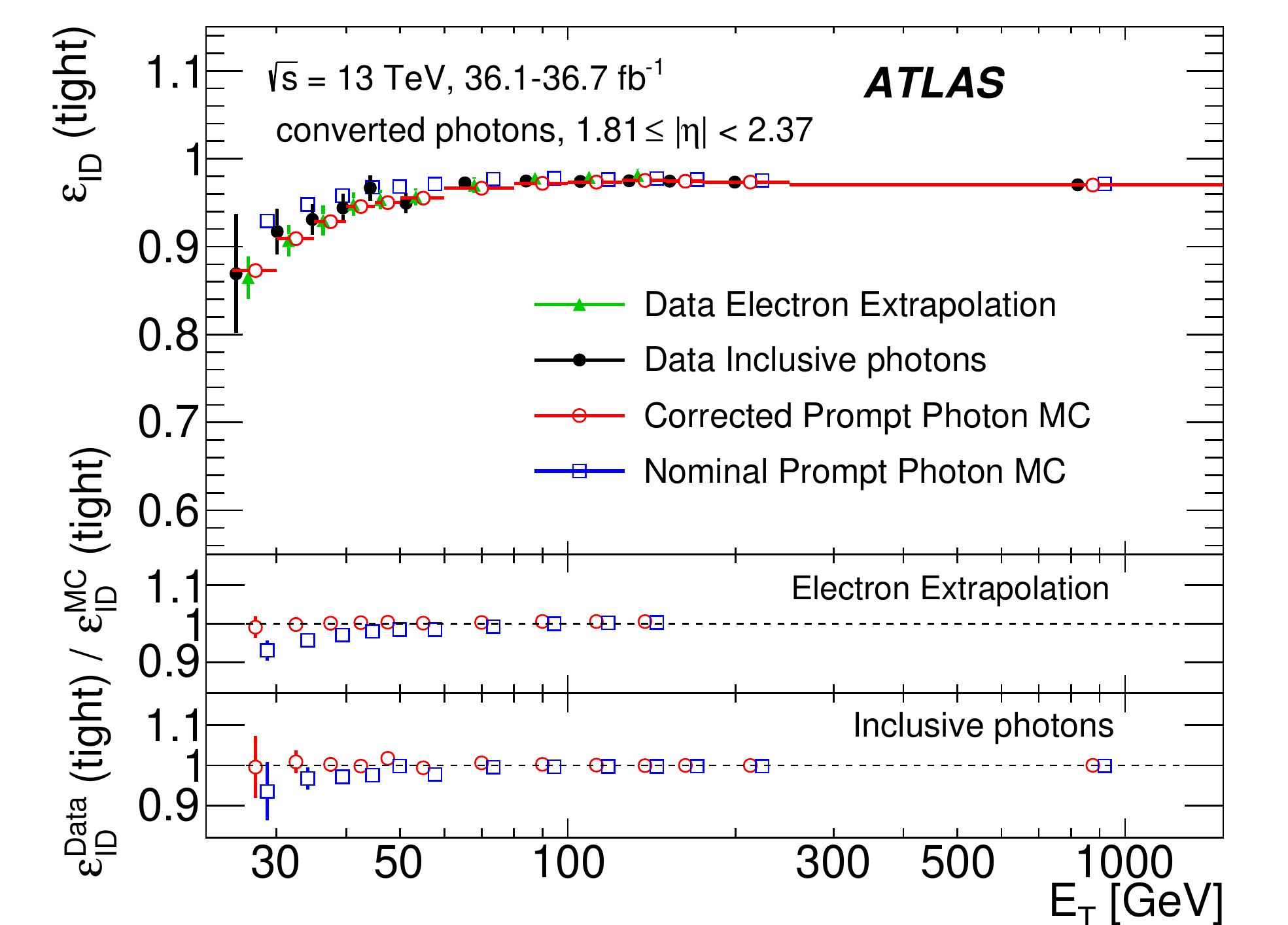}}
   \caption{Comparison of the measurements of the data-driven identification efficiency for converted photons obtained using the electron extrapolation and inclusive photon methods with
   the predictions from prompt-photon+jet simulation as a function of photon $\ET$, for the four
     pseudorapidity intervals (a) $|\etacal|<0.6$, (b) $0.6\leq|\etacal|<1.37$,
     (c) $1.52 \leq |\etacal|<1.81$, and (d) $1.81\leq|\etacal|<2.37$.
     Predictions are shown for both the nominal simulation and with the corrections described in Section~\ref{sec:DataMC:FF}.
     The bottom panels show the ratio of the data-driven values to the
     MC predictions (referred to as scale factors in the text).
     The last bin extends to 1.5 \TeV.
   } 
   \label{fig:eemm_eff_conv_vs_mc} 
\end{figure}
The three methods show excellent agreement over the energy ranges where their measurements overlap, with photon \effID\ values ranging from 50--60\% at $\ET=10\,\GeV$, to 95--99\% (unconverted) and 88--96\% (converted) for photons with $\ET$ above 250~\GeV\ . The \effID\ values obtained from the inclusive photon and electron extrapolation methods can be compared directly, since they both apply to photons from the $\gamma$+jet  prompt-photon production process. The radiative $Z$ method applies to the photon sample obtained from the $Z \rightarrow \ell\ell\gamma$ process, with a different \effID\ value due to differences in the photon kinematics within each measurement bin, and because of the impact of photons from fragmentation processes in $\gamma$+jet production, which are absent from $Z \rightarrow \ell\ell\gamma$. Fragmentation photons are typically identified with lower efficiency, due to higher hadronic activity around the photon candidate. However, they only constitute a small fraction of the total sample, about 10\% at low \ET\ and a few percent at higher \ET~\cite{PERF-2013-04}, and their impact does not lead to significant differences between the \effID\ values measured by the different methods. 

Efficiency scale factors (SF) are computed as the ratios of the data-driven values of \effID\ to the values obtained in simulation. To account for the differences in photon properties discussed above, simulated $\gamma$+jet production is used for the electron extrapolation and inclusive photon methods, while $Z \rightarrow \ell\ell\gamma$ simulation is used for the radiative $Z$ method. Results from the radiative $Z$ method are shown in Figures~\ref{fig:radZ_eff_unconv_vs_mc} and~\ref{fig:radZ_eff_conv_vs_mc} for unconverted and converted photons respectively, while those from the other two methods are shown in Figures~\ref{fig:eemm_eff_unconv_vs_mc} and~\ref{fig:eemm_eff_conv_vs_mc}. The SF values are all compatible with unity, and do not show significant trends as a function of $\ET$.

To increase the measurement precision in the ranges where measurements from different methods are available, a combination of the three measurements is performed. The SF values are combined rather than the \effID\ directly, in order to account for the small differences discussed above in the composition of the photon samples which may affect the measured \effID\ value.
The combination is performed using the BLUE algorithm~\cite{Lyons,Valassi}: the combined scale factor in each measurement bin computed as a linear combination of the per-method scale factors, with coefficients that are obtained from the values of their statistical and systematic uncertainties and their correlations.

Statistical uncertainties are taken to be uncorrelated since the methods are applied to disjoint datasets. Systematic uncertainties in material modelling, which affect both the radiative $Z$ and inclusive photon method, are expected to have correlated effects in these two measurements. However, since these methods provide measurements in largely disjoint photon $\ET$ intervals, the correlation assumption has a negligible impact on the combined result. All sources of statistical and systematic uncertainties are therefore assumed to be uncorrelated.
The combined results are shown in Figures~\ref{fig:combSF_unconv} and~\ref{fig:combSF_conv} for unconverted and converted photons respectively.
The central values are again compatible within uncertainties, which are at the level of 1\% or less in the region $30 < \ET < 150\,\gev$, as well as at higher \ET excepting the region $1.52 \leq |\etacal|< 1.81$ where they reach 4\% at $\ET = 1.5\,\tev$. For $\ET < 30\,\gev$ the uncertainties rise due to the limited size of the photon samples, reaching 25\% at $\ET = 10\,\gev$. 

\begin{figure}
   \centering
   \includegraphics[width=.9\textwidth]{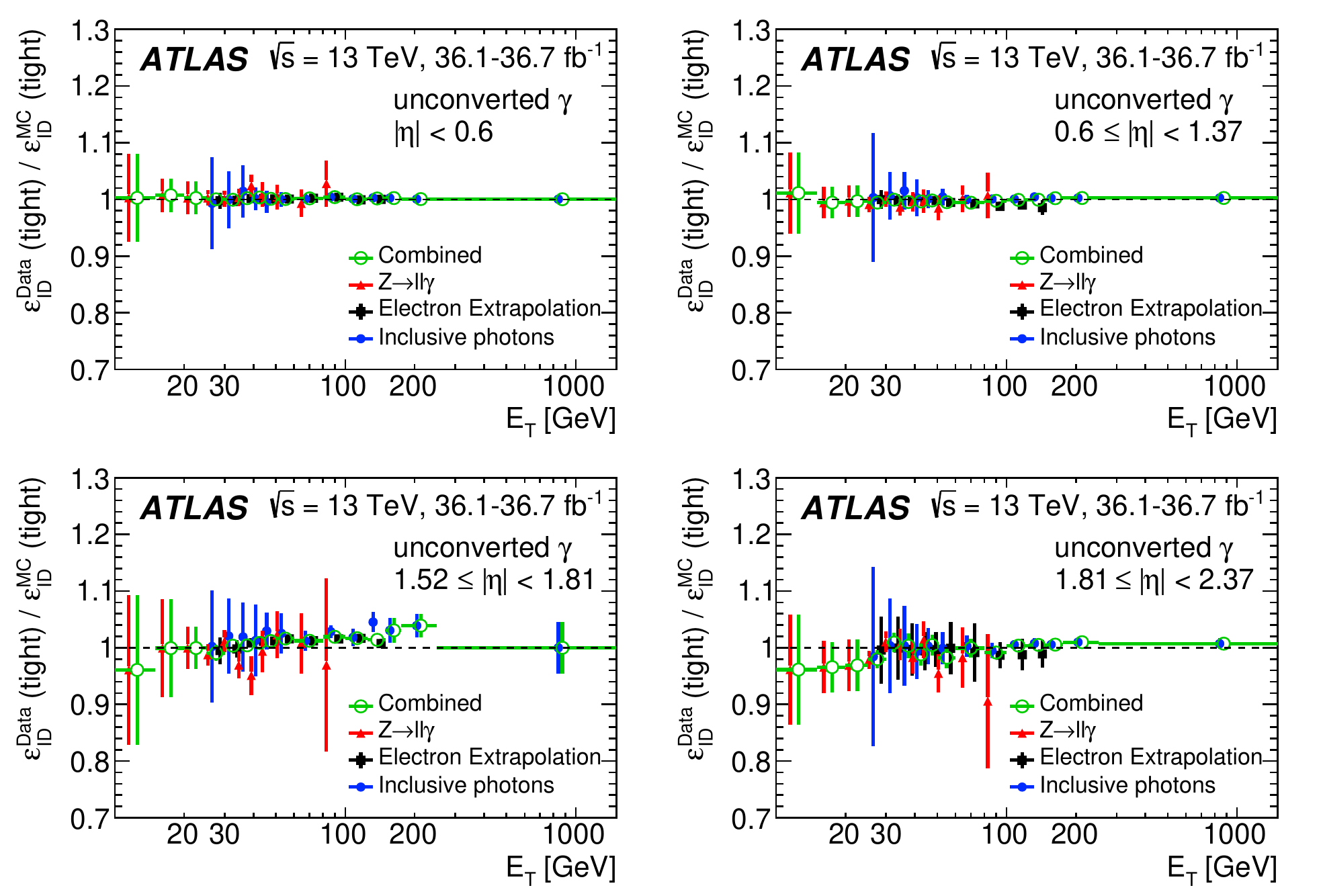}
   \caption{Efficiency scale factors (SF) for each method and their combination for unconverted photons.
     The last bin extends to 1.5 \TeV.
   }
   \label{fig:combSF_unconv}
\end{figure}

\begin{figure}
   \centering
   \includegraphics[width=.9\textwidth]{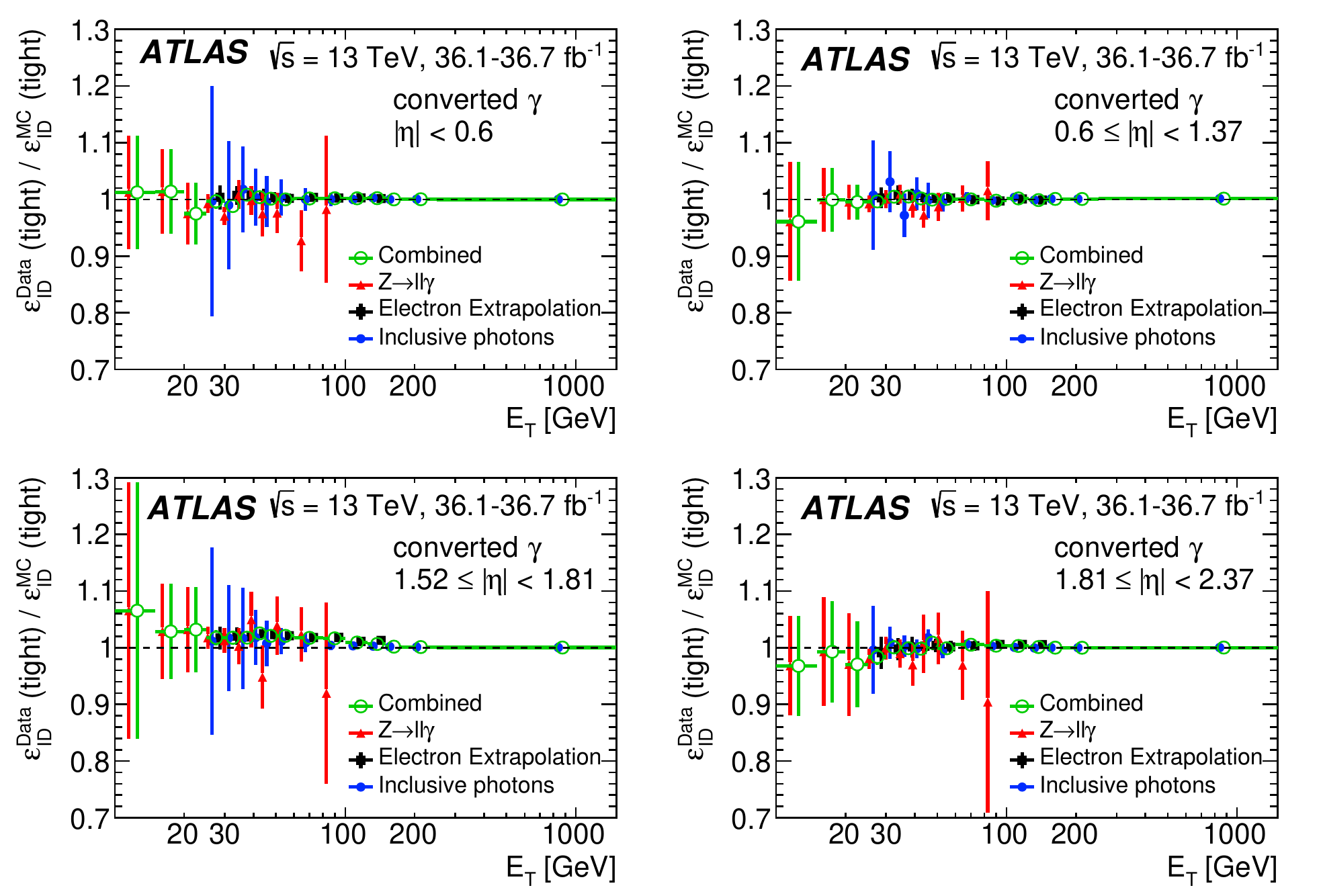}
   \caption{Efficiency scale factors (SF) for each method and their combination for converted photons.
     The last bin extends to 1.5 \TeV.
   }
   \label{fig:combSF_conv} 
\end{figure}

\subsection{Dependency of the photon identification efficiency on pile-up}
\label{sec:PileUp}

The presence of additional $pp$ interactions together with the hard-scattering process is expected to affect the photon isolation efficiency. To describe this effect, the simulated MC events are reweighted to reproduce the distribution of $\mu$ observed in data.  Thus, if the dependence of the identification efficiency on $\mu$ is correctly described by simulation, there is no need to account for it in analyses using photons.

The dependency of the photon identification efficiency on $\mu$ is measured using photon candidates originating from the $Z\to\ell\ell\gamma$ radiative decay process, following the procedure described in Section~\ref{sec:measurements:RadZ} and using only data collected in 2016. The measurement is restricted to candidates with transverse momentum $\ET>20~\GeV$ for which background contamination is lower, and no background subtraction procedure is performed. The requirement $\ET~<~40~\GeV$ is also applied to avoid large statistical uncertainties. Figure~\ref{fig:eff_RadZ_pileup} shows the evolution of photon identification efficiency for these events as a function of $\mu$ in different regions of pseudorapidity, as measured in data and in MC simulation. A clear dependency is observed, with the photon identification efficiency dropping by about 10-15\% when going from $\mu \sim 5$ to $\mu \sim 40$, depending on the photon candidate's pseudorapidity and conversion status.  The ratio of the efficiency dependence on $\mu$ in data to that in simulation is modelled with a linear dependence on $\mu$ and found to be compatible with a constant value within the measurement uncertainties in all detector regions for both the unconverted and converted photons. \footnote{Any constant difference from unity in the data/MC ratio is accounted for by the efficiency scale factors.} Similar results are obtained with lower precision from the other measurement methods.

\begin{figure}
  \centering
  \includegraphics[width=0.48\textwidth]{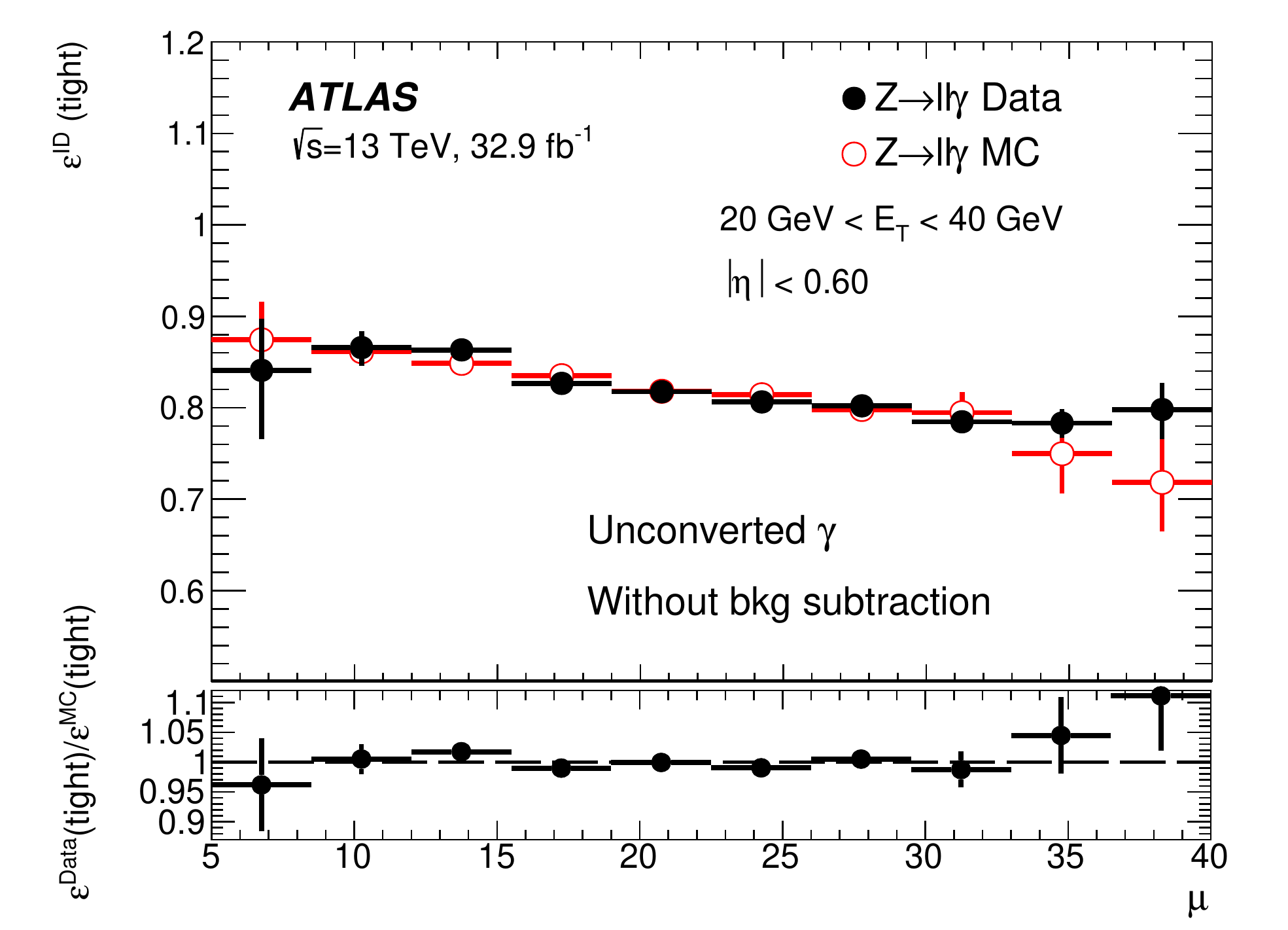}
  \includegraphics[width=0.48\textwidth]{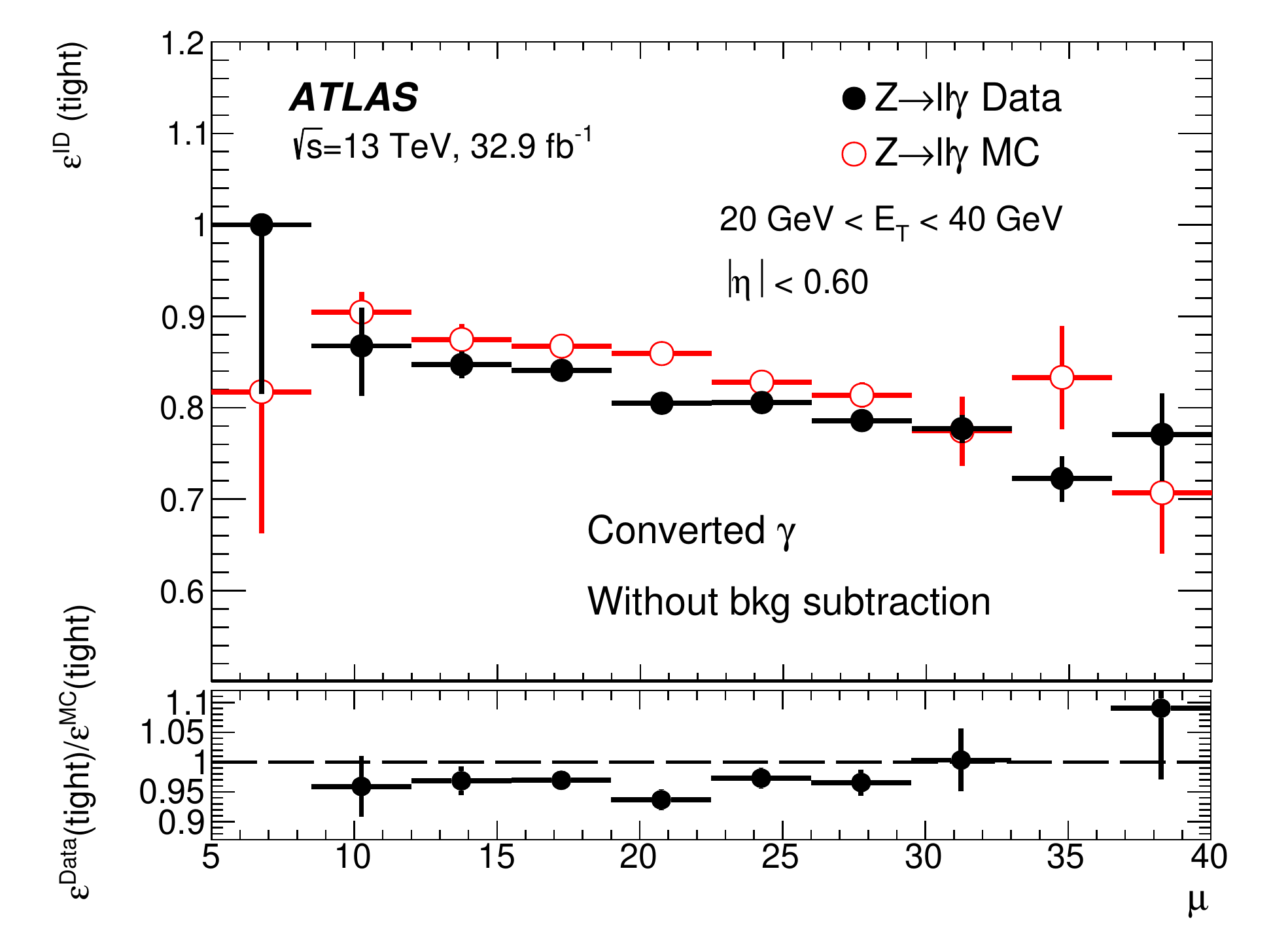} 
  \includegraphics[width=0.48\textwidth]{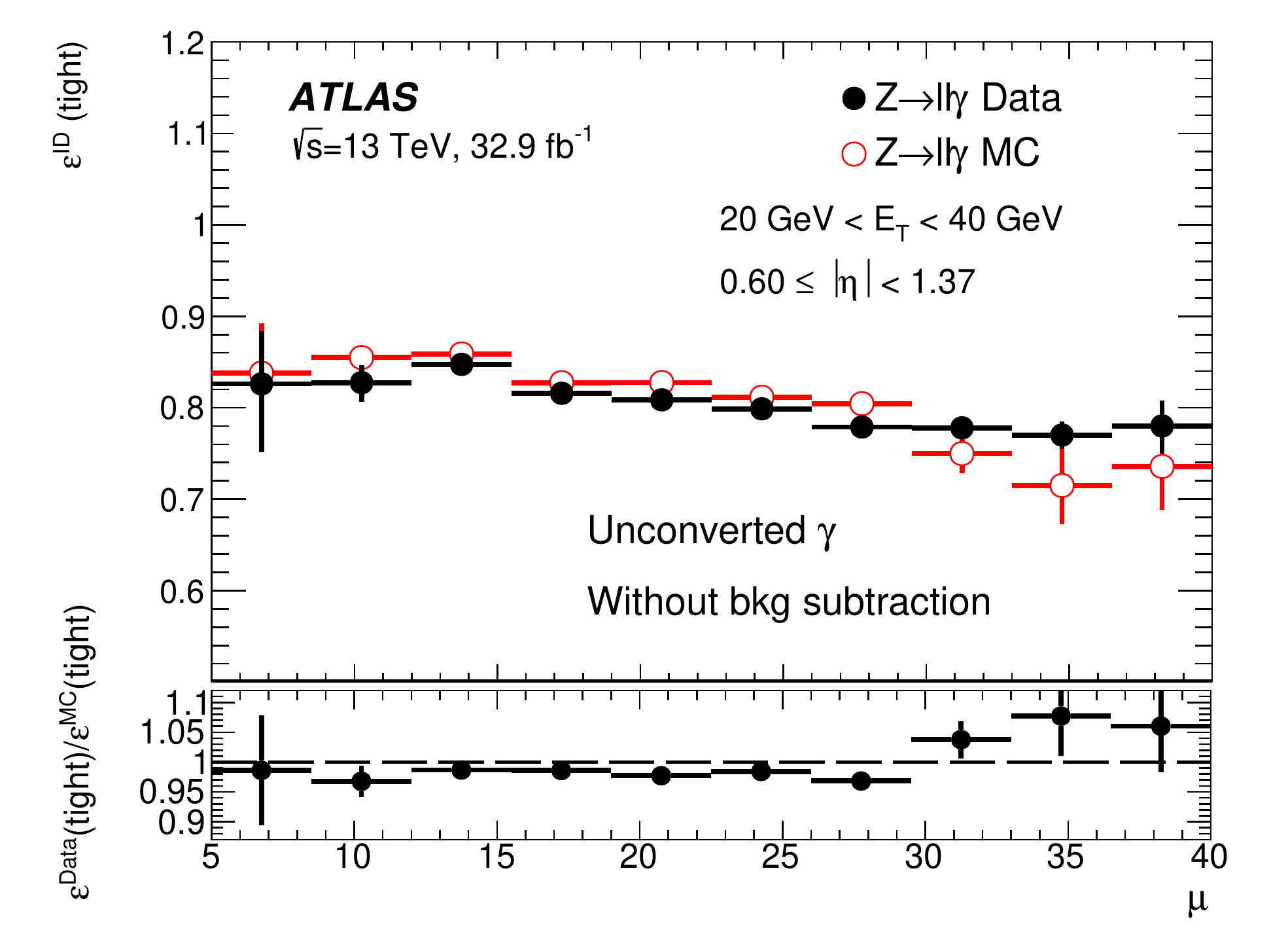}
  \includegraphics[width=0.48\textwidth]{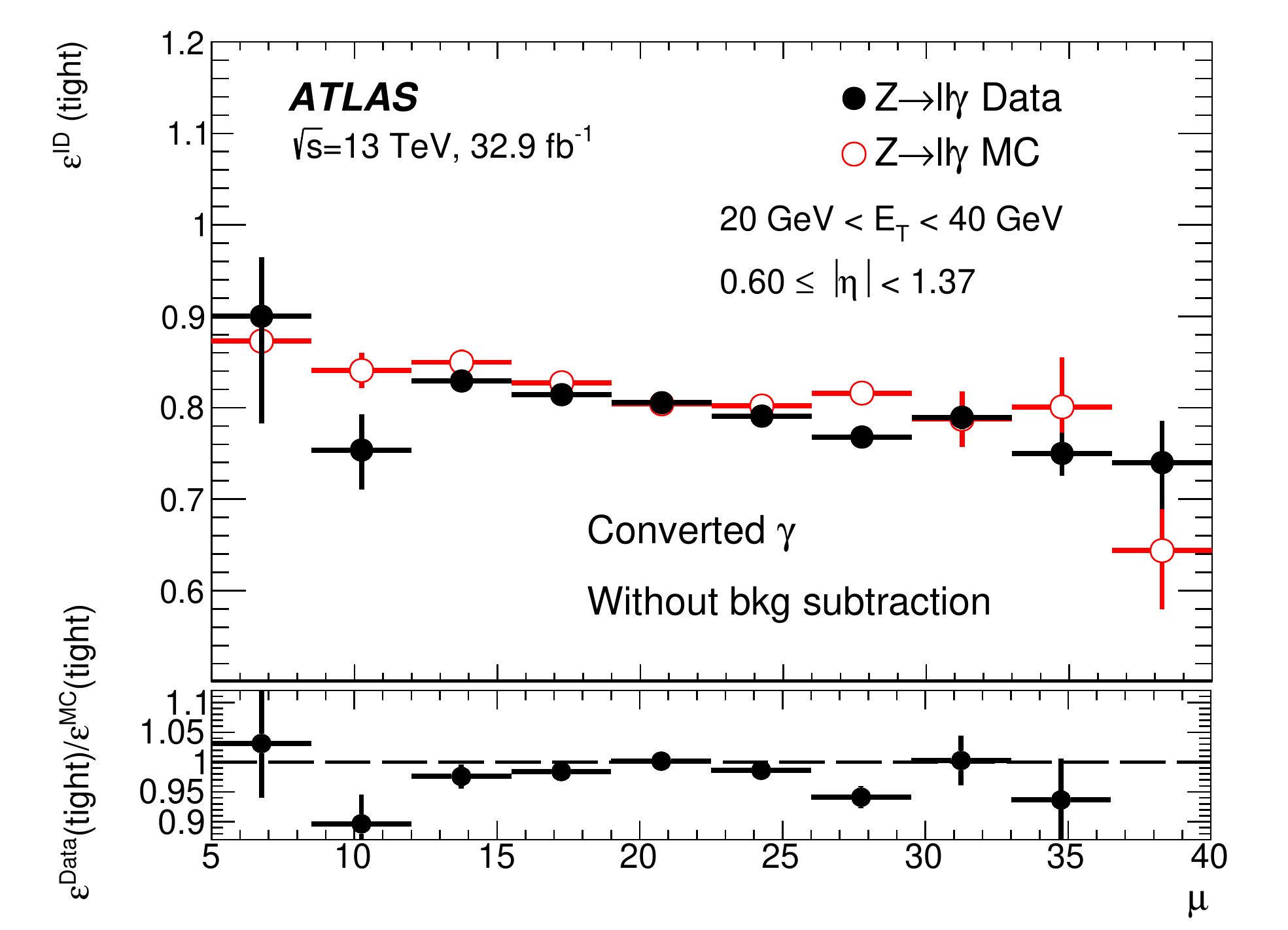} 
  \includegraphics[width=0.48\textwidth]{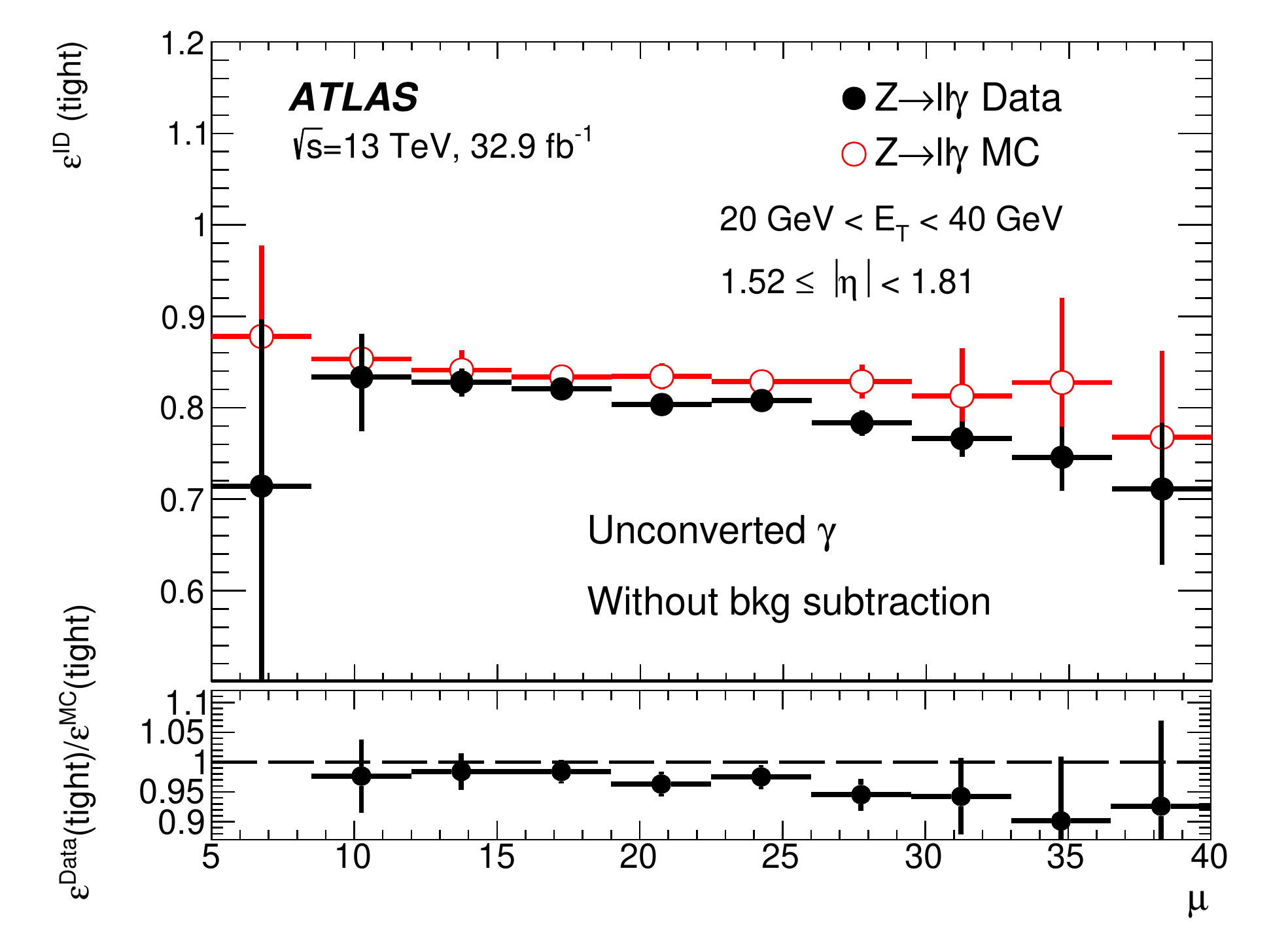}
  \includegraphics[width=0.48\textwidth]{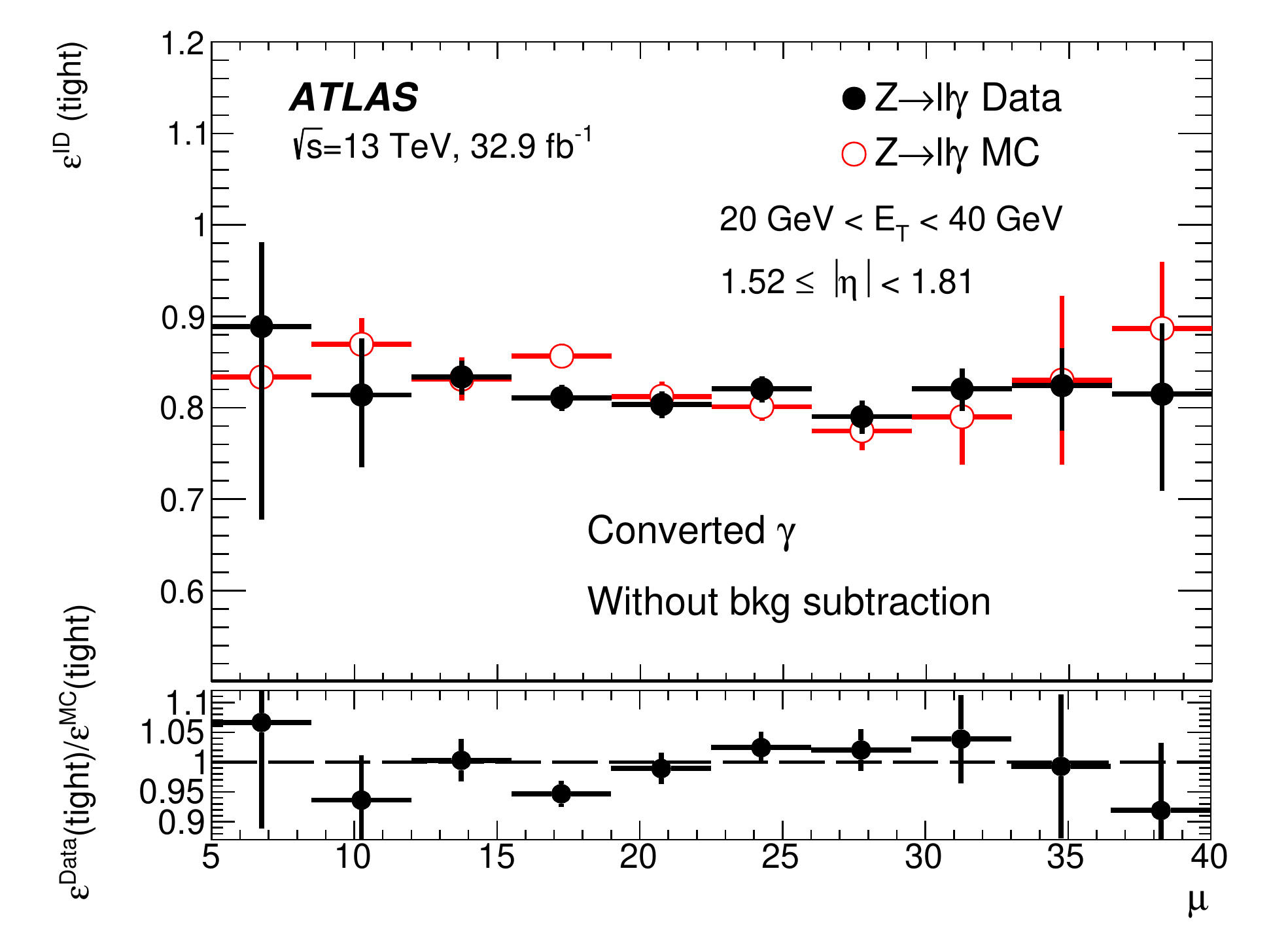} 
  \includegraphics[width=0.48\textwidth]{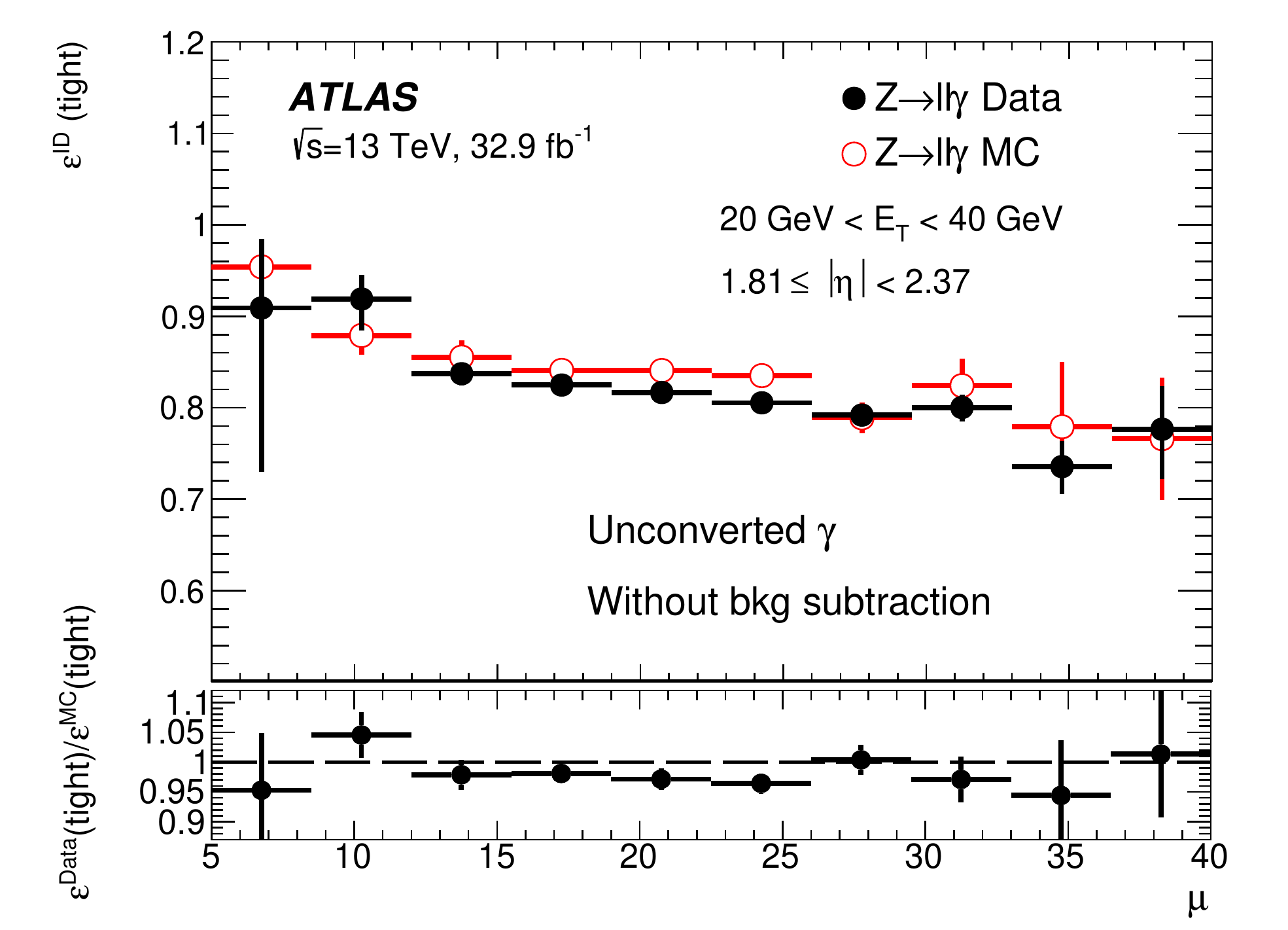}
  \includegraphics[width=0.48\textwidth]{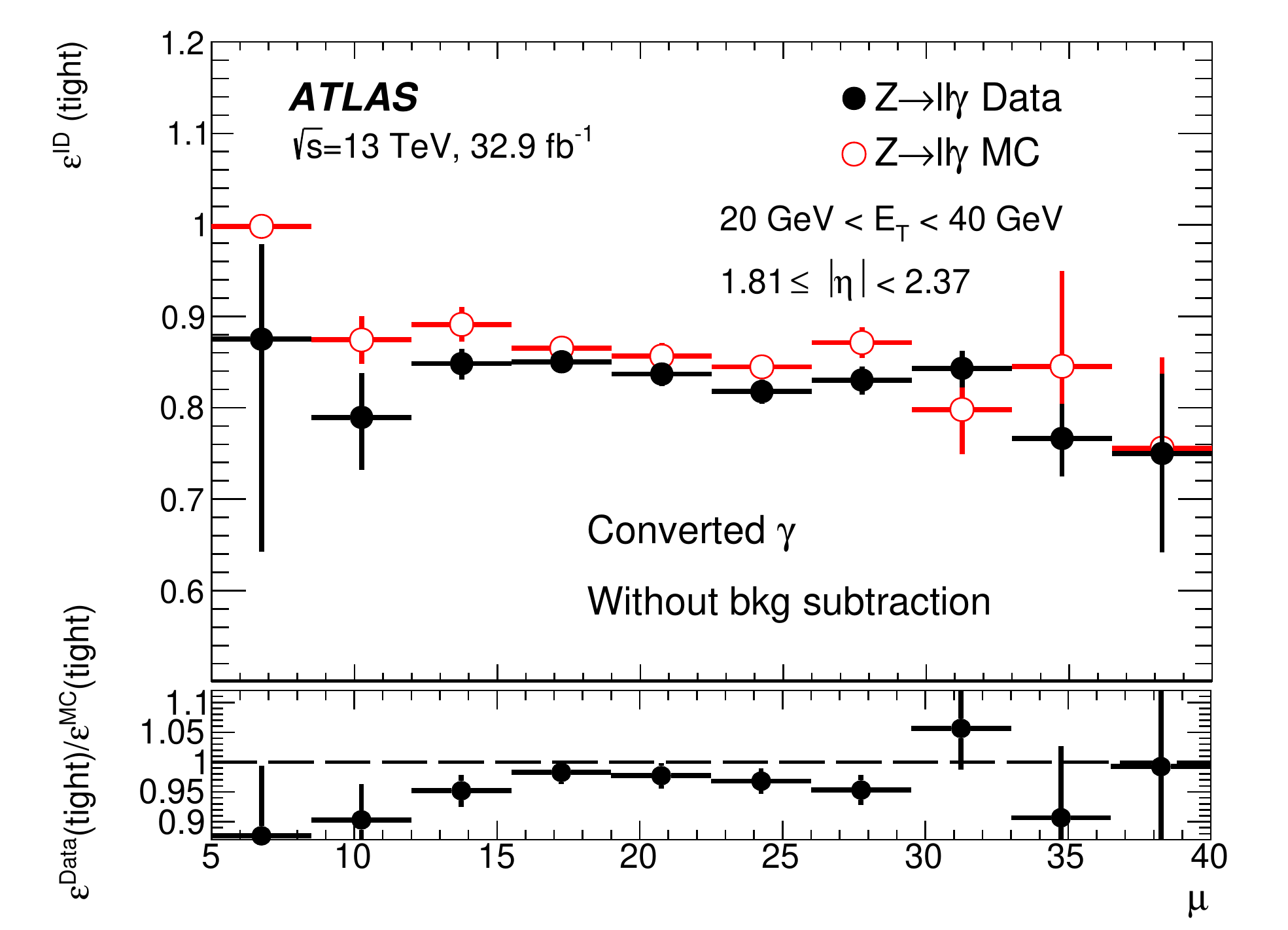} 
  \caption{Photon identification efficiencies as a function of the number of $pp$ interactions per bunch crossing $\mu$ for reconstructed unconverted photons (left) and converted photons (right), in four pseudorapidity regions.}
  \label{fig:eff_RadZ_pileup}
\end{figure}

\subsection{Dependency of the photon identification efficiency on isolation}
\label{sec:PIDvsIso}

All measurements of the photon identification efficiency discussed in Section~\ref{sec:measurements} are performed for isolated photons satisfying the loose isolation selection described in Section~\ref{sec:PhotonIso}. The dependency of the identification efficiency is also studied as a function of the isolation criterion, using photon candidates from radiative $Z\to\ell\ell\gamma$ decays, and following the same procedure discussed in Section~\ref{sec:measurements:RadZ}. Due to the high purity of these photon candidates, it is possible to perform the measurement for the tight isolation selection described in Section~\ref{sec:PhotonIso}, or for an isolation requirement using only the calorimeter requirement of the tight selection. The results obtained are in agreement with the nominal ones within the respective systematic and statistical uncertainties.

\section{Measurement of the $e\to\gamma$ fake-rate}
\label{sec:EGammaFakeRate}

Although the photon and the electron reconstruction algorithms are designed to reduce the misidentification of electrons as photons (see Section~\ref{sec:PhotonReco} for details) a residual small fraction of electrons can still be reconstructed as photon candidates and pass the photon identification requirements. Since the reconstruction efficiency of an electromagnetic cluster is about 100\,\%, the fraction of electrons wrongly reconstructed and misidentified as photon candidates is mainly due to tracking inefficiency or bad matching of the track with the electromagnetic cluster reconstructed in the calorimeter.

The electron-to-photon fake-rate is defined as the ratio of the probability of wrongly reconstructing and identifying an electron as a photon to the probability of correctly reconstructing an electron:
\begin{equation*}
  F_{e\to\gamma} \equiv  \frac{\epsilon(e^\mathrm{truth}\rightarrow\gamma^\mathrm{reco}) \epsilon_\gamma }{\epsilon(e^\mathrm{truth}\rightarrow e^\mathrm{reco}) \epsilon_e} 
  \equiv
  \frac{\rho}{1-\rho} \frac{\epsilon_\gamma}{\epsilon_e}
  \, ,
  \label{eq:fake_rate_reco}
\end{equation*}
where $\epsilon_\gamma$ and $\epsilon_e$ are the identification efficiency for the photon and the electron respectively and $\rho$ is the fraction of true electrons reconstructed as photon candidates.  The fake-rate $F_{e\to\gamma}$ can be used to estimate the background from single-electrons misidentified as photons in analyses using photons: a control region is defined using the same selection as for the signal region, except that the photon reconstruction and identification criteria are replaced by the electron selection criteria given below. The misidentified electron background yield is then obtained as the product of $F_{e\to\gamma}$ with the observed electron yield in this control region.

The fake-rate $F_{e\to\gamma}$ is measured in data from the numbers of electron--positron candidates from $Z\to ee$ decays that are reconstructed as $ee$ or $e\gamma$ pair. Such yields are estimated by comparing the observed $ee$ and $e\gamma$ invariant mass spectra around the $Z$ boson peak, after having subtracted the respective background contributions, estimated from the sidebands around the $Z$ boson peak.  Events are required to contain at least two opposite-sign electron candidates, or an electron and a photon candidate, both having a transverse momentum greater than $25\,\GeV$ and pseudorapidity $|\eta| < 2.37$, excluding the transition region $1.37\leq|\eta|<1.52$. Electron candidates are required to satisfy the electron tight identification criterion and the electron tight isolation criterion \cite{ATLAS-CONF-2016-024}. Photon candidates are required to meet the tight identification criterion and the \emph{loose} isolation criterion, both discussed in Section~\ref{sec:PhotonIso}. When both an electron and a photon candidate are selected within a cone of radius $\Delta R=0.4$, the photon is kept and the electron discarded. Events with $m_{ee}$ or $m_{e\gamma} \in [86,96]\,\GeV$ are selected. If multiple $ee$ and/or $e\gamma$ pairs in the same event pass the previous requirements, only the pair with invariant mass closest to the $Z$ boson world-average mass is kept \cite{PhysRevD.86.010001}. The fake-rate is computed as a function of the transverse momentum and absolute value of the pseudorapidity of the electron candidate.  The measured fake-rate is corrected for the different background contamination in the $ee$ and $e\gamma$ samples. The background contamination is estimated by a combined signal-plus-background fit of the $ee$ or $e\gamma$ invariant mass distribution, where the resonant signal is modelled by a double-sided Crystal Ball function \cite{Gaiser:1982yw} and the continuous background by an exponential function or a Bernstein polynomial, depending on the $\eta$ bin. In order to avoid bias associated with the spectra distortion introduced by the $\pt$ binning, the background is estimated in $\pt/m_{ee,e\gamma}$ bins.

The following systematic uncertainties in the measured $e\to\gamma$ fake-rate are considered: the variation of the invariant mass range used to select the dielectron and electron--photon pair candidates, from $[88.5,93.5]$ \GeV\ to $[83.5,98.5]$ \GeV, leads to a $\sim$ 5\,\% variation of the measured fake-rate; the bias introduced by calibrating the energy of an electron as a photon is found to be $\sim$ 4\,\%.  The impact of the background contributions, estimated by ignoring these components or by increasing them by 100\,\%, results in a $\sim$ 3\,\% variation in the fake-rate.

Comparing the fake-rate measured in MC simulation to the true value yields good agreement in all regions of $\eta$ and $\pt$, with differences of at most 5\,\%.

The integrated $e\to\gamma$ fake-rate, averaged over $\eta$ and for $\pt > 25\,\GeV$, is 0.015 $\pm$ 0.001 for converted photons and 0.030 $\pm$ 0.002 for unconverted photons. Examples of the fake-rate values measured as a function of $\pT$ in two different $\eta$ bins are shown in Figure~\ref{fig:EGammaFakeRate_vs_pt} for converted and unconverted photons. The fake-rate amounts to a few percent in the central region and increases in the more forward ones, reaching a value of 10--20\% for converted photons in the most extreme cases. Figure~\ref{fig:EGammaFakeRate_results} shows the measured fake-rates as a function of $\eta$, averaged over $\pt$ for $\pt > 25\,\GeV$, and compared to the MC predictions. The fake-rate of electrons reconstructed as converted (unconverted) photons is 2--4\,\% (1--2\,\%) in the barrel and 4--12\,\% (2--4\,\%) in the endcaps. The absolute uncertainty, dominated by the systematic component, is typically around 0.2\,\%. The MC simulation generally underestimates the fake-rate by values ranging from a few~\% to about 50\,\% in the most extreme case (converted photons, central region).

\begin{figure}
  \centering
  \includegraphics[width=.49\textwidth]{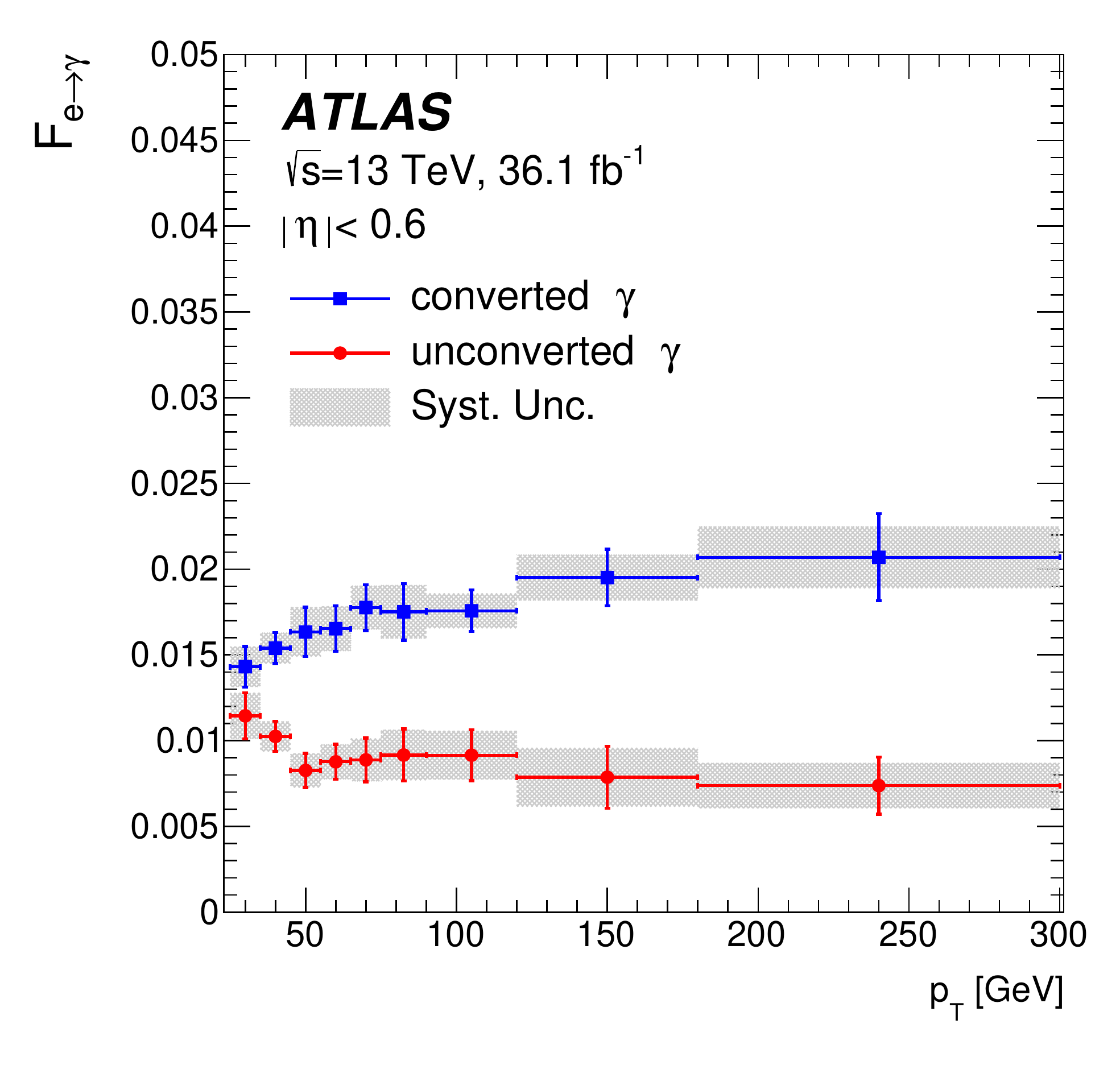}
  \includegraphics[width=.49\textwidth]{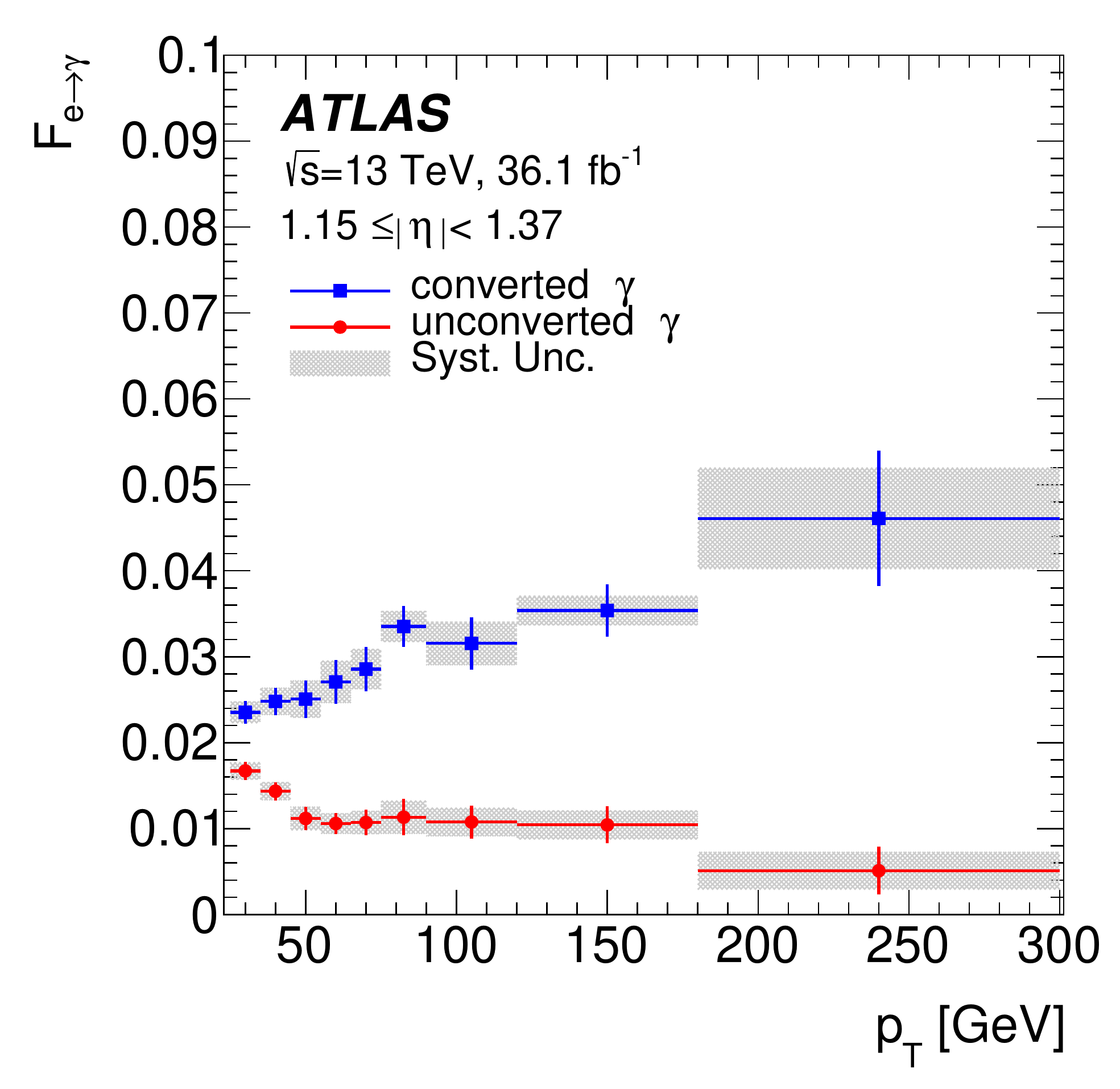} \\
  \includegraphics[width=.49\textwidth]{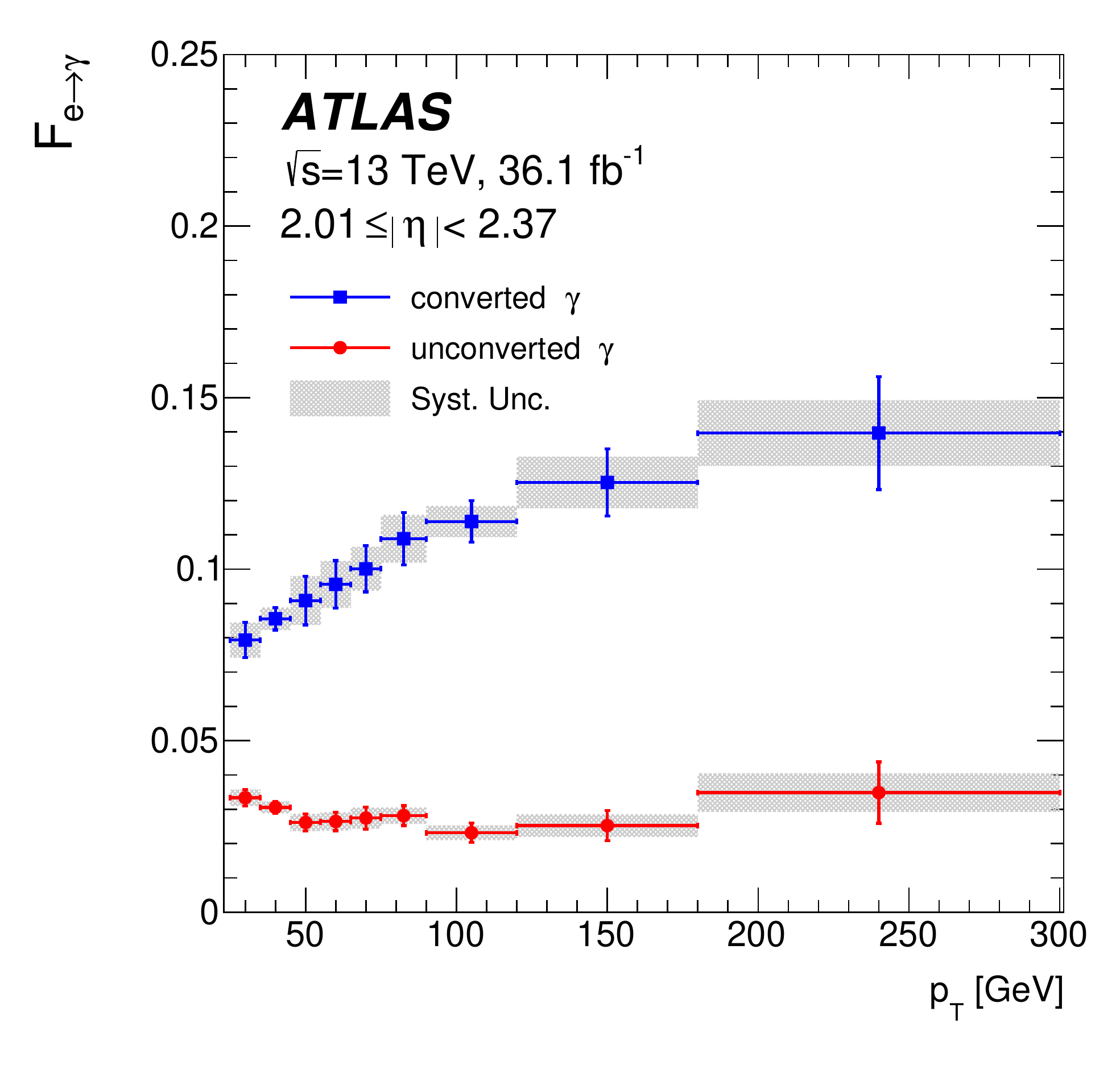}
  \hspace{.49\textwidth}
  \caption{Evolution of the $e\to\gamma$ fake-rate as a function of
    $\pt$ for electrons reconstructed as converted (blue squares) and
    unconverted (red circles) photons in the $|\eta|$ bins 0.0--0.6,
    1.15--1.37 and 2.01--2.37.  Symbols with error bars correspond to
    the results with their total uncertainties, and gray shaded areas
    correspond to systematic uncertainties.}
  \label{fig:EGammaFakeRate_vs_pt}
\end{figure}

\begin{figure}
  \centering
  \includegraphics[width=.48\textwidth]{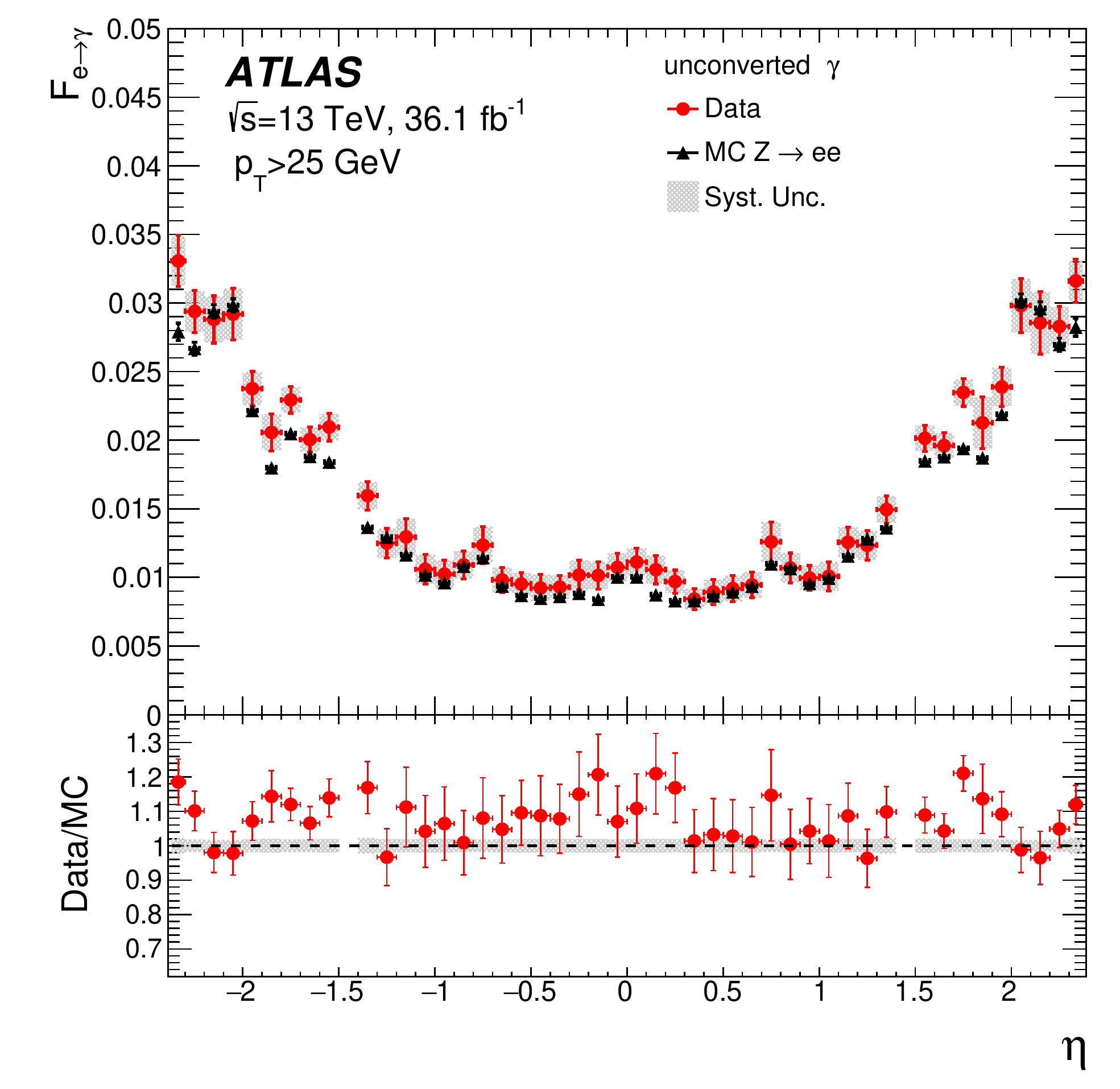}
  \includegraphics[width=.48\textwidth]{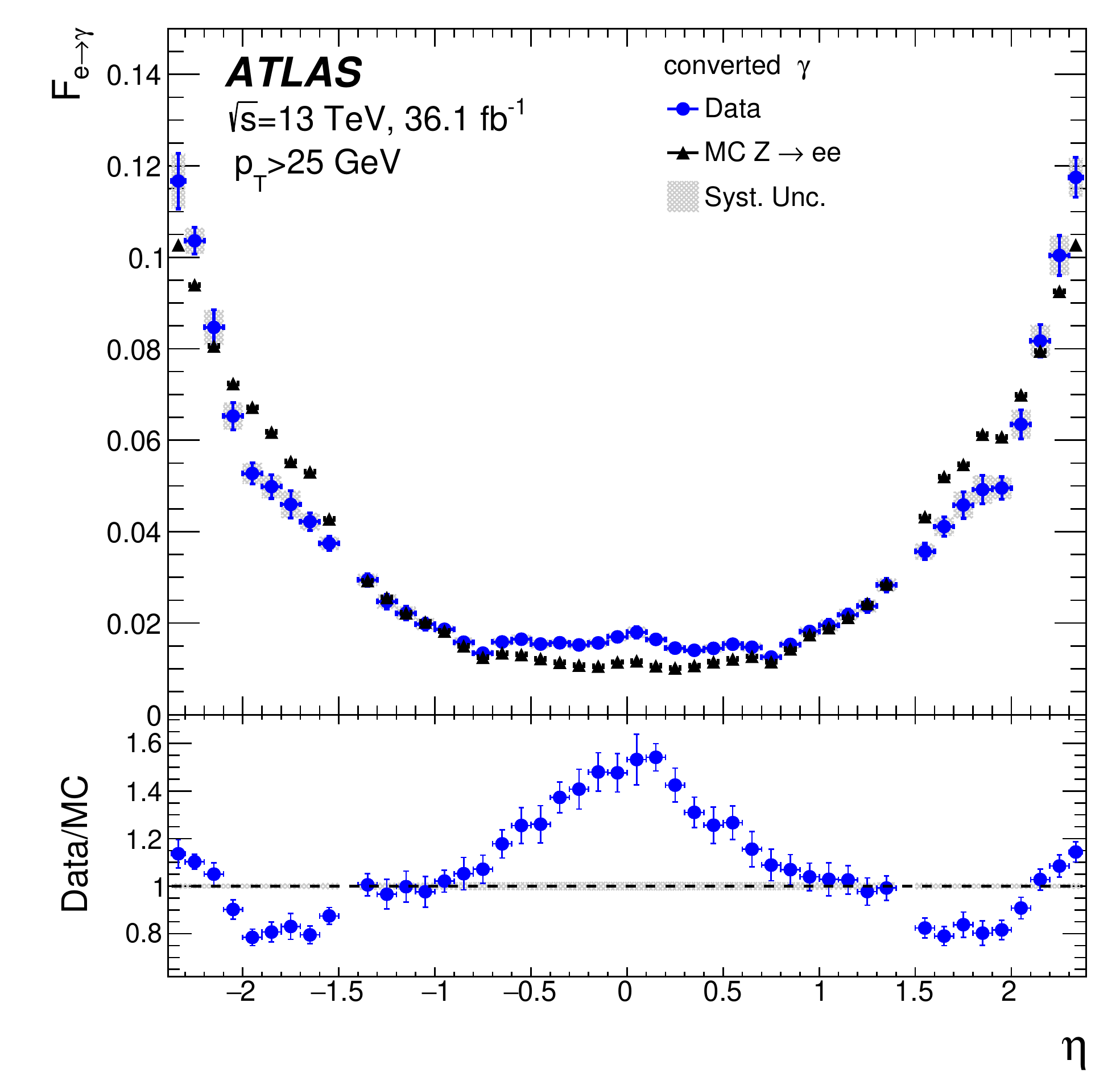}
  \caption{Evolution of the $e\to\gamma$ fake-rate as a function of
    $\eta$ for electrons reconstructed as unconverted (left) and
    converted (right) photon candidates, in data (circles) and $Z\rightarrow ee$
    MC events (black triangles).  Symbols with error bars correspond to the results with
    their total uncertainties, and gray shaded areas correspond to
    systematic uncertainties.  The bottom plot shows the ratio of the
    two measurements. The error bars correspond to the scaled data
    uncertainty and the gray dashed area corresponds to the relative
    MC uncertainty.}
  \label{fig:EGammaFakeRate_results}
\end{figure}

\section{Measurement of the conversion reconstruction efficiency}
\label{sec:ConvRecoEff}

The photon identification selections described in Section~\ref{sec:PhotonID} are optimised separately for unconverted and converted photons, since differences in electromagnetic shower development between the two types lead to different distributions for the discriminating variables listed in Table~\ref{tab:IsEM-DV}. Differences between the true conversion status of a photon and the one determined in reconstruction, described in Section~\ref{sec:PhotonReco}, can thus lead to the use of a suboptimal selection, if they  are not properly described in the simulation samples used for the optimization.

Photon conversion reconstruction efficiencies are therefore measured in data using a clean photon sample obtained from $Z \rightarrow \mu\mu\gamma$ decays. The selection closely follows the one used for the determination of photon identification efficiency using radiative $Z$ events, described in Section~\ref{sec:measurements:RadZ}, but requires in addition the \textit{legacy} photon isolation requirement (see Section~\ref{sec:PhotonIso}). 

Photon conversions are probed through the ratio $E_1/E_2$ of energy deposited in the first and second layers of the EMC, which has generally lower values for the later-developing showers of unconverted photons than for the electrons resulting from conversions. Binned fits in $E_1/E_2$ are performed using templates describing true unconverted photons, true converted photons and background. The fits are performed simultaneously in samples of reconstructed unconverted and converted photon samples, and split into the same four $|\etacal|$ bins as for the photon identification measurements. Signal shapes are obtained from simulated $Z \rightarrow \mu\mu\gamma$ events, requiring the presence of a true converted or unconverted photon. Background levels are estimated by performing in each $E_1/E_2$ bin a template fit to the distribution of the invariant mass $m_{\mu\mu\gamma}$ of the $\mu\mu\gamma$ system, over the range $76 < m_{\mu\mu\gamma} < 106\,\gev$. The signal shape is obtained from $Z \rightarrow \mu\mu\gamma$ simulation, and modelled by the sum of a Crystal Ball function~\cite{Gaiser:1982yw} and a Gaussian distribution. The background is modelled using a second-order Chebychev polynomial, which is found to describe well the $m_{\mu\mu\gamma}$ shape in simulated $Z \rightarrow \mu\mu$ + jets production. The resulting background yield in each $E_1/E_2$ bin is used to describe the background shape in the $E_1/E_2$ template fit. Figure~\ref{fig:conv_templates} shows the result of the $E_1/E_2$ fit for converted photons in the region $|\etacal| < 0.6$, and the $m_{\mu\mu\gamma}$ fit used to obtain the background yield in the bin $E_1/E_2 < 0.2$ for the same photon selection.

\begin{figure}
  \centering
  \includegraphics[width=0.48\textwidth]{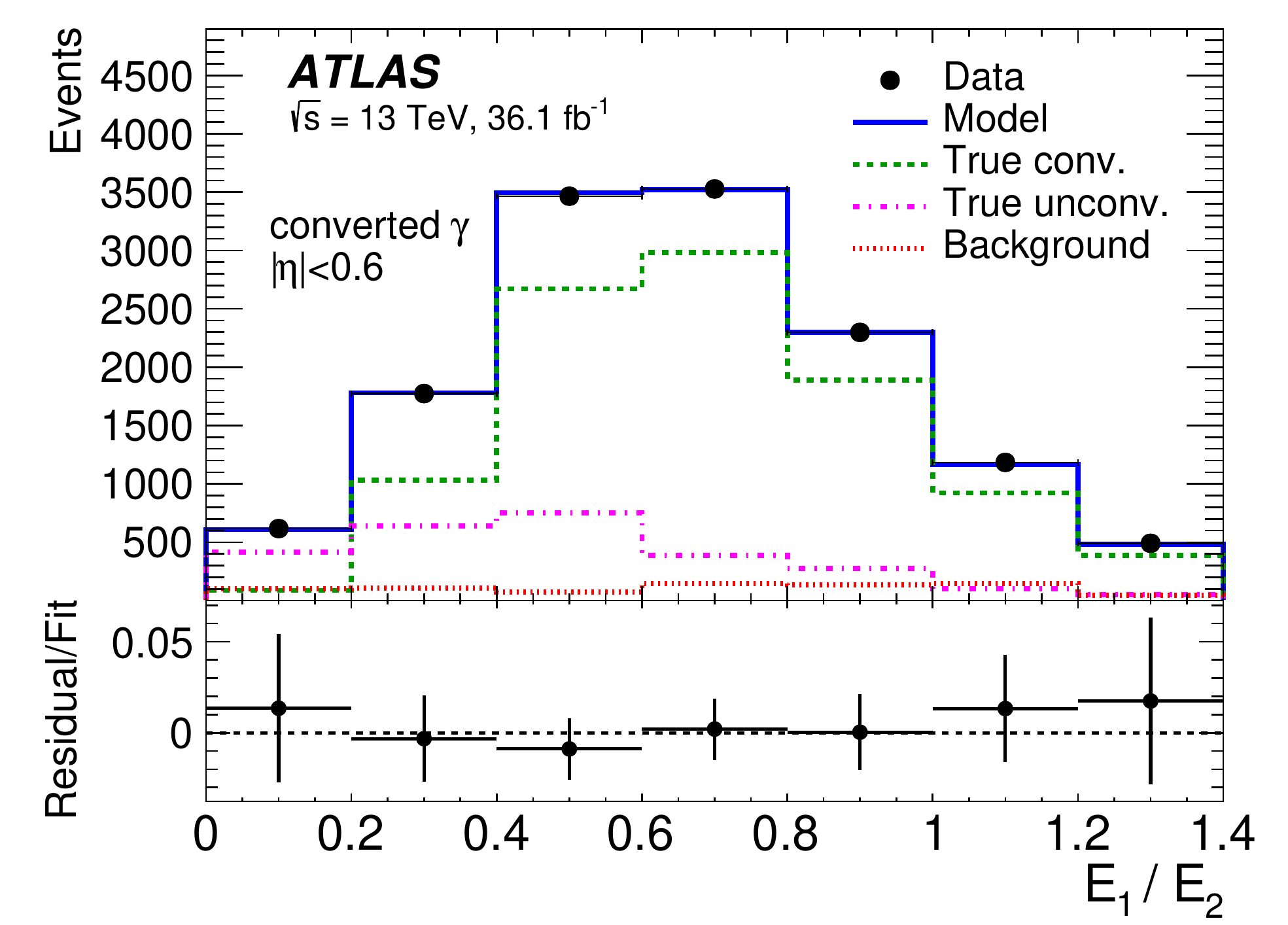}
  \includegraphics[width=0.48\textwidth]{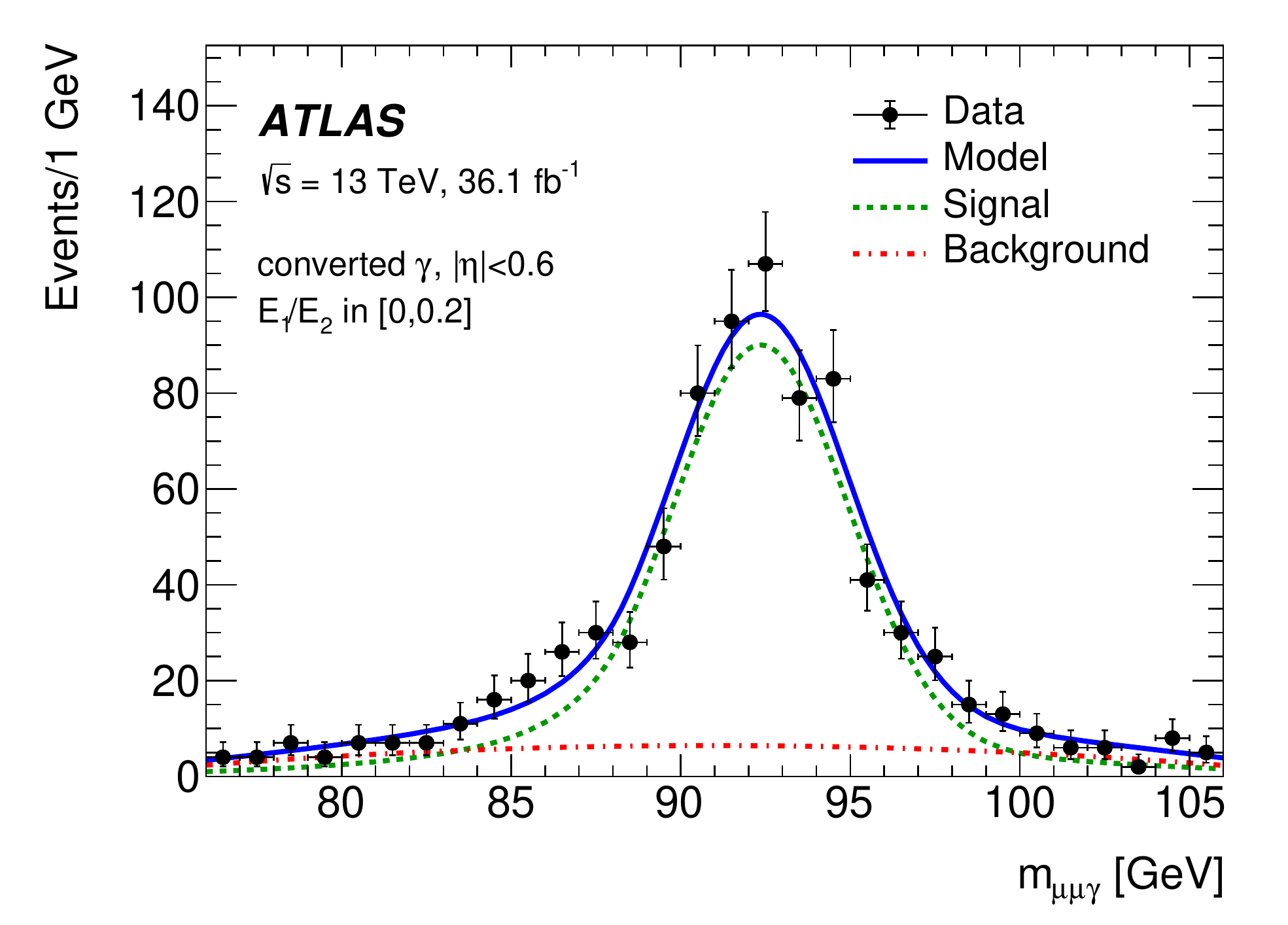}
  \hspace{0.5cm}
  \caption{Left: Distribution of $E_1/E_2$ for converted photons with $|\etacal| < 0.6$ in data (black points), for true converted photons (dashed green line), unconverted photons (dashed pink line), and background (dashed red line). The full model after the fit to data is also shown (solid blue line). Right: Distribution of $m_{\mu\mu\gamma}$ for converted photons with $|\etacal| < 0.6$ in data (black points), together with the model fit to data (blue line) and its signal (dashed green line) and background (dashed red line) components.}
  \label{fig:conv_templates}
\end{figure}

The true unconverted and converted photon yields in each bin are expressed in terms
of the fraction $f_{\text{total}}^{\text{conv}}$ of all photons that are reconstructed as converted,
the fraction $f_{\text{fake}}^{\text{conv}}$ of true unconverted photons that are reconstructed as converted,
and the fraction $f_{\text{reco}}^{\text{conv}}$ of true converted photons that are reconstructed as converted,
Results are shown in Table~\ref{tab:conv_results} in bins of $|\etacal|$.
\begin{table}[tp]
  \centering
  \caption{Results from data and MC simulation
    of the $f_{\text{total}}^{\text{conv}}$ fraction of all photons that are reconstructed as converted,
    of the $f_{\text{fake}}^{\text{conv}}$ fraction of true unconverted photons that are reconstructed as converted,
    and of the $f_{\text{reco}}^{\text{conv}}$ probability for a true converted photon to be reconstructed as converted,
    in bins of $|\etacal|$. }
  \begin{tabular}{l|cc|cc|cc}
  \hline \hline
  \multirow{2}{*}{$|\etacal|$ region}  & \multicolumn{2}{c|}{$f_{\text{total}}^{\text{conv}}$} & \multicolumn{2}{c|}{$f_{\text{fake}}^{\text{conv}}$} & \multicolumn{2}{c}{$f_{\text{reco}}^{\text{conv}}$} \\
  & data & MC  & data & MC & data & MC  \\
  \hline \hline
       $|\etacal| < 0.6$  & $0.215\pm0.014$ & 0.185 & $0.053\pm0.007$ & 0.053 & $0.731\pm0.040$ & 0.882 \\
 $0.6  \leq  |\etacal| < 1.37$ & $0.309\pm0.021$ & 0.328 & $0.036\pm0.007$ & 0.029 & $0.708\pm0.043$ & 0.778 \\
$1.52  \leq  |\etacal| < 1.81$ & $0.438\pm0.031$ & 0.439 & $0.001\pm0.009$ & 0.027 & $0.812\pm0.052$ & 0.907 \\
$1.81  \leq  |\etacal| < 2.37$ & $0.536\pm0.014$ & 0.521 & $0.003\pm0.006$ & 0.008 & $0.544\pm0.014$ & 0.614 \\
  \hline \hline
  \end{tabular}
  \label{tab:conv_results}
\end{table}
Systematic uncertainties include uncertainties in the signal shapes from variations in the amount of material upstream of the calorimeter (following the method described in Section~\ref{sec:measurements:RadZ}) and from the use of an analytical function to describe the signal shapes, as well as statistical uncertainties in the simulation sample. Their combined impact on the measurement of $f_{\text{reco}}^{\text{conv}}$ ranges from $1.3\%$ to $4.9\%$. depending on the $\eta$ bin. Statistical uncertainties range from $0.6\%$ to $1.5\%$.

\section{Conclusion}
\label{sec:Conclusion}

The reconstruction of photon candidates measured in the ATLAS detector is improved in several ways for the 2015 and 2016 Run 2 data-taking periods at the LHC with $pp$ collisions at $\sqrt{s}$ = 13 \TeV, building on what was implemented for Run 1. A more robust reconstruction of photon conversions is introduced to cope with the higher pile-up conditions. The photon identification criteria, based on rectangular cuts on calorimetric variables, are also tuned to reduce the dependence of the identification efficiency on pile-up.

The efficiency of the \emph{tight} identification criteria is measured from from 36.1 fb$^{-1}$ to 36.7 fb$^{-1}$ of $pp$ collision data for isolated photon candidates, using three independent methods in different photon \ET\ ranges. The three measurements agree within their uncertainties in the overlapping \ET\ ranges, and are combined. The identification efficiency increases from 45–60\% at \ET\ = 10 GeV to 95–98\% for \ET\ > 100 \GeV, depending on the pseudorapidity region of the detector and on the photon candidate conversion status. The nominal MC simulation of prompt-photons in ATLAS predicts significantly higher identification efficiency values than those measured in some regions of the phase space, particularly at low \ET\ and high $|\eta|$. A simulation with shower shapes corrected for the average shifts observed relative to the data better reproduces the identification efficiencies measured in the entire \ET\ and $\eta$ range accessible by the data-driven methods. The residual difference between the efficiencies in data and in the corrected simulation are taken into account by computing data-to-MC efficiency scale factors. These factors differ from one by up to 3--4\% at \ET\ = 10 \GeV\ and at most by 1--2\% above \ET\ = 40 \GeV, with an uncertainty decreasing from about 10\% at \ET\ = 10 \GeV\ to <1\%--2\% at high \ET. The dependence of the identification efficiency on pile-up is measured and found to be well described by simulation. The dependence of the identification efficiency on the photon isolation criterion is similarly measured, and found to be negligible compared with the measurement uncertainties.

The probability of reconstructing an electron as a photon candidate is also measured in data, and compared with the predictions from simulation. It is found to range from a few \% in the central region and increase in the more forward ones, reaching a value of 10--20\% for converted photons in the most extreme cases. The MC simulation generally underestimates the fake rate by values ranging from a few \% to about 50\% in the most extreme case (converted photons, central region). 

The efficiency of the reconstruction of photon conversions is measured using a sample of photon candidates from $Z\to\mu\mu\gamma$ events, exploiting the properties of the ratio of energies deposited in the first and second longitudinal layers of the ATLAS electromagnetic calorimeter. The efficiency is found to be compatible with the prediction of the MC simulation.

\section*{Acknowledgements}

We thank CERN for the very successful operation of the LHC, as well as the
support staff from our institutions without whom ATLAS could not be
operated efficiently.

We acknowledge the support of ANPCyT, Argentina; YerPhI, Armenia; ARC, Australia; BMWFW and FWF, Austria; ANAS, Azerbaijan; SSTC, Belarus; CNPq and FAPESP, Brazil; NSERC, NRC and CFI, Canada; CERN; CONICYT, Chile; CAS, MOST and NSFC, China; COLCIENCIAS, Colombia; MSMT CR, MPO CR and VSC CR, Czech Republic; DNRF and DNSRC, Denmark; IN2P3-CNRS, CEA-DRF/IRFU, France; SRNSFG, Georgia; BMBF, HGF, and MPG, Germany; GSRT, Greece; RGC, Hong Kong SAR, China; ISF and Benoziyo Center, Israel; INFN, Italy; MEXT and JSPS, Japan; CNRST, Morocco; NWO, Netherlands; RCN, Norway; MNiSW and NCN, Poland; FCT, Portugal; MNE/IFA, Romania; MES of Russia and NRC KI, Russian Federation; JINR; MESTD, Serbia; MSSR, Slovakia; ARRS and MIZ\v{S}, Slovenia; DST/NRF, South Africa; MINECO, Spain; SRC and Wallenberg Foundation, Sweden; SERI, SNSF and Cantons of Bern and Geneva, Switzerland; MOST, Taiwan; TAEK, Turkey; STFC, United Kingdom; DOE and NSF, United States of America. In addition, individual groups and members have received support from BCKDF, CANARIE, CRC and Compute Canada, Canada; COST, ERC, ERDF, Horizon 2020, and Marie Sk{\l}odowska-Curie Actions, European Union; Investissements d' Avenir Labex and Idex, ANR, France; DFG and AvH Foundation, Germany; Herakleitos, Thales and Aristeia programmes co-financed by EU-ESF and the Greek NSRF, Greece; BSF-NSF and GIF, Israel; CERCA Programme Generalitat de Catalunya, Spain; The Royal Society and Leverhulme Trust, United Kingdom. 

The crucial computing support from all WLCG partners is acknowledged gratefully, in particular from CERN, the ATLAS Tier-1 facilities at TRIUMF (Canada), NDGF (Denmark, Norway, Sweden), CC-IN2P3 (France), KIT/GridKA (Germany), INFN-CNAF (Italy), NL-T1 (Netherlands), PIC (Spain), ASGC (Taiwan), RAL (UK) and BNL (USA), the Tier-2 facilities worldwide and large non-WLCG resource providers. Major contributors of computing resources are listed in Ref.~\cite{ATL-GEN-PUB-2016-002}.

\printbibliography

\clearpage 
 
\begin{flushleft}
{\Large The ATLAS Collaboration}

\bigskip

M.~Aaboud$^\textrm{\scriptsize 34d}$,    
G.~Aad$^\textrm{\scriptsize 99}$,    
B.~Abbott$^\textrm{\scriptsize 125}$,    
O.~Abdinov$^\textrm{\scriptsize 13,*}$,    
B.~Abeloos$^\textrm{\scriptsize 129}$,    
D.K.~Abhayasinghe$^\textrm{\scriptsize 91}$,    
S.H.~Abidi$^\textrm{\scriptsize 164}$,    
O.S.~AbouZeid$^\textrm{\scriptsize 39}$,    
N.L.~Abraham$^\textrm{\scriptsize 153}$,    
H.~Abramowicz$^\textrm{\scriptsize 158}$
% ...
\bigskip
\\

% ...
$^{13}$Institute of Physics, Azerbaijan Academy of Sciences, Baku; Azerbaijan.\\
% ...
$^{34}$$^{(a)}$Facult\'e des Sciences Ain Chock, R\'eseau Universitaire de Physique des Hautes Energies - Universit\'e Hassan II, Casablanca;$^{(b)}$Centre National de l'Energie des Sciences Techniques Nucleaires (CNESTEN), Rabat;$^{(c)}$Facult\'e des Sciences Semlalia, Universit\'e Cadi Ayyad, LPHEA-Marrakech;$^{(d)}$Facult\'e des Sciences, Universit\'e Mohamed Premier and LPTPM, Oujda;$^{(e)}$Facult\'e des sciences, Universit\'e Mohammed V, Rabat; Morocco.\\
% ...
$^{39}$Niels Bohr Institute, University of Copenhagen, Copenhagen; Denmark.\\
% ...
$^{91}$Department of Physics, Royal Holloway University of London, Egham; United Kingdom.\\
% ...
$^{99}$CPPM, Aix-Marseille Universit\'e, CNRS/IN2P3, Marseille; France.\\
% ...
$^{125}$Homer L. Dodge Department of Physics and Astronomy, University of Oklahoma, Norman OK; United States of America.\\
% ...
$^{129}$LAL, Universit\'e Paris-Sud, CNRS/IN2P3, Universit\'e Paris-Saclay, Orsay; France.\\
% ...
$^{153}$Department of Physics and Astronomy, University of Sussex, Brighton; United Kingdom.\\
% ...
$^{158}$Raymond and Beverly Sackler School of Physics and Astronomy, Tel Aviv University, Tel Aviv; Israel.\\
% ...
$^{164}$Department of Physics, University of Toronto, Toronto ON; Canada.\\
% ...
$^{a}$ Also at Borough of Manhattan Community College, City University of New York, NY; United States of America.\\
$^{b}$ Also at California State University, East Bay; United States of America.\\
$^{c}$ Also at Centre for High Performance Computing, CSIR Campus, Rosebank, Cape Town; South Africa.\\
$^{d}$ Also at CERN, Geneva; Switzerland.\\
$^{e}$ Also at CPPM, Aix-Marseille Universit\'e, CNRS/IN2P3, Marseille; France.\\
% ...
$^{*}$ Deceased

\end{flushleft}



\end{document}